\documentclass[a4paper]{JHEP3}

\usepackage{graphicx}
\usepackage{amsmath}
\usepackage{cite}
\usepackage{multirow}
\usepackage{array}
\usepackage{color}
\usepackage{caption}
\usepackage{subcaption}
\allowdisplaybreaks 
\let\normalcolor\relax

\def \stop{\tilde t_1}
\def \nn{\nonumber \\}
\newcommand{\Order}[2]{\mathcal{O}\left(\alpha_s^{#1}\, \alpha^{#2} \right)}
\newcommand{\psec}[3]{\sigma^{\mbox{\tiny \,#1,\,#2}}_{#3}}
\newcommand{\xsec}[2]{\sigma^{\mbox{\tiny \,#1}}_{#2}}

\title{
Top-squark pair production at the LHC: 
a complete  analysis at next-to-leading order
}

\author{Jan Germer\thanks{Now at PR\"UFTECHNIK Condition Monitoring GmbH.}  \\
Max-Planck-Institut f\"ur Physik, F\"ohringer Ring 6, D-80805 M\"unchen, Germany; \\
Email: \email{germer@mpp.mpg.de}}

\author{Wolfgang Hollik  \\
Max-Planck-Institut f\"ur Physik, F\"ohringer Ring 6, D-80805 M\"unchen, Germany; \\
Email: \email{hollik@mpp.mpg.de}}

\author{Jonas M. Lindert \\ 
Physik-Institut, Universit\"at Z\"urich,
Wintherturerstrasse 190, CH-8057 Z\"urich,
Switzerland; 
E-mail: \email{lindert@physik.uzh.ch}
}    

\author{Edoardo Mirabella \\ 
Max-Planck-Institut f\"ur Physik, F\"ohringer Ring 6, D-80805 M\"unchen, Germany; \\
E-mail: \email{mirabell@mpp.mpg.de}
}    

\preprint{MPP-2014-109, \; ZU-TH 15/14, \;
LPN14-064}

\abstract{ 
We present a complete next-to-leading order study of top-squark pair production at the LHC, including QCD and EW corrections. The calculation is performed within the Minimal Supersymmetric Standard Model and numerical results are presented for parameter regions compatible with the observed Higgs boson.
We employ the most recent parton distribution functions  including QED corrections and we find NLO EW corrections to the inclusive stop-pair production cross section up to $25 - 30\%$ compared to the leading-order prediction. Besides corrections to inclusive cross sections also important kinematic distributions are investigated.
}

\begin{document}

\section{Introduction}
The discovery of a signal in the Higgs boson searches at the
LHC~\cite{Aad:2012tfa,Chatrchyan:2012ufa} has triggered 
extensive studies of the properties of this particle, in particular of
its mass,  couplings and  spin~\cite{LHCHiggsCrossSectionWorkingGroup:2012nn,Heinemeyer:2013tqa}. 
The experimental results  are in agreement with Standard Model (SM) predictions. 
Nevertheless, experimental evidence for  beyond the Standard Model
(BSM) physics may still be found in the future run, 
 in  particular for the minimal supersymmetric extension of the Standard Model (MSSM), 
where the value of the Higgs boson  mass can naturally be explained.

The MSSM is an appealing BSM scenario allowing for precise
quantitative predictions which can be tested in the LHC experiments.
To date,   the searches at the LHC have investigated various final states and signatures~\cite{AtlasPage,CmsPage},
without  finding deviations from the SM,  
thus  setting limits in several regions of the MSSM parameter space.   
Such searches are based on 
direct production of supersymmetric (SUSY)  partners of the SM particles, 
in particular on processes producing  
strongly-interacting SUSY particles.   
The importance of these production channels is reflected in the extensive efforts spent in 
improving their theoretical predictions, including not only   the leading-order (LO) 
contributions~\cite{Kane:1982hw,Harrison:1982yi,Reya:1984yz,Dawson:1983fw,Baer:1985xz},
 but also  the  next-to-leading-order (NLO) QCD~\cite{Beenakker:1994an,Beenakker:1995fp,Beenakker:1996ch, Beenakker:1997ut,GoncalvesNetto:2012yt}, 
 the tree-level  electroweak 
 (EW)~\cite{Bozzi:2005sy,Alan:2007rp,Hollik:2007wf,Bornhauser:2007bf,Hollik:2008vm,Bornhauser:2009ru,Arhrib:2009sb,Germer:2010vn}
 and the 
 $\mathcal{O}(\alpha^2_s \alpha)$~\cite{Hollik:2007wf,Beccaria:2008mi,Hollik:2008yi,Hollik:2008vm,Mirabella:2009ap,Germer:2010vn,Germer:2011an}
corrections to the cross sections.
The large radiative corrections in the threshold region have been  resummed at the next-to-leading logarithmic  
(NLL) accuracy~\cite{Kulesza:2008jb,Kulesza:2009kq,Beenakker:2009ha,Beenakker:2010nq,Beenakker:2011fu}, and
resummation of soft-gluon and Coulomb corrections has been performed within an effective field theory 
approach~\cite{Beneke:2009nr,Beneke:2010da,Falgari:2012hx,Falgari:2012sq}.
Resummations at the next-to-next-to-leading logarithmic  (NNLL) accuracy have been performed as 
well~\cite{Beenakker:2011sf,  Broggio:2013cia,Pfoh:2013iia, Beenakker:2013mva, Beenakker:2014sma}.
Gluino~\cite{Kauth:2009ud}  and stop~\cite{Kim:2014yaa} bound states
as well as  bound-state effects in  gluino-gluino and squark-gluino 
production~\cite{Hagiwara:2009hq,Kauth:2011vg,Kauth:2011bz} have been  studied.
Approximate  next-to-next-to-leading-order (NNLO) predictions for quark--anti-squark production
are available~\cite{Langenfeld:2009eg,Langenfeld:2010vu, Langenfeld:2012ti,Broggio:2013uba}. 
The  decay rate of squarks is known at NLO, including both the QCD~\cite{Beenakker:1996dw, Djouadi:1996wt,Beenakker:1996de} and the EW corrections~\cite{Guasch:2001kz,Guasch:2002ez}. The decay 
rate of the gluino  in a quark-squark pair  is known at NLO QCD as well~\cite{Beenakker:1996dw,Beenakker:1996de}.
Furthermore, phenomenological studies have been performed including systematically NLO QCD corrections to the production and the subsequent decay of stop--anti-stop~\cite{Boughezal:2012zb,Boughezal:2013pja} 
and squark--squark pairs~\cite{Hollik:2012rc,Hollik:2013xwa}.  
In the case of squark--squark production, the NLO QCD corrections have
been matched with parton showers in Ref.~\cite{Gavin:2013kga}.

\begin{figure}[t]
\centering
\includegraphics[width=7.4cm,height=6.3cm]{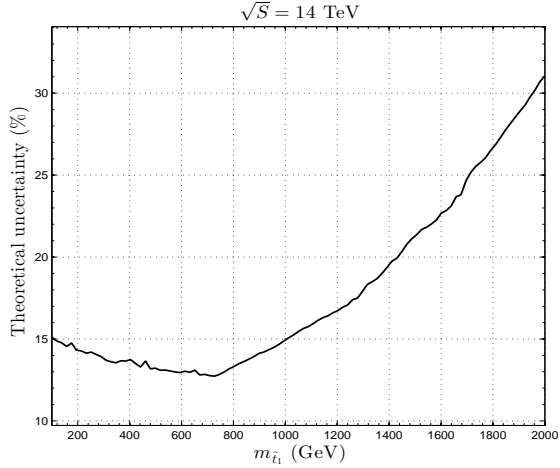}
\caption[.]{Theoretical uncertainty  on the total cross section  for $\stop \tilde t_1^\ast$ production at  the LHC. It has been computed by using the code
{\tt NLL-fast}~\cite{Beenakker:2011fu} and the procedure described in Ref.~\cite{Kramer:2012bx}.
}
\label{fig:Error14}
 \end{figure}

Experimentally the production of third generation squarks is different from those of the other generations. 
Final-state bottom and top quarks from the decays of the squarks can be distinguished from their light-flavor counterparts.  
This leads to characteristic signatures which are extensively used  
in the experimental searches for SUSY at the LHC~\cite{Aad:2014mha,Aad:2014qaa,Aad:2013ija,Chatrchyan:2013mya,Chatrchyan:2013xna,CMS-PAS-SUS-13-004}. 
These searches assume a dominant decay channel and put limits on the mass of the light stop, $\stop$, and of the decay products. For instance
the analysis in Ref.~\cite{CMS-PAS-SUS-13-004} puts a lower bound on the $\stop$ one, $m_{\stop}\geq 750$ GeV, assuming  $100 \%$
branching ratio into a massless neutralino $\tilde\chi^0_1$.\footnote{This bound does not hold if 
$m_{\stop}  - m_{\tilde \chi^0_1} \simeq m_t, m_{\rm W}+m_b$. These regions 
in the parameter space are not yet  covered by the experimental searches, due to the large  $t\bar t$ background.}  
In the experimental analyses signal events are usually produced at LO and the total cross section 
is normalized to NLO QCD + NLL accuracy using the codes {\tt
  Prospino}~\cite{Beenakker:1996ed} and {\tt  NLL-fast}~\cite{Beenakker:2011fu}, 
along the lines of Ref.~\cite{Kramer:2012bx}. The theoretical
uncertainty on the generated signal is obtained by varying the parton distribution function (PDF) sets 
and the renormalization/factorization scale. For a hadronic
center-of-mass  energy $\sqrt{S}= 14$~TeV,  the obtained uncertainty is depicted in Fig.~\ref{fig:Error14}.  It is  
below $15 \%$ ($20 \%$) for masses  below $1000$ (1400)~GeV.  It is worth to mention that this procedure
does not provide an accurate estimation  of the  impact of the higher order contributions. Indeed, the latter 
may depend strongly on the kinematics of the final state changing  the shapes of differential  distributions.
Higher-order contributions to the decays of the produced particles can have 
substantial effects as well~\cite{Boughezal:2012zb,Boughezal:2013pja,Hollik:2012rc,Hollik:2013xwa}.

In this paper we analyze the numerical impact of the NLO EW corrections to the total cross section of the hadronic 
pair-production of the lightest top-squark $\tilde t_1$,
\begin{align}
p \, p \, \to  \, \tilde t_1 \,  \tilde t_1^\ast \, , 
\end{align}
by presenting the first phenomenological study combining NLO QCD and NLO EW corrections.   
A similar study of the NLO EW corrections has already been published in~\cite{Hollik:2007wf}.
There, results were presented for a handful of Snowmass points~\cite{Allanach:2002nj} and for benchmark slopes 
passing along the SPS1a$'$ scenario~\cite{AguilarSaavedra:2005pw}. 
NLO EW corrections to  inclusive cross sections were found to be small, below $5\%$, 
while  differential distributions were found to be affected by up to $20\%$ in their tails. 
In view of the developments over the recent years,
updated and improved analyses are considered appropriate.    
Firstly, the previous study focuses purely on the EW corrections without  including the (dominant) NLO QCD corrections. 
Only a consistent treatment of NLO QCD and NLO EW including common PDFs allows for a quantitative comparison.  
Secondly, in the MSSM the mass of the lightest (SM-like) Higgs boson is not a free parameter, but
depends crucially on the top-squark masses  and the mixing parameter $X_t=A_t - \mu / \tan\beta$.  
Consequently, the experimental 
signal for a Higgs boson with a mass around $125$~GeV 
yields a strong constraint for the scalar-top sector
that has to be taken into account in any phenomenological studies. 
In general, the Higgs-mass constraint requires heavy stops and/or large
mixing in the stop sector.  
Finally, a consistent computation at NLO EW accuracy requires
a PDF set which includes both the photon  structure function and the 
QED contributions to the evolution  equations. When the study in Ref.~\cite{Hollik:2007wf} was performed, 
the {\tt MRST2004QED} set~\cite{Martin:2004dh} was the only one fulfilling this requirement.
This set is affected by large uncertainties that have been significantly reduced in the 
newly available  {\tt NNPDF2.3QED} PDF set~\cite{Ball:2013hta}. Here, latest data from deep inelastic
scattering and from the LHC are included. It is worth to investigate how the new PDF 
alters the EW corrections to the production of the lightest top-squark.

Our study mainly focuses on the NLO corrections to the  inclusive stop--anti-stop production cross section, including 
corrections of electroweak origin. This analysis can be directly used in experimental analyses to estimate the theoretical uncertainty related to the missing EW corrections. Besides corrections to inclusive cross sections we also consider the impact of the electroweak corrections on several kinematic distributions, which may be significant in specific SUSY scenarios~\cite{Hollik:2007wf}.

%
%

This paper is structured as follows. Section~\ref{sec:comp} reviews the calculation of the total cross section
of stop--anti-stop  pair production,  including both NLO QCD and NLO EW contributions.
Section~\ref{sec:numerics} is dedicated to the presentation of the numerical results, 
followed by our conclusions in Section~\ref{sec:conclusions}.

\section{Computational details}
\label{sec:comp}
The leading-order contribution to the hadronic cross section for the production 
of stop--anti-stop pairs  is of the $\mathcal{O}\left  (\alpha_s^2\right )$,
described as follows,
\begin{align}
\xsec{LO}{pp \, \to \,  \stop \stop^\ast}  = \psec{2}{0}{gg \, \to \,  \stop \stop^\ast} + \sum_q \;  \psec{2}{0}{q \bar q \, \to \,  \stop \stop^\ast} \, ,
\label{Eq:LO}
\end{align}
indicating the perturbative order $\Order{a}{b}$  of the
partonic processes $X$ contributing 
to the total hadronic cross section as $\sigma^{a,b}_{X}$. The sum in Eq.~(\ref{Eq:LO}) 
runs over the  four lightest  quark flavors, $q= u,d,c,s$. 
We neglect the bottom-initiated partonic processes, since they are suppressed by the bottom-quark parton distribution function and turn out to 
be small~\cite{Beenakker:2010nq, Germer:2011an}.\footnote{Large contributions from the
  $b\bar{b}$ channel quoted in Ref.~\cite{Arhrib:2009sb} are a
  consequence of resonant Higgs-boson 
  exchange in the $s$-channel,  as well as 
  Higgs--$b\bar{b}$ Yukawa couplings enhanced by the choice of a
  negative value for the parameter $\mu$. In our analyses we do not consider resonant  Higgs bosons 
  or negative values of $\mu$, which are in general
  disfavored by the measured value of the anomalous magnetic moment of
  the muon~\cite{Stockinger:2006zn}. } 
  The partonic cross sections for the $gg$  and $q \bar q$ channel can be found  in Eq.~(3) and Eq.~(4) of \cite{Beenakker:1997ut}, respectively.  They  depend 
  only on the mass of the produced top-squarks.
\medskip
 
The NLO  QCD contributions, of  $\mathcal{O}\left (\alpha_s^3 \right)$,  are given by
\begin{align}
\label{Eq:dQCD}
\Delta \xsec{QCD}{pp \, \to \,  \stop \stop^\ast} \, = 
& \, \psec{3}{0}{gg \, \to \,  \stop \stop^\ast}  + \psec{3}{0}{gg \, 
  \to \,  \stop \stop^\ast g}     \\
 & +\, \sum_q \; \Big (   \psec{3}{0}{q \bar q \, \to \,  \stop \stop^\ast}  +  \psec{3}{0}{q \bar q \, \to \,  \stop \stop^\ast g}  
   +   \psec{3}{0}{q g \, \to \,  \stop \stop^\ast q} +
 \psec{3}{0}{\bar q g \, \to \,  \stop \stop^\ast \bar q}   \Big ) \, . \nonumber
\end{align}
They have been  computed in Ref.~\cite{Beenakker:1997ut} and are implemented in the public code {\tt Prospino}~\cite{Beenakker:1996ed}.  
The NLO QCD corrections depend on the 
masses of the top squarks and of the gluino, and on the stop mixing angle.

\medskip

The EW contributions arise at
$\mathcal{O}\left (\alpha^2 \right )$, $\mathcal{O}\left (\alpha_s  \alpha \right )$ 
and  $\mathcal{O}\left (\alpha^2_s \alpha \right )$; they
can be separated into four different channels according to the partonic initial states,
\begin{align}
\Delta \xsec{EW}{pp \, \to \,  \stop \stop^\ast}  &=     \Delta \sigma^{\mbox{\tiny EW }  gg}_{pp \, \to \,  \stop \stop^\ast} +  \Delta \sigma^{\mbox{\tiny EW } q \bar q}_{pp \, \to \,  \stop \stop^\ast}
+  \Delta \sigma^{\mbox{\tiny EW } q g}_{pp \, \to \,  \stop \stop^\ast} +   \Delta \sigma^{\mbox{\tiny EW }g \gamma}_{pp \, \to \,  \stop \stop^\ast} \, , 
\label{Eq:dEW}
\end{align}
with
\begin{align}
\Delta \sigma^{\mbox{\tiny EW }  gg}_{pp \, \to \,  \stop \stop^\ast}  &=  \;  \psec{2}{1}{gg \, \to \,  \stop \stop^\ast}  +  \psec{2}{1}{gg \, \to \,  \stop \stop^\ast \gamma} \, , \nn
 \Delta \sigma^{\mbox{\tiny EW } q \bar q}_{pp \, \to \,  \stop \stop^\ast} &= \sum_q \, \left ( \,     \psec{0}{2}{q \bar q \, \to \,  \stop \stop^\ast} +
   \psec{2}{1}{q \bar q \, \to \,  \stop \stop^\ast } + \psec{2}{1}{q \bar q \, \to \,  \stop \stop^\ast g}   +  \psec{2}{1}{q \bar q \, \to \,  \stop \stop^\ast \gamma} \, \right )   \, , \nn
 \Delta \sigma^{\mbox{\tiny EW } q g}_{pp \, \to \,  \stop \stop^\ast} &=  \sum_q \, \left ( \,     \psec{2}{1}{q g \, \to \,  \stop \stop^\ast q} +  \psec{2}{1}{\bar q g \, \to \,  \stop \stop^\ast \bar q} \, \right  )   \, , \nn
 \Delta \sigma^{\mbox{\tiny EW }g \gamma}_{pp \, \to \,  \stop \stop^\ast}   &=   \; \psec{1}{1}{g \gamma \, \to \,  \stop \stop^\ast}   \, .
\label{Eq:NLOEWchannels}
\end{align}
In principle these electroweak terms depend on the full set of MSSM parameters. 
We   compute them  by using {\tt  FeynArts}~\cite{Hahn:2000kx,Hahn:2006qw}
and {\tt FormCalc}~\cite{Hahn:2006qw,Hahn:2001rv}, together with 
{\tt LoopTools}~\cite{Hahn:2006qw} for  the numerical evaluation of
the one-loop integrals.
Ultraviolet divergences  are cancelled by renormalization 
at the electroweak one-loop level, along
the lines of Ref.~\cite{Germer:2011an}.
Infrared  and collinear singularities are handled by using mass regularization and are computed by 
using the double cut-off phase-space-slicing method~\cite{Yennie:1961ad,Weinberg:1965nx,Baier:1973ms}, 
as described in Ref.~\cite{Germer:2010vn}. The initial-state collinear singularities of gluonic 
(photonic) origin are factorized and absorbed in the parton distribution functions  by using  
the $\overline{\mbox{MS}}$ (DIS) scheme.  
Our computation has been  numerically checked  against the results presented  in~\cite{Hollik:2007wf}. 
The EW contributions 
have been implemented in the code {\tt SusyHell}, a (to be public) Monte Carlo integrator 
for the production of colored SUSY particles at the LHC. 

\medskip

The inclusive cross section for stop--anti-stop production, complete at NLO, 
is obtained by summing the LO cross section, 
Eq.~(\ref{Eq:LO}),  and the NLO contributions, Eqs.~(\ref{Eq:dQCD}) and~(\ref{Eq:dEW}).
For a discussion of the QCD and EW effects separately, it is convenient to define
the individual and summed cross-section parts as follows,
\begin{subequations}
\begin{align}
\xsec{NLO QCD}{pp \, \to \,  \stop \stop^\ast}  &= \xsec{LO}{pp \, \to \,  \stop \stop^\ast}  +\Delta \xsec{QCD}{pp \, \to \,  \stop \stop^\ast}   \label{Eq:NLOQCD} \, , \\[0.5ex]
\xsec{NLO EW}{pp \, \to \,  \stop \stop^\ast}  &= \xsec{LO}{pp \, \to \,  \stop \stop^\ast}  + \Delta \xsec{EW}{pp \, \to \,  \stop \stop^\ast}  \, ,  \label{Eq:NLOEW} \\[0.5ex]
\xsec{NLO}{pp \, \to \,  \stop \stop^\ast}  &= \xsec{LO}{pp \, \to \,  \stop \stop^\ast}  + \Delta \xsec{QCD}{pp \, \to \,  \stop \stop^\ast}  +\Delta \xsec{EW}{pp \, \to \,  \stop \stop^\ast}  \, ,
\end{align}
\label{Eq:NLO}
\end{subequations}
to be used in the numerical studies of the next section.

\section{Numerical analysis}
\label{sec:numerics}

The numerical values of the Standard Model input parameters are chosen according to 
\begin{align}
m_{\rm Z} &=  91.1876 \mbox{ GeV} \, , &  m_{\rm W} &=  80.425 \mbox{ GeV}  \, 
, \nn
m_t &=  173.2 \mbox{ GeV}          \, ,         &  m_{b}^{\overline{\rm 
MS}}(m_{\rm Z}) &=  2.94 \mbox{ GeV}  \, , \nn
\alpha^{-1} &= 137.036  \, , & \alpha_s(m_{{\rm Z}})  &= 0.119 \, .
\end{align}
The results presented in this section are computed for a
hadronic center-of-mass energy of $\sqrt{S} = 14$ TeV.   
For the numerical evaluation of the hadroinc cross sections, we use 
the  {\tt NNPDF2.3QED} PDF set~\cite{Ball:2013hta}.  The latter has been implemented 
in {\tt Prospino} and {\tt SusyHell} through the {\tt LHAPDF} 
interface~\cite{Bourilkov:2006cj}, which also automatically 
accounts for the consistent evolution of the strong coupling constant. The 
factorization scale $\mu_{\rm F}$ and the renormalization scale $\mu_{\rm R}$ are 
set to a common value, equal to the mass of the lightest top-squark,
 $\mu_{\rm F} = \mu_{\rm R} = m_{\tilde t_1}$.

In the following we consider various phenomenological MSSM 
scenarios characterized  by ten TeV-scale free parameters,
 \begin{align}
m_{A_0} \, ,     && 
\tan{\beta} \, ,  &&  
X_t  \, ,             &&  
\mu  \, ,            &&  
M_2  \, ,           && 
m_{\tilde g}  \, ,               &&
M_{\tilde q_{1,2}} \, ,     && 
M_{\tilde \ell_{1,2}}  \, , &&
M_{\tilde q _{ 3}} \, ,      &&   
M_{\tilde \ell_3}\,  , 
\label{Eq:Inputs}
 \end{align}
from which the individual soft-breaking parameters
are obtained in the following simplified way:
\begin{align}
M_{\tilde t,  \mbox {\tiny L}} &=   M_{\tilde t ,  \mbox {\tiny R}}  =   
M_{\tilde b ,  \mbox {\tiny R}} =  M_{\tilde q_3}  \, , &
M_{\tilde f,  \mbox {\tiny L}} &=   M_{\tilde f ,  \mbox {\tiny R}}  =  
M_{\tilde q_{1,2}} \, ,   &(f &= u,d,c,s)  \, ,  \nn
M_{\tilde \tau , \mbox {\tiny L}} &=   M_{\tilde \tau ,  \mbox {\tiny R}}  =   
M_{\tilde \nu_\tau ,  \mbox {\tiny R}} =  M_{\tilde \ell_3}    \, ,  &
M_{\tilde f,  \mbox {\tiny L}} &=   M_{\tilde f ,  \mbox {\tiny R}}   = 
M_{\tilde \ell_{1,2}} \, ,   &(f &= e,\mu ,\nu_e,\nu_\mu) \, ,  \nn 
M_1 & = \frac{5}{3}  \frac{   s^2_{\mbox{\tiny W}}  }{   c^2_{\mbox{\tiny W}}    
 } \, M_2 \, ,   &
%
A_f & =  X_t - \frac{\mu}{\tan \beta}  \, ,    &(f &= e,\mu,\tau,u,d,c,s,b,t) \,  ;
\end{align}
thereby,  for $M_1$ gaugino-mass unification at the GUT scale is assumed.
From these soft parameters we calculate the physical spectrum using 
tree-level relations. The only exception is the physical mass of 
the Higgs bosons, computed by using the code 
{\tt FeynHiggs 2.10}~\cite{Heinemeyer:1998np,Heinemeyer:1998yj,Frank:2006yh,Hahn:2009zz,Hahn:2013ria}. 


\subsection{Total cross section}
For our numerical evaluation we consider several benchmark scenarios as defined below.
Within these scenarios we perform various one-dimensional scans over 
the parameters listed in~(\ref{Eq:Inputs}).
In these scans we focus on regions allowed by limits from the
Higgs sector, i.e. we require a Higgs-state close to the observed one, with couplings
compatible with the observed rates in the Higgs search channels. 
The compatibility between the considered  models and the experimental results 
has been checked by using the 
codes  {\tt HiggsBounds}~\cite{Bechtle:2008jh,Bechtle:2011sb,Bechtle:2013gu,Bechtle:2013wla}
 and {\tt HiggsSignals}~\cite{Bechtle:2013xfa}.  They compute a $\chi^2$ measure  
from the predictions of the model and the measured Higgs rates and masses.
From this measure a $p$-value is estimated, testing the 
consistency between the model and the data.  As a practical rejection criterion, 
we discard a model point if the corresponding $p$-value is below $0.0027$, 
i.e.\ we require consistency at the three-sigma level.

\begin{figure}[t]
\begin{subfigure}[b]{0.5\textwidth}
\includegraphics[width=7.4cm,height=6.3cm]{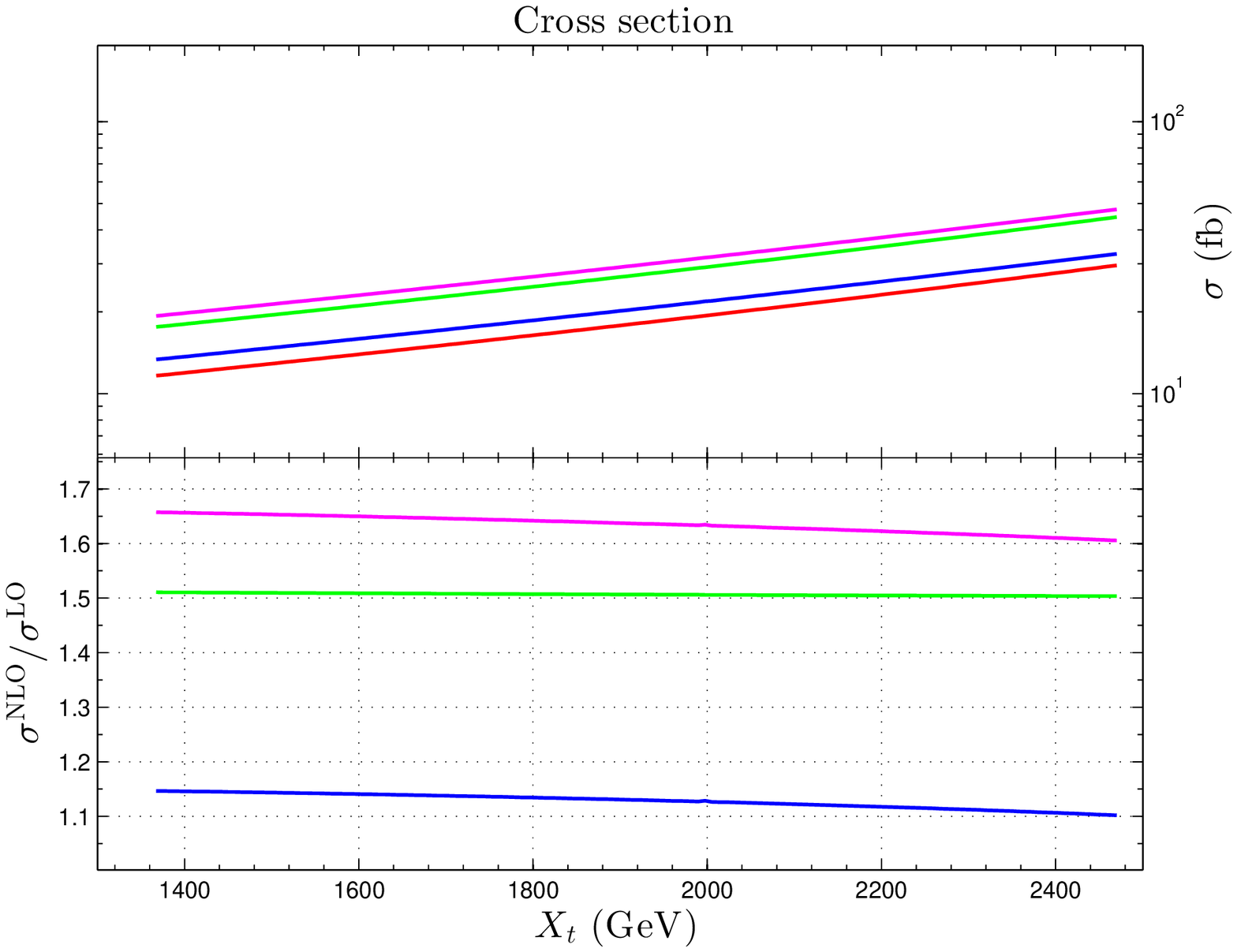}
\caption{}
\end{subfigure}
\begin{subfigure}[b]{0.5\textwidth}
\includegraphics[width=7.4cm,height=6.3cm]{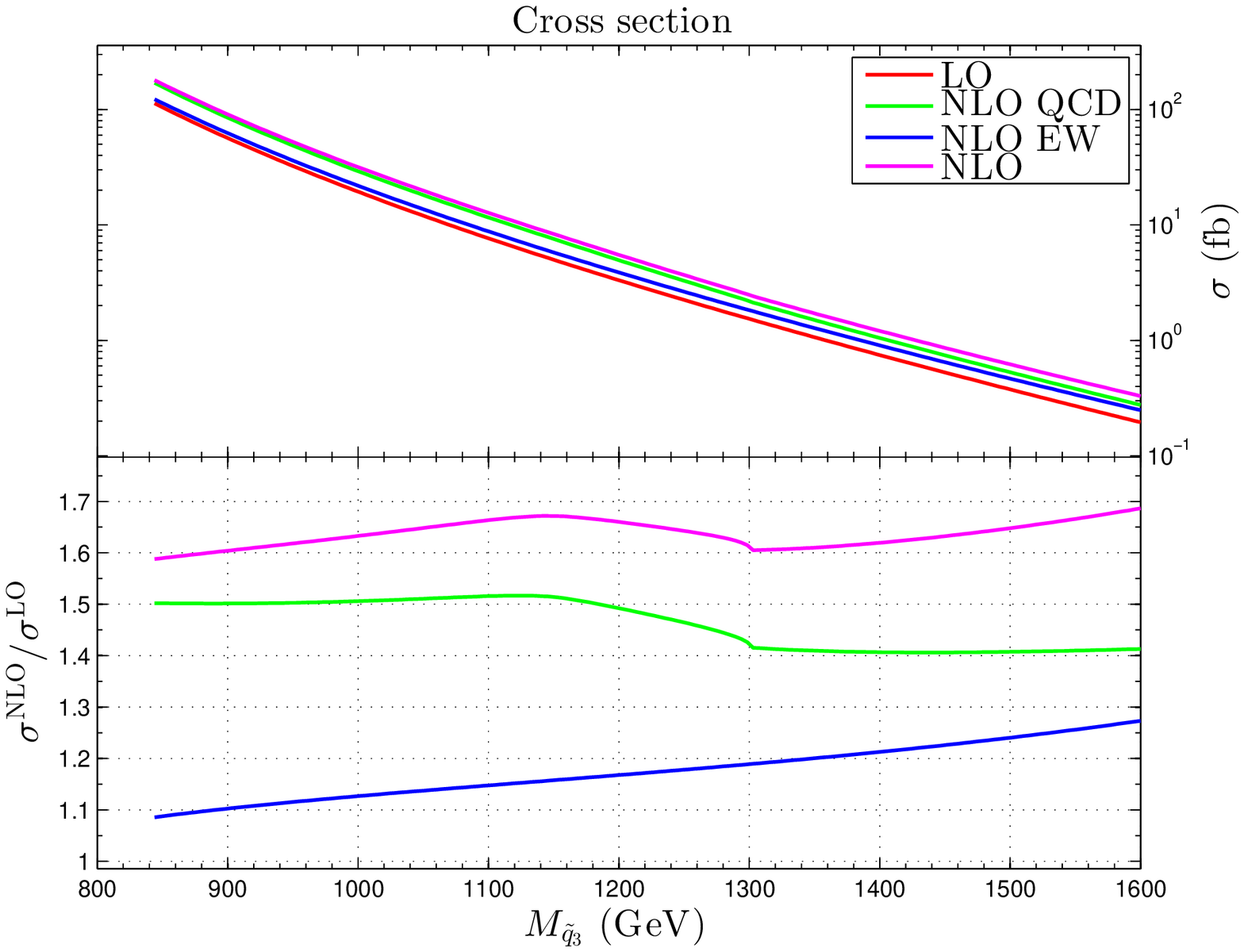}
\caption{}
\end{subfigure}
\phantom{pic}  \\
\begin{subfigure}[b]{0.5\textwidth}
\includegraphics[width=7.4cm,height=6.3cm]{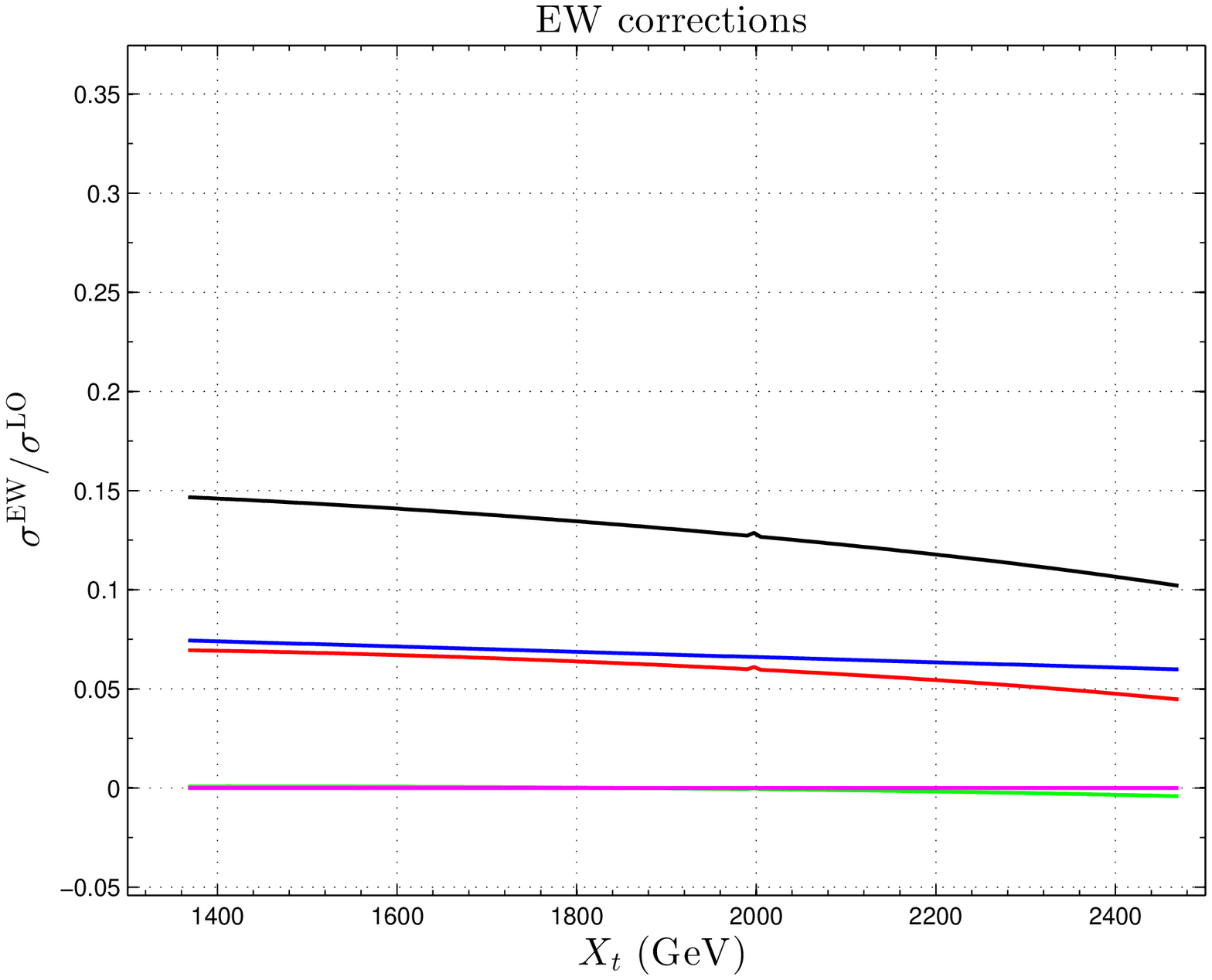}
\caption{}
\end{subfigure}
\begin{subfigure}[b]{0.5\textwidth}
\includegraphics[width=7.4cm,height=6.3cm]{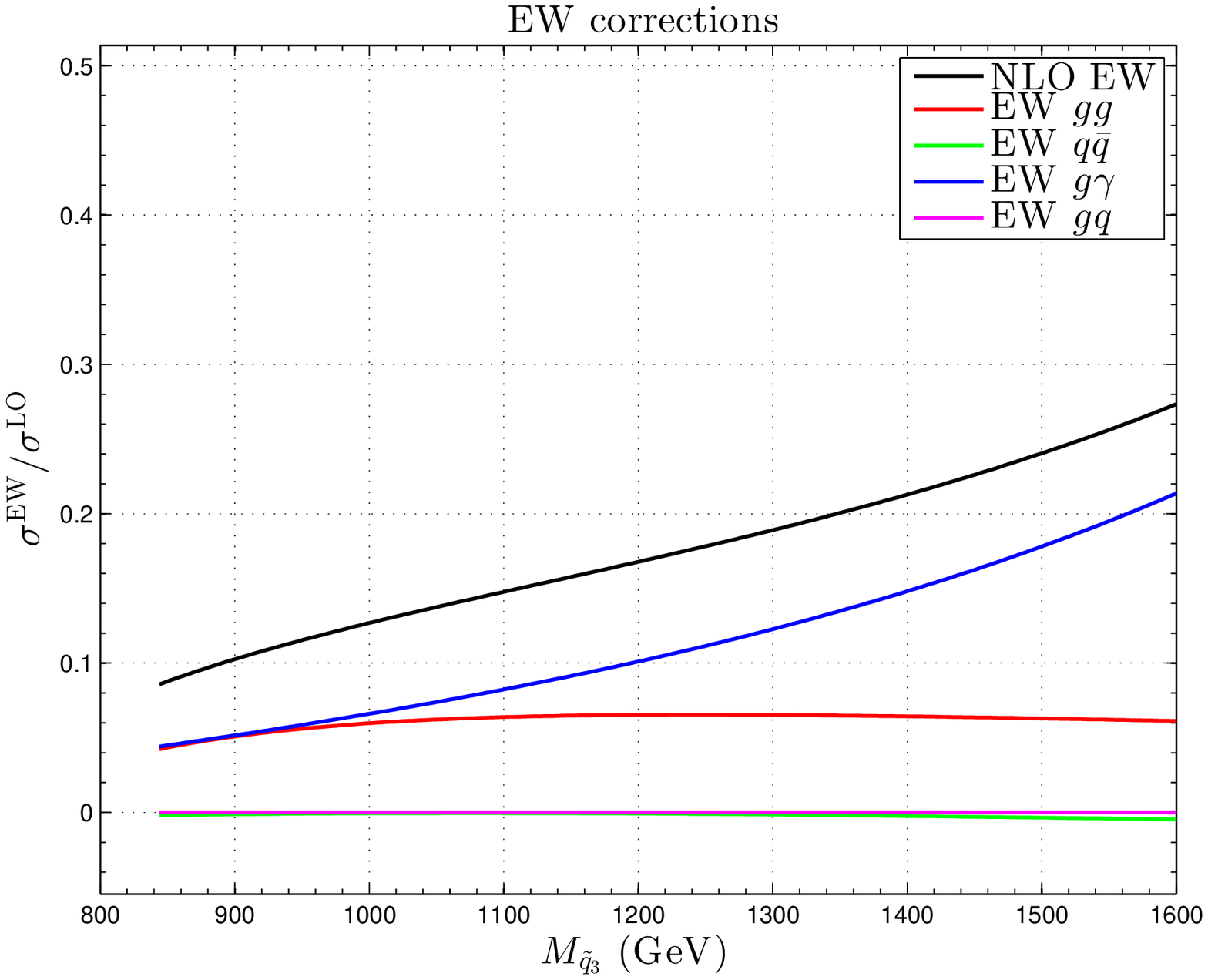}
\caption{}
\end{subfigure}
\phantom{pic}  \\
\begin{subfigure}[b]{0.5\textwidth}
\includegraphics[width=7.4cm,height=6.3cm]{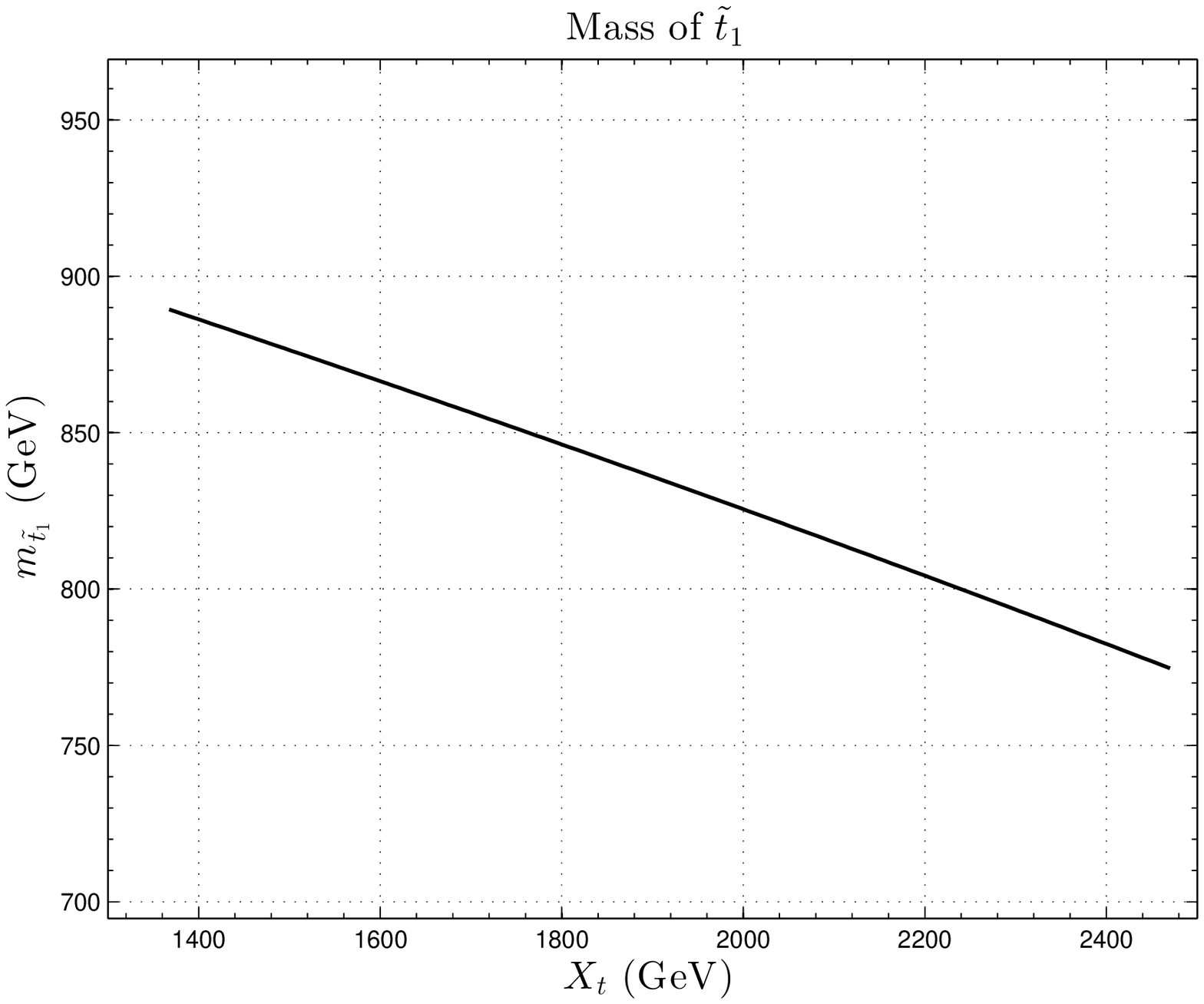}
\caption{}
\end{subfigure}
\begin{subfigure}[b]{0.5\textwidth}
\includegraphics[width=7.4cm,height=6.3cm]{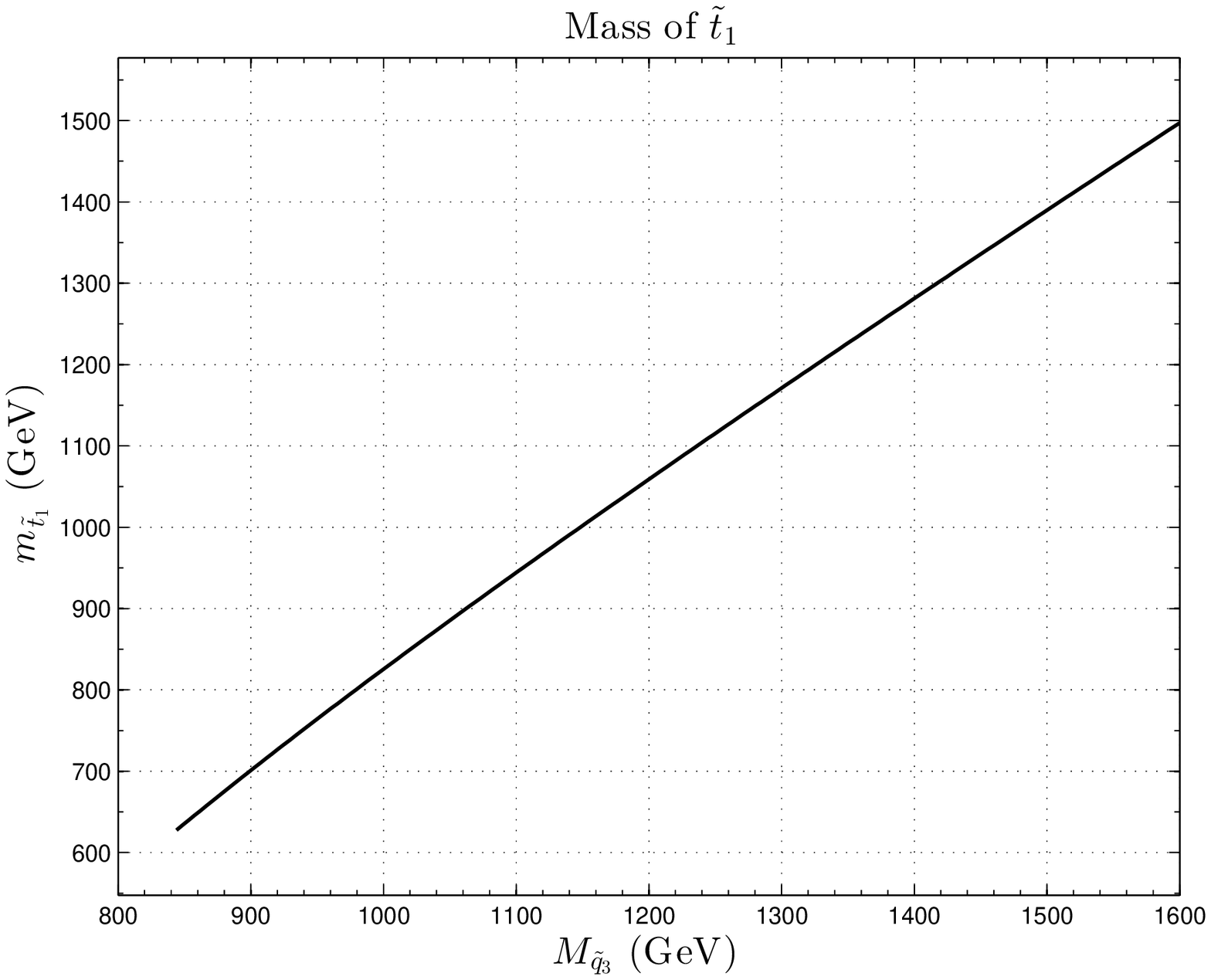}
\caption{}
\end{subfigure}
\caption[.]{Left (Right) panels: scans over  $X_t$  ($M_{\tilde q_3}$)   in the light-Higgs scenario. The value of the parameters fixed in the scans are collected in Table~\ref{Tab:Bench}(a).}
\label{fig:LightHiggs1}
 \end{figure}

\begin{figure}[t]
\begin{subfigure}[b]{0.5\textwidth}
\includegraphics[width=7.4cm,height=6.3cm]{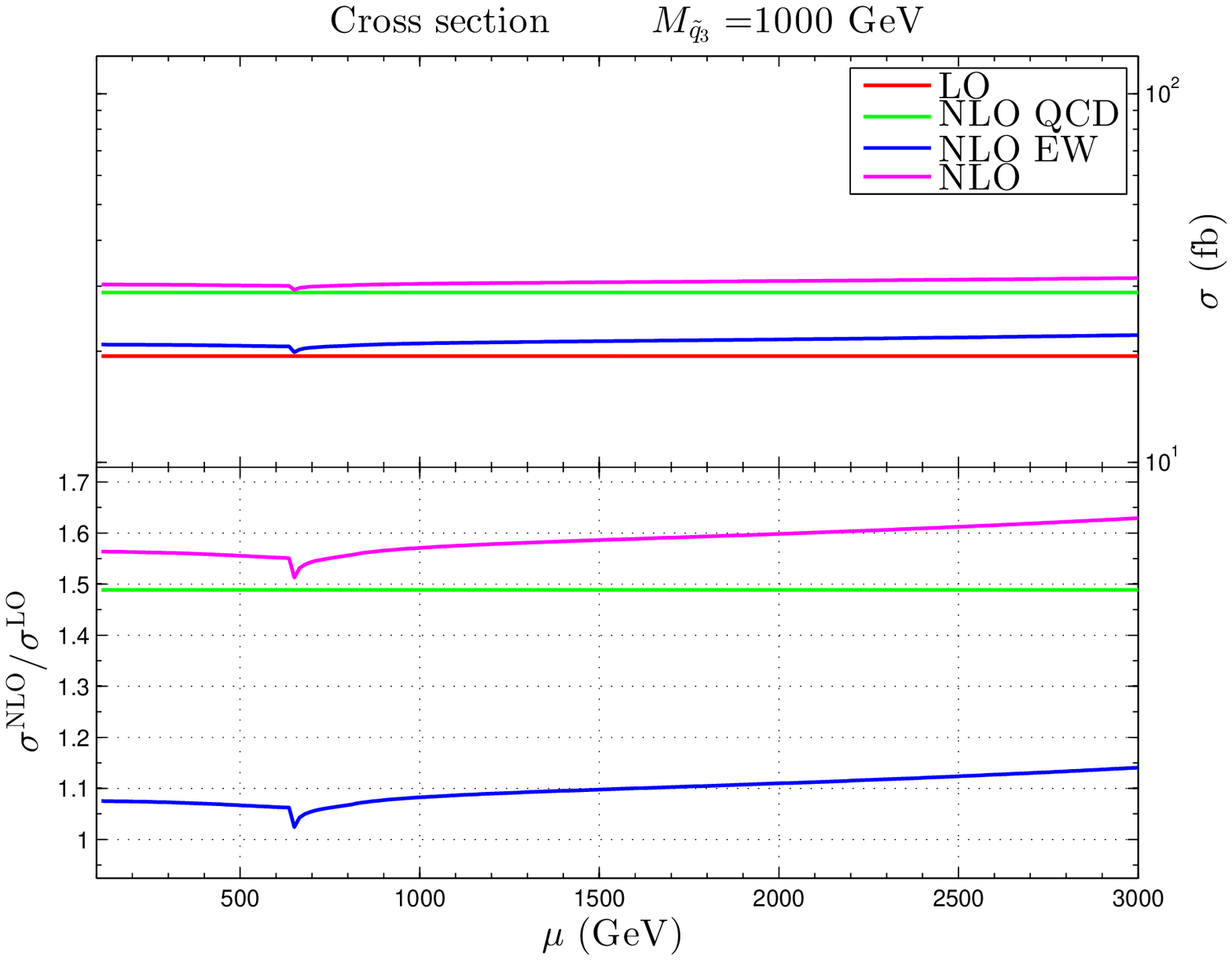}
\caption{}
\end{subfigure}
\begin{subfigure}[b]{0.5\textwidth}
\includegraphics[width=7.4cm,height=6.3cm]{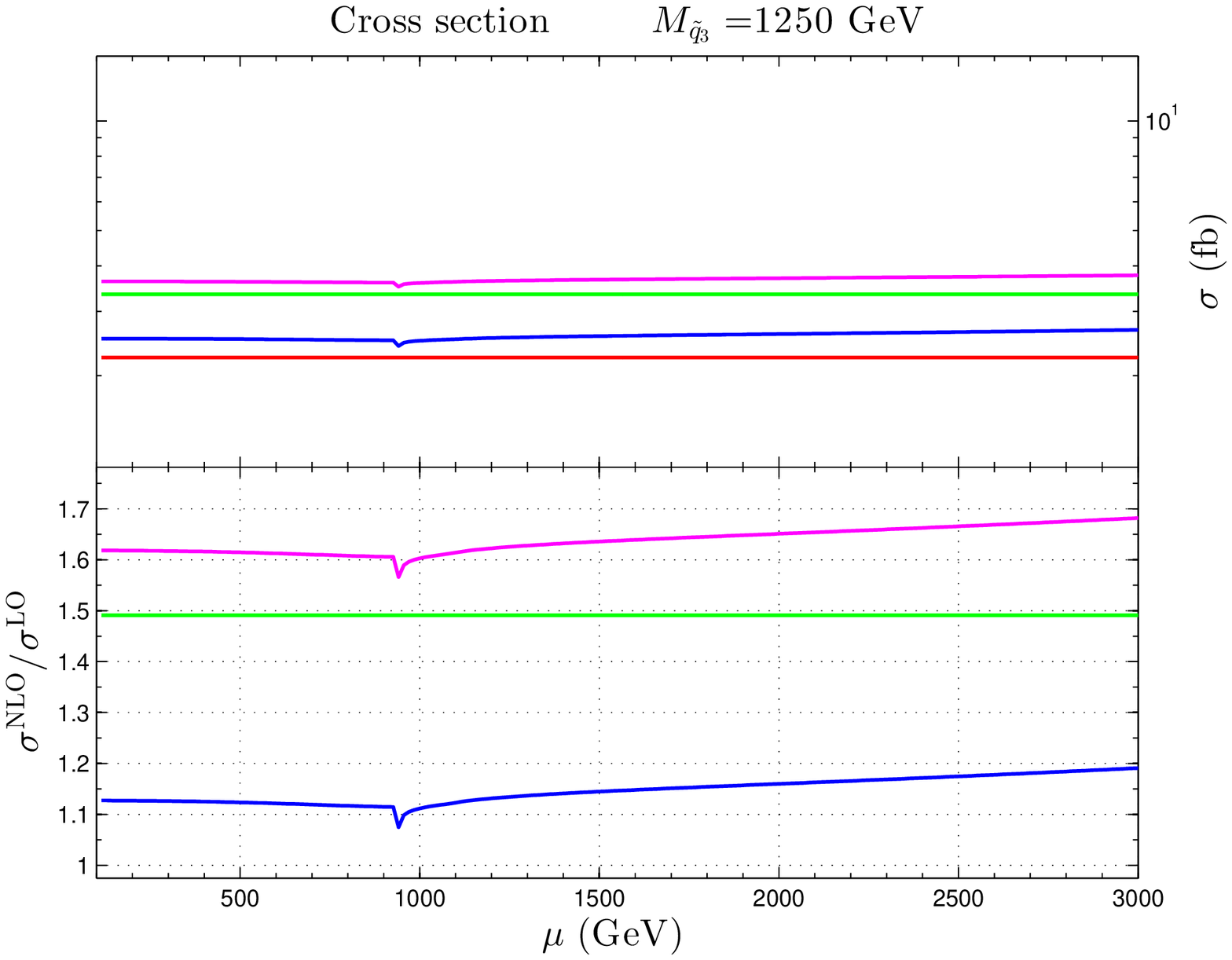}
\caption{}
\end{subfigure}
\phantom{pic}  \\
\begin{subfigure}[b]{0.5\textwidth}
\includegraphics[width=7.4cm,height=6.3cm]{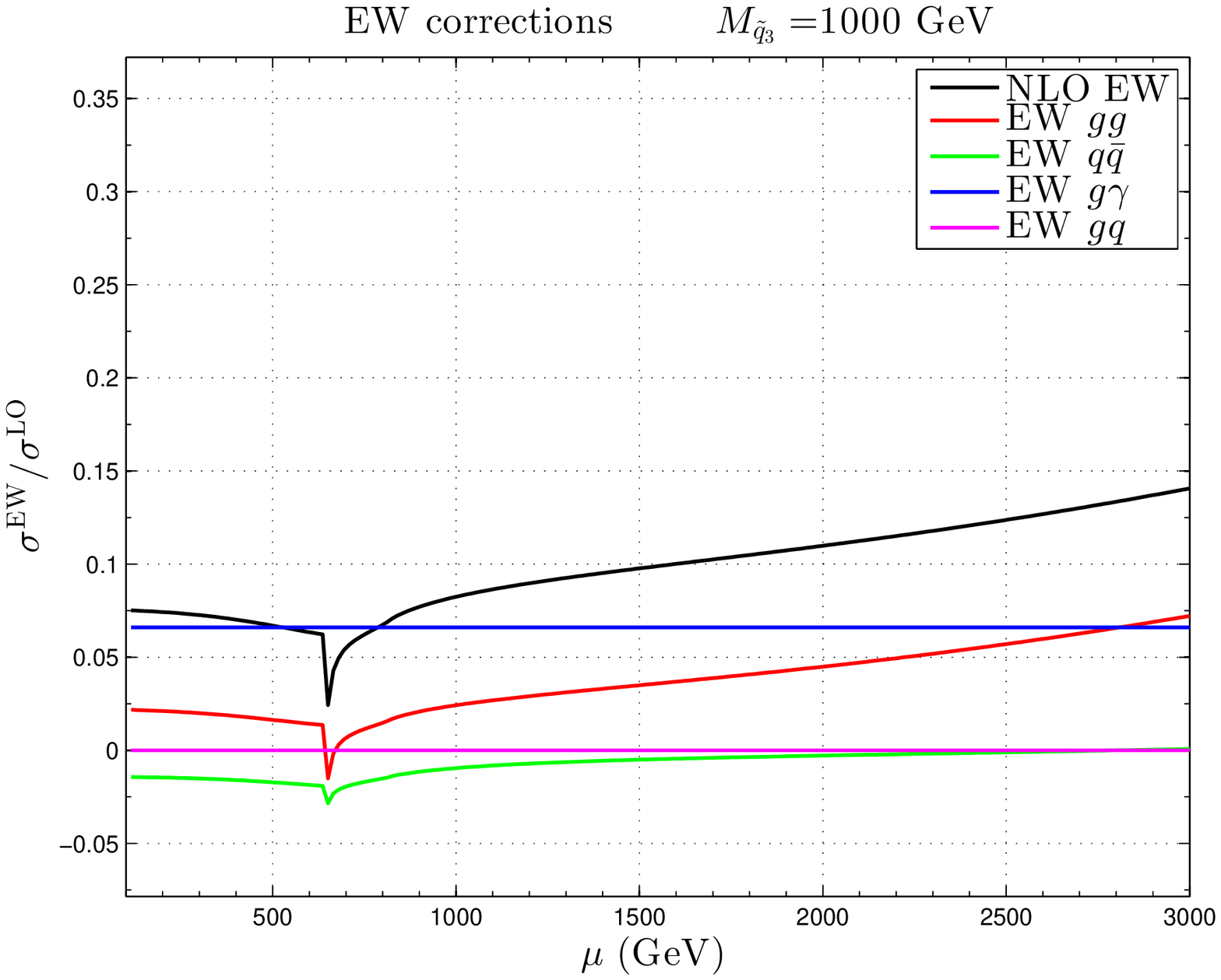}
\caption{}
\end{subfigure}
\begin{subfigure}[b]{0.5\textwidth}
\includegraphics[width=7.4cm,height=6.3cm]{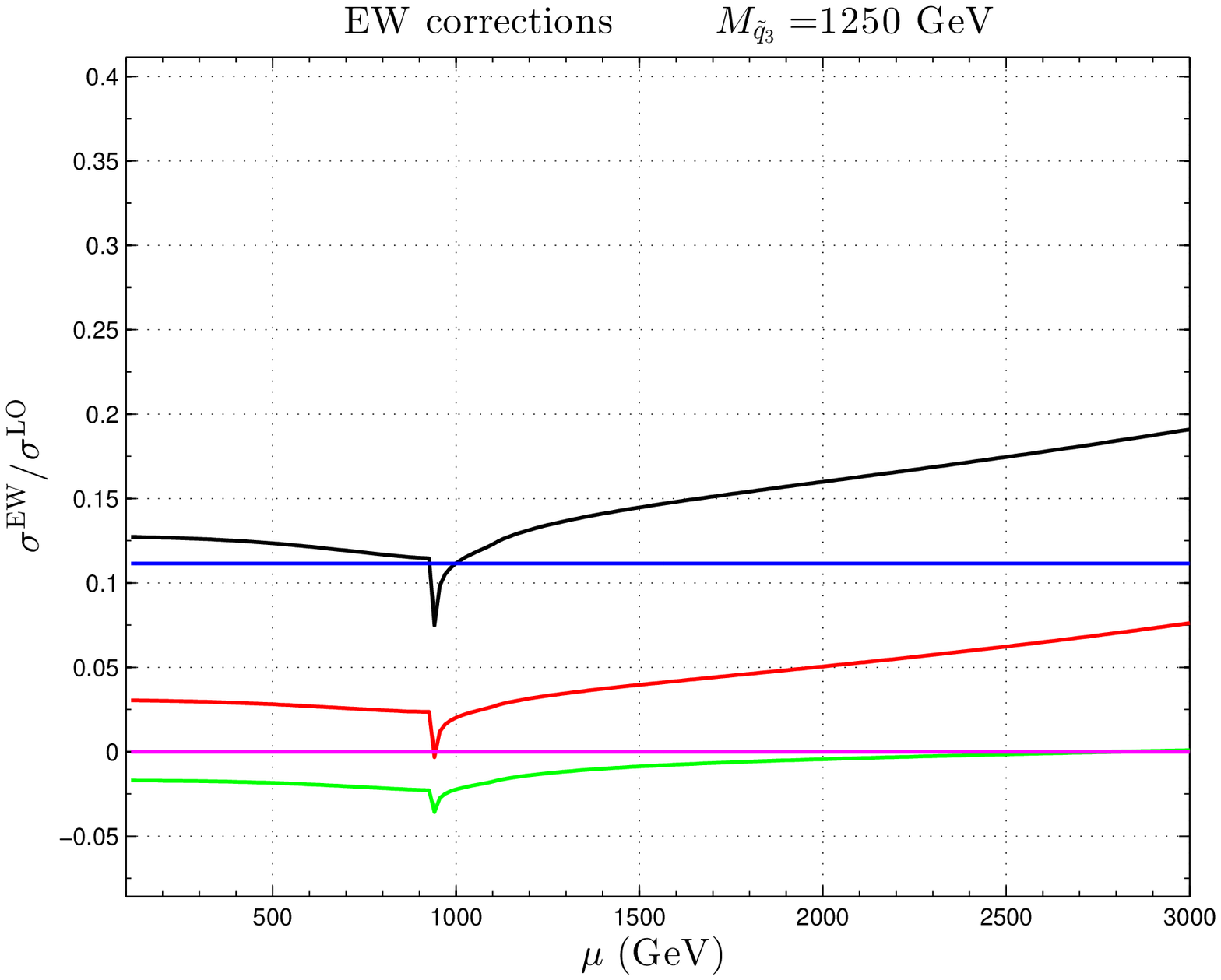}
\caption{}
\end{subfigure}
\phantom{pic}  \\
\begin{subfigure}[b]{0.5\textwidth}
\includegraphics[width=7.4cm,height=6.3cm]{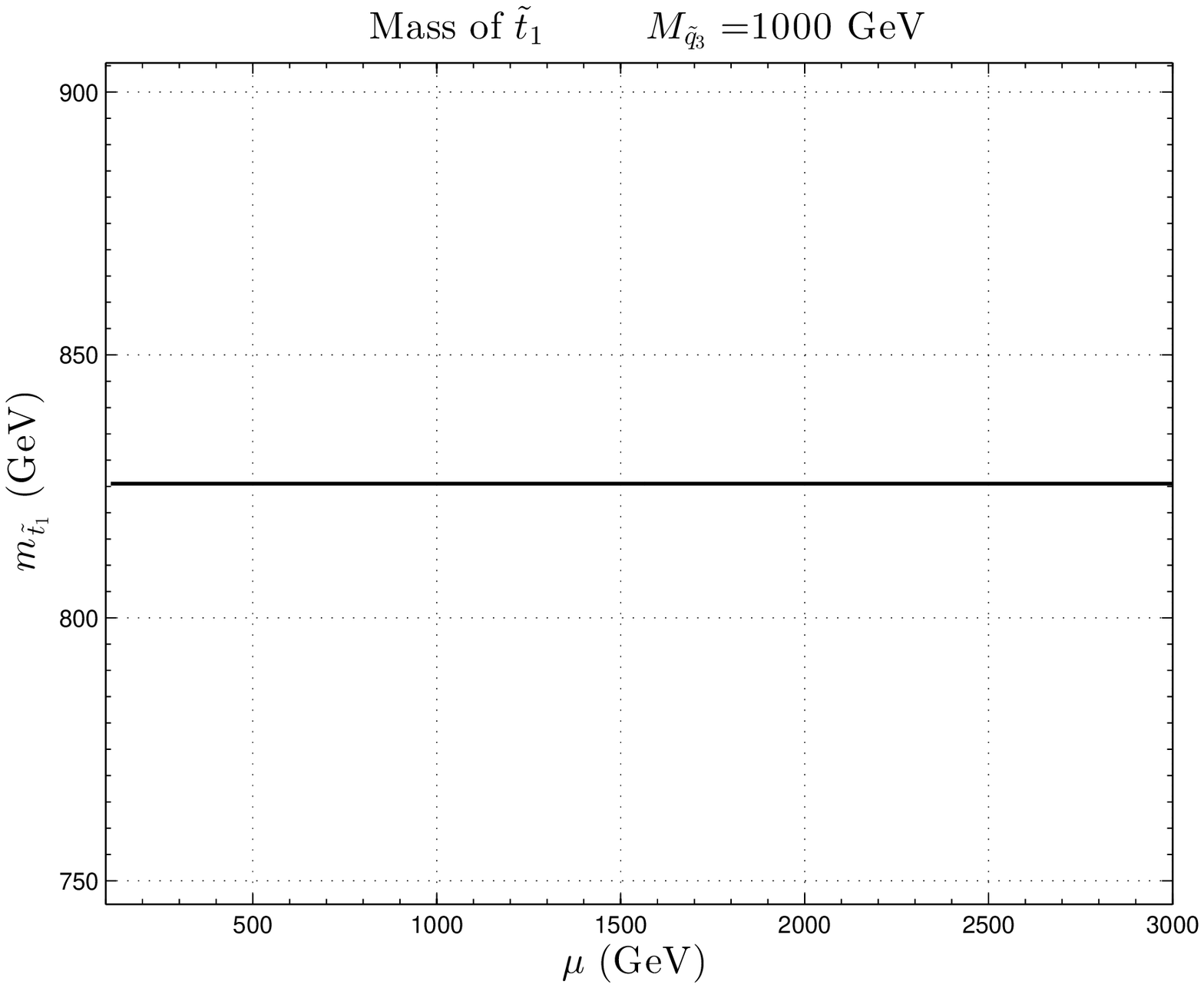}
\caption{}
\end{subfigure}
\begin{subfigure}[b]{0.5\textwidth}
\includegraphics[width=7.4cm,height=6.3cm]{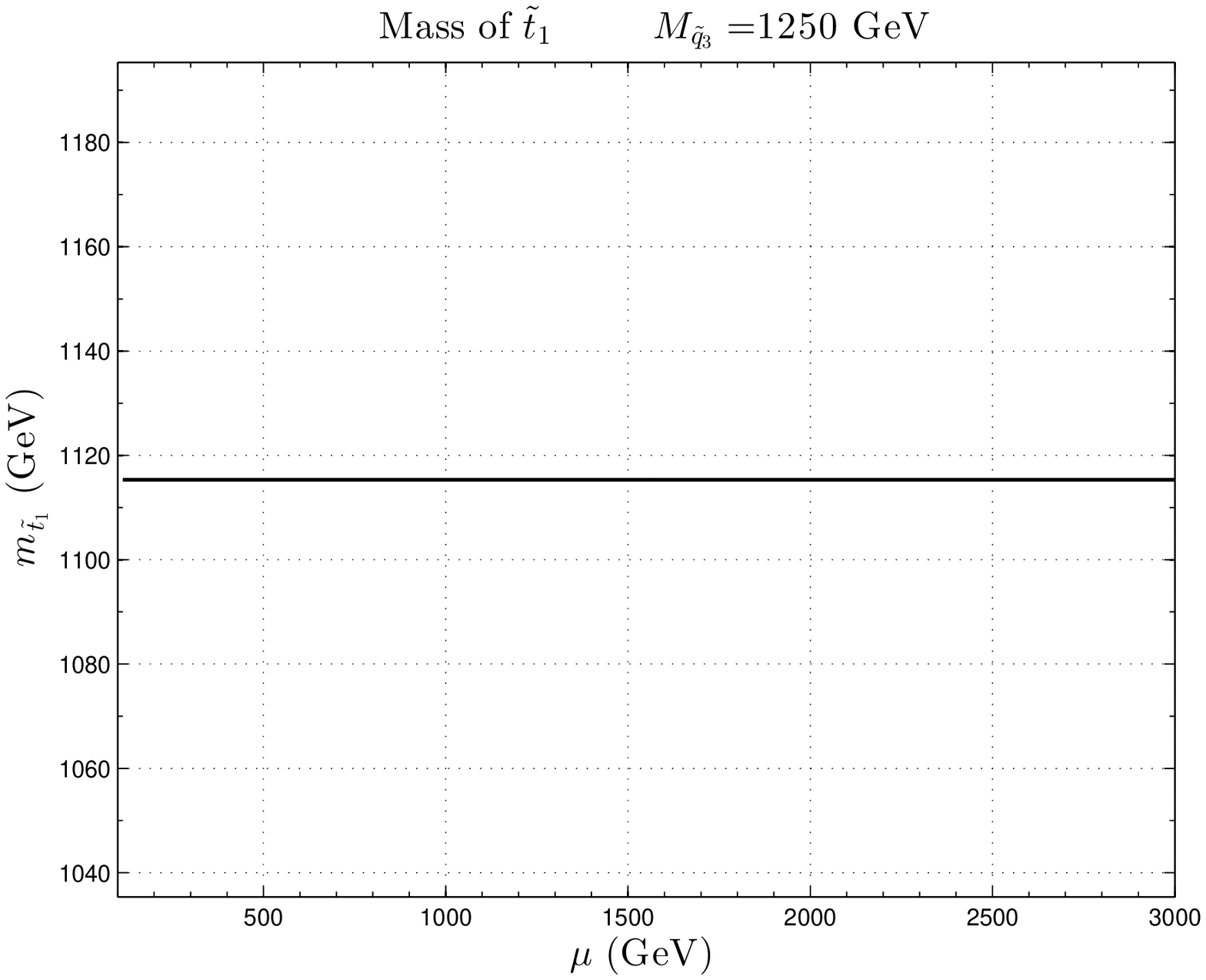}
\caption{}
\end{subfigure}
\caption[.]{Scans over  $\mu$ in the light-Higgs scenario.  In the left (right)
panel  $M_{\tilde q_3}$ is set equal to $1000$ ($1260$)~GeV. Table~\ref{Tab:Bench}(a)
collect the value of the  other  parameters.}
\label{fig:LightHiggs3}
 \end{figure}

\begin{figure}[t]
\begin{subfigure}[b]{0.5\textwidth}
\includegraphics[width=7.4cm,height=6.3cm]{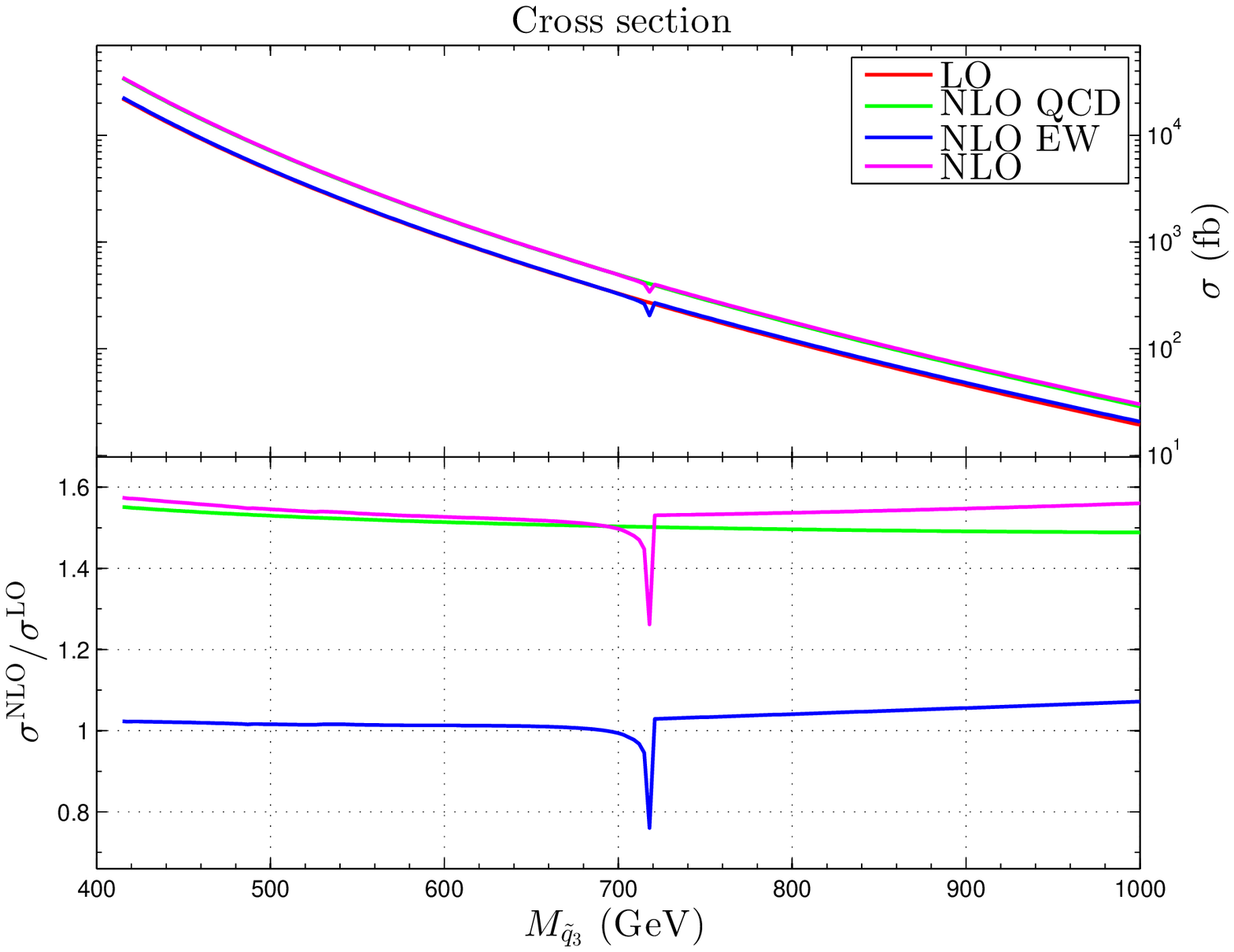}
\caption{}
\end{subfigure}
\begin{subfigure}[b]{0.5\textwidth}
\includegraphics[width=7.4cm,height=6.3cm]{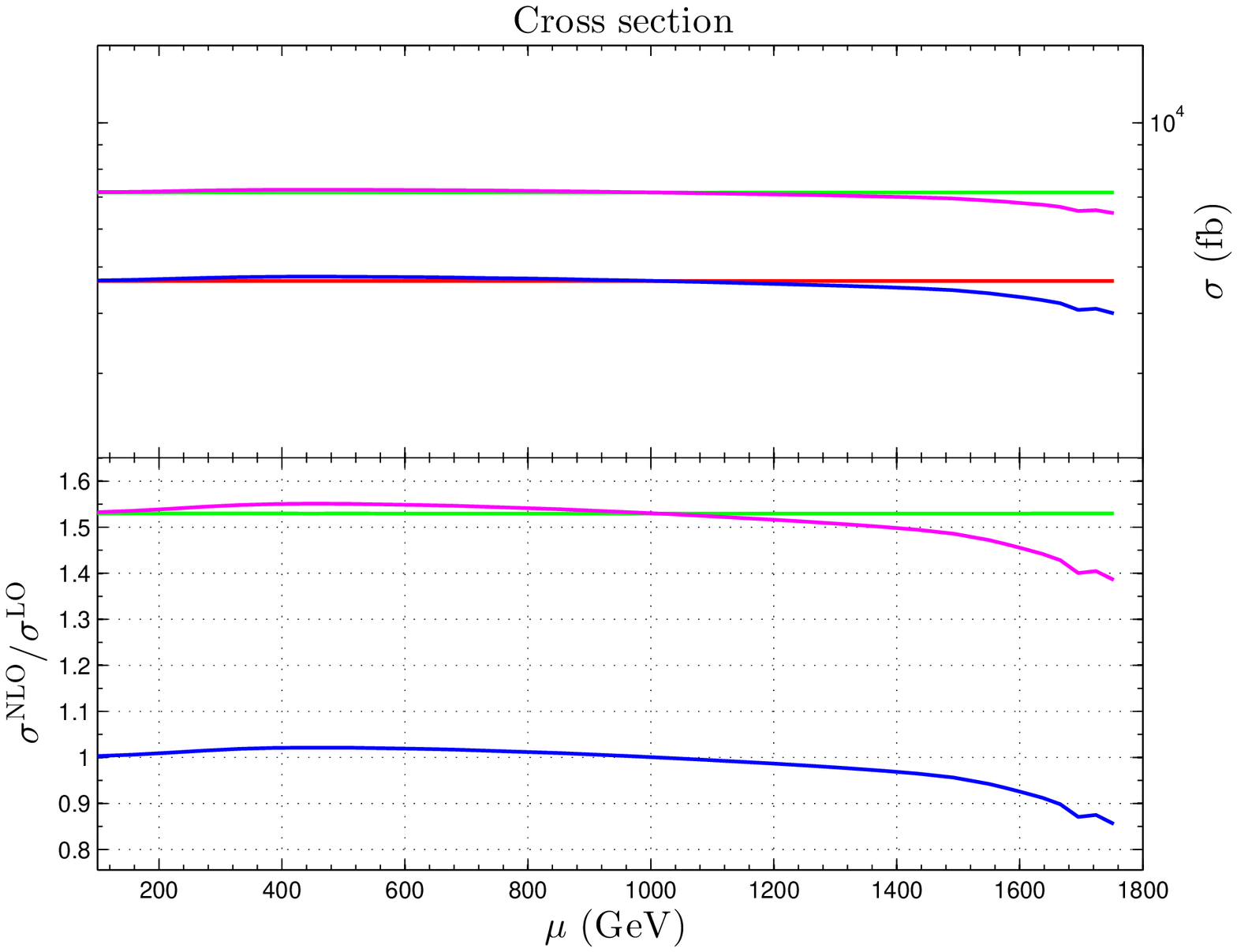}
\caption{}
\end{subfigure}
\phantom{pic}  \\
\begin{subfigure}[b]{0.5\textwidth}
\includegraphics[width=7.4cm,height=6.3cm]{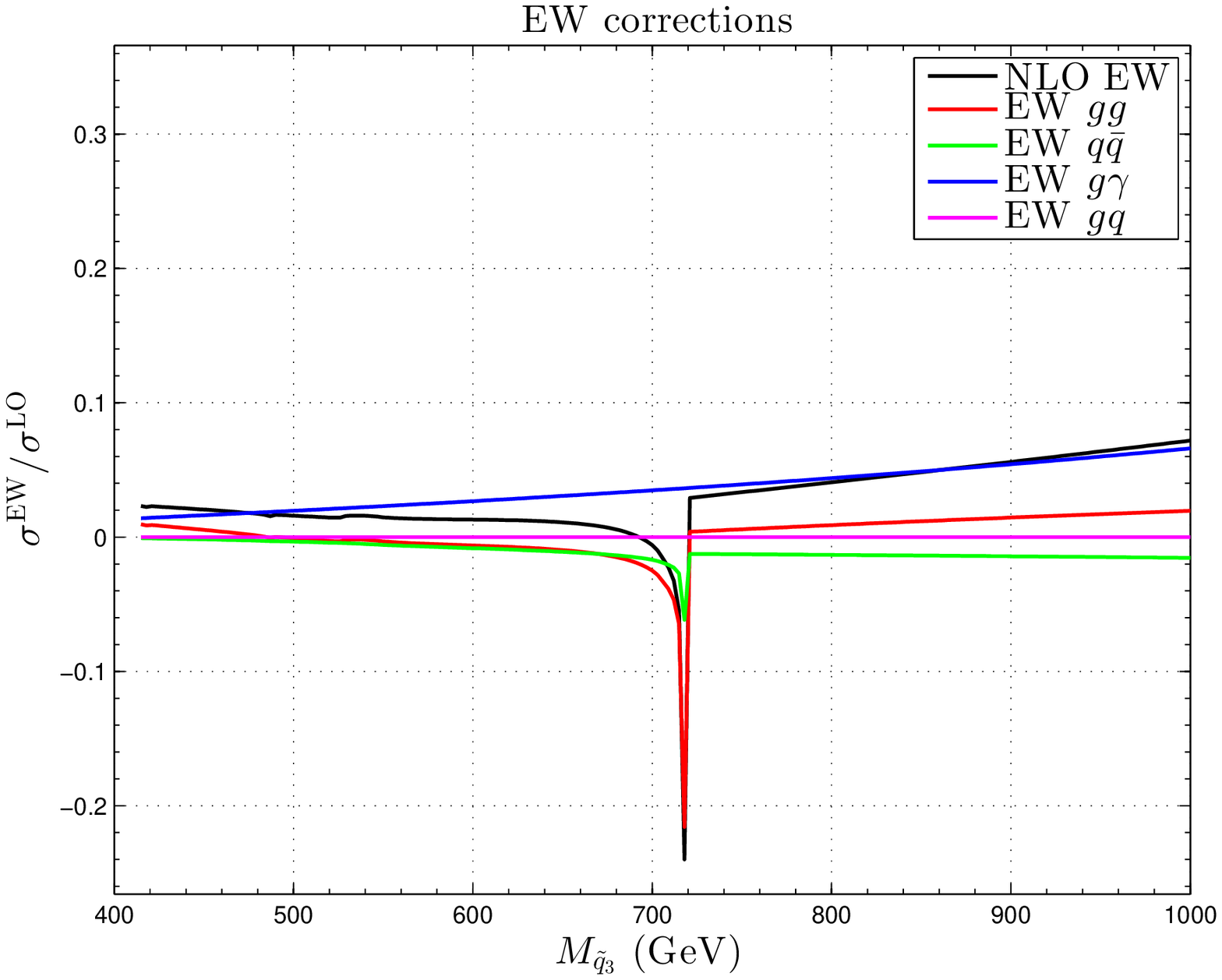}
\caption{}
\end{subfigure}
\begin{subfigure}[b]{0.5\textwidth}
\includegraphics[width=7.4cm,height=6.3cm]{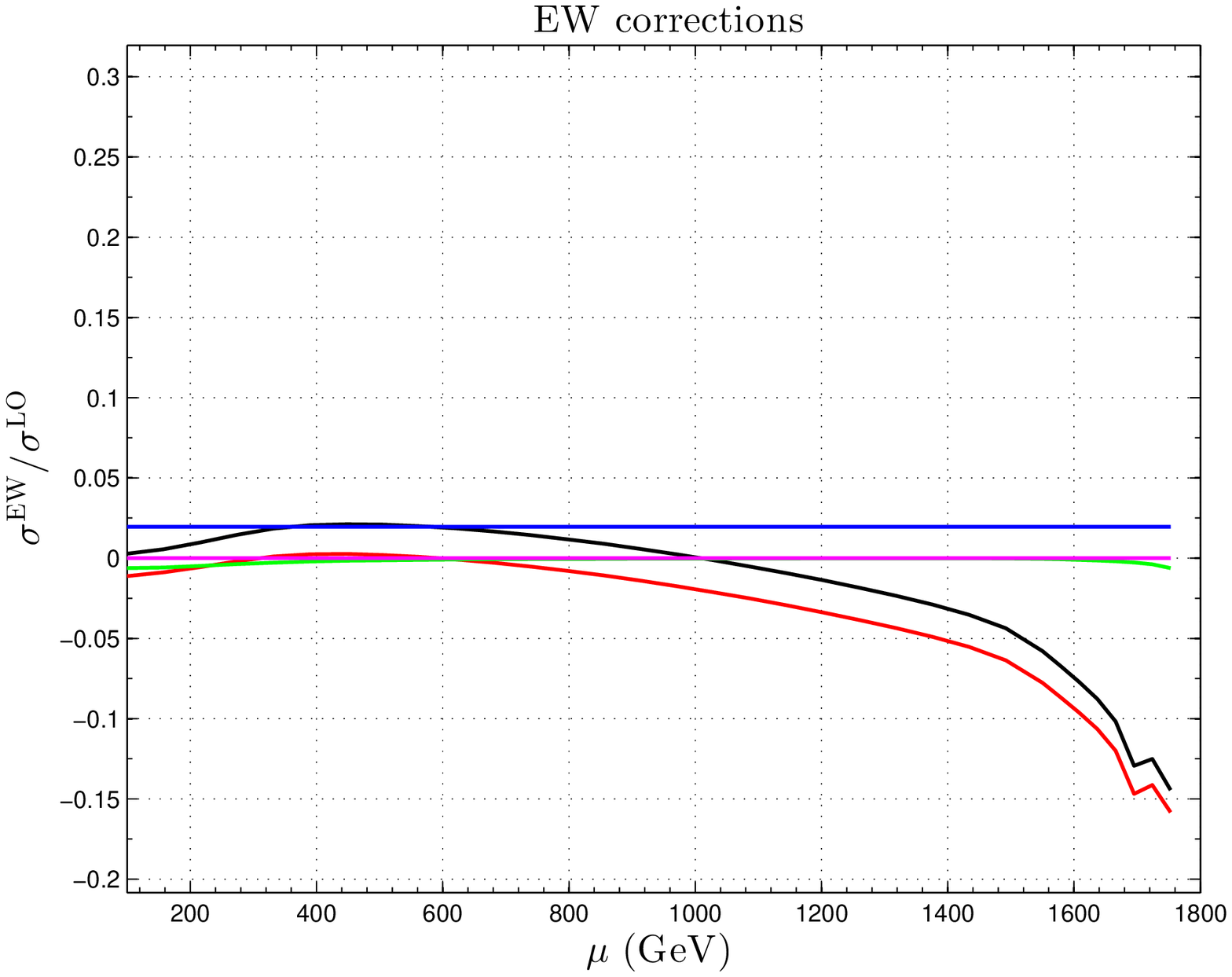}
\caption{}
\end{subfigure}
\phantom{pic}  \\
\begin{subfigure}[b]{0.5\textwidth}
\includegraphics[width=7.4cm,height=6.3cm]{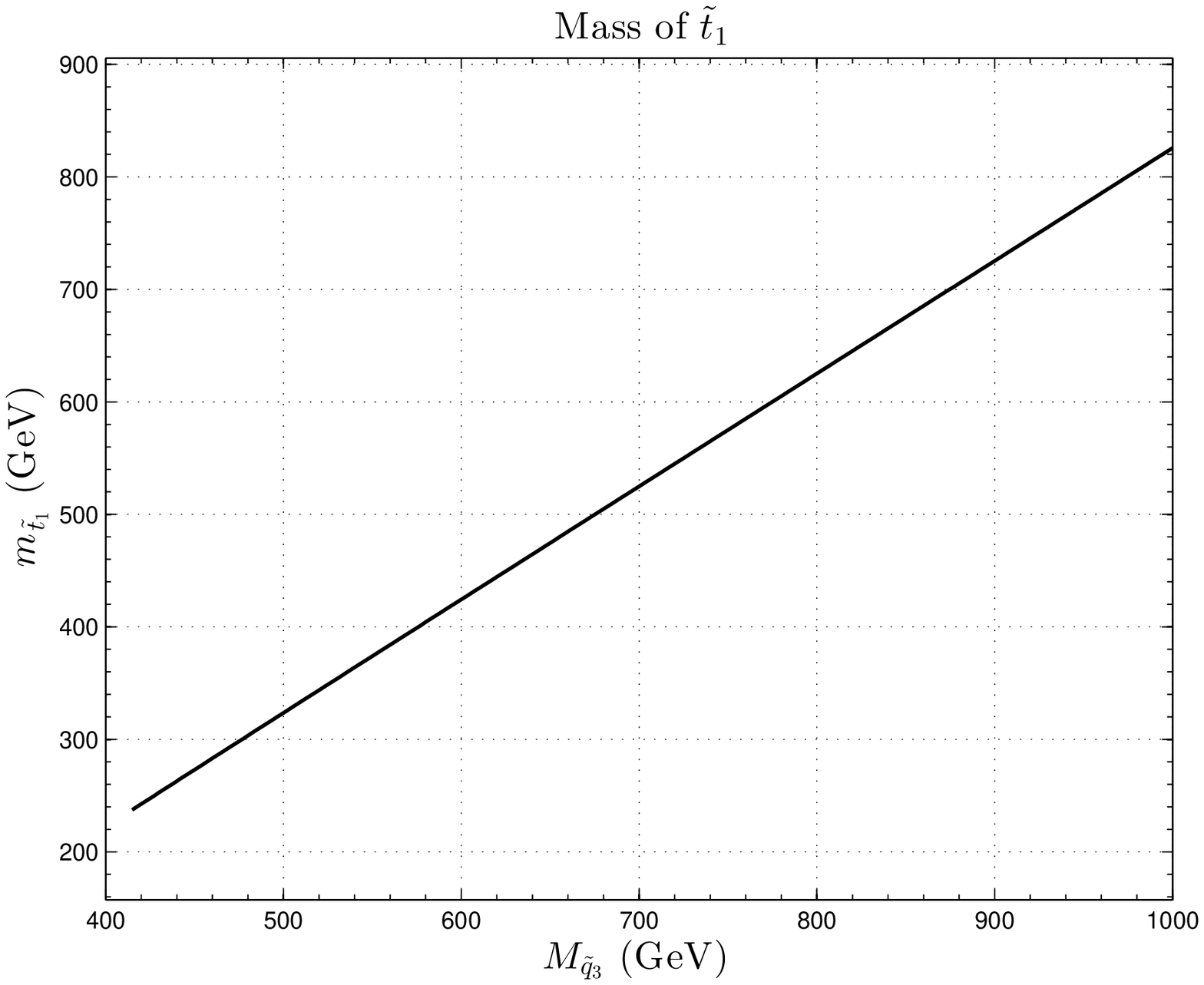}
\caption{}
\end{subfigure}
\begin{subfigure}[b]{0.5\textwidth}
\includegraphics[width=7.4cm,height=6.3cm]{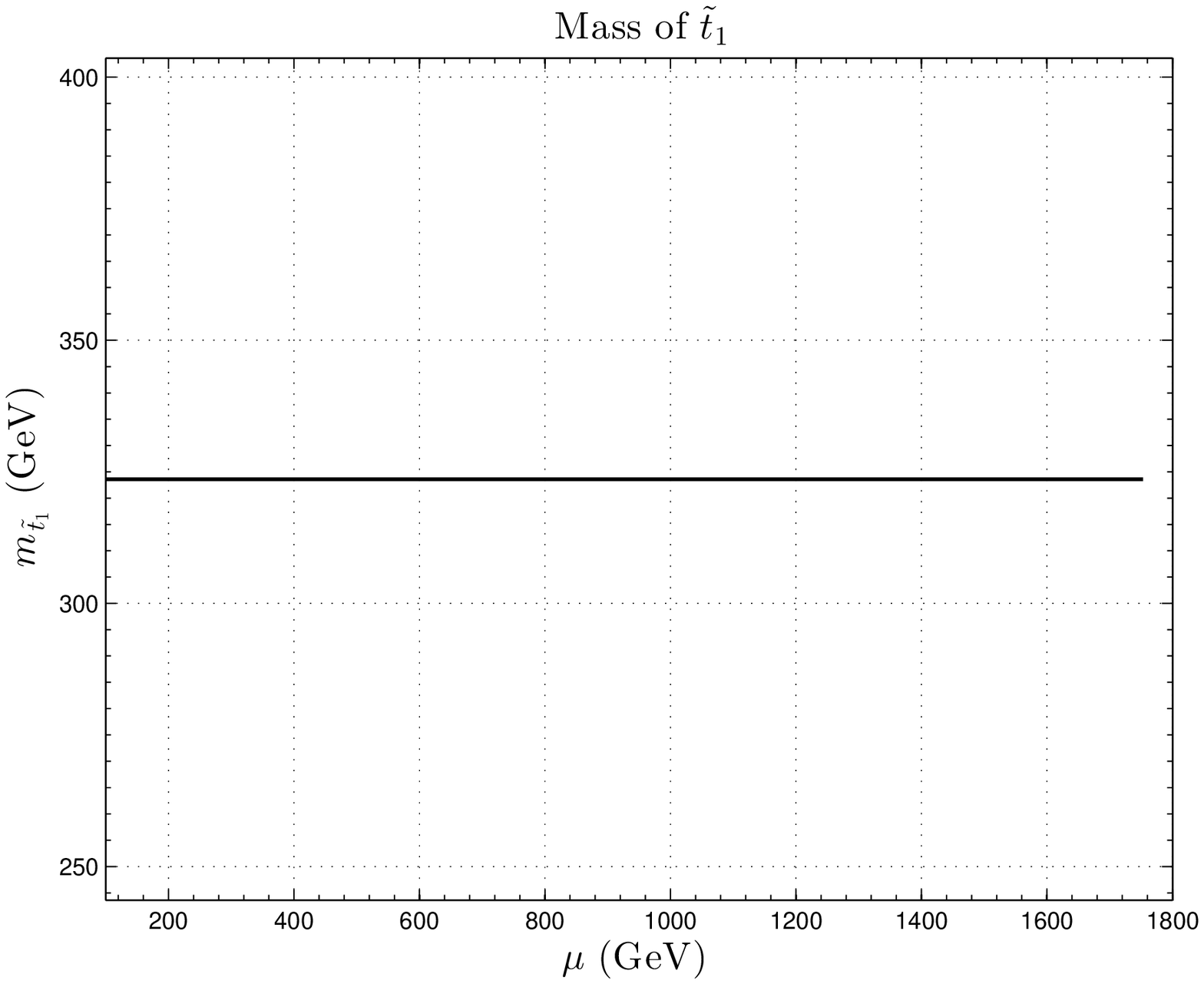}
\caption{}
\end{subfigure}
\caption[.]{Left (Right) panels: scans over  $M_{\tilde q_3}$  ($\mu$) in the light-stop scenario. The value of the  parameters
not involved in the scans are collected in Table~\ref{Tab:Bench}(b).}
\label{fig:LightStop}
 \end{figure}

\begin{figure}[t]
\begin{subfigure}[b]{0.5\textwidth}
\includegraphics[width=7.4cm,height=6.3cm]{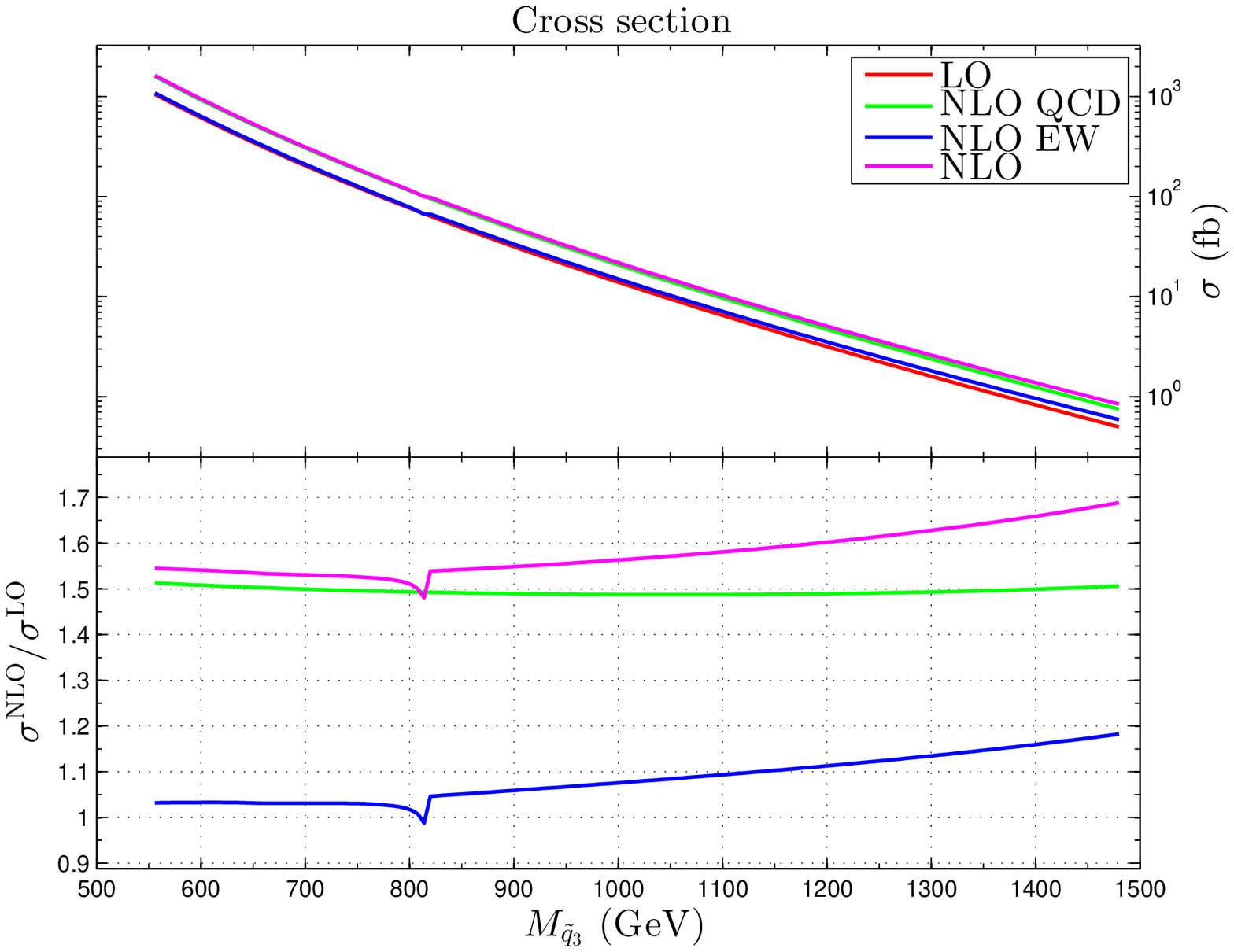}
\caption{}
\end{subfigure}
\begin{subfigure}[b]{0.5\textwidth}
\includegraphics[width=7.4cm,height=6.3cm]{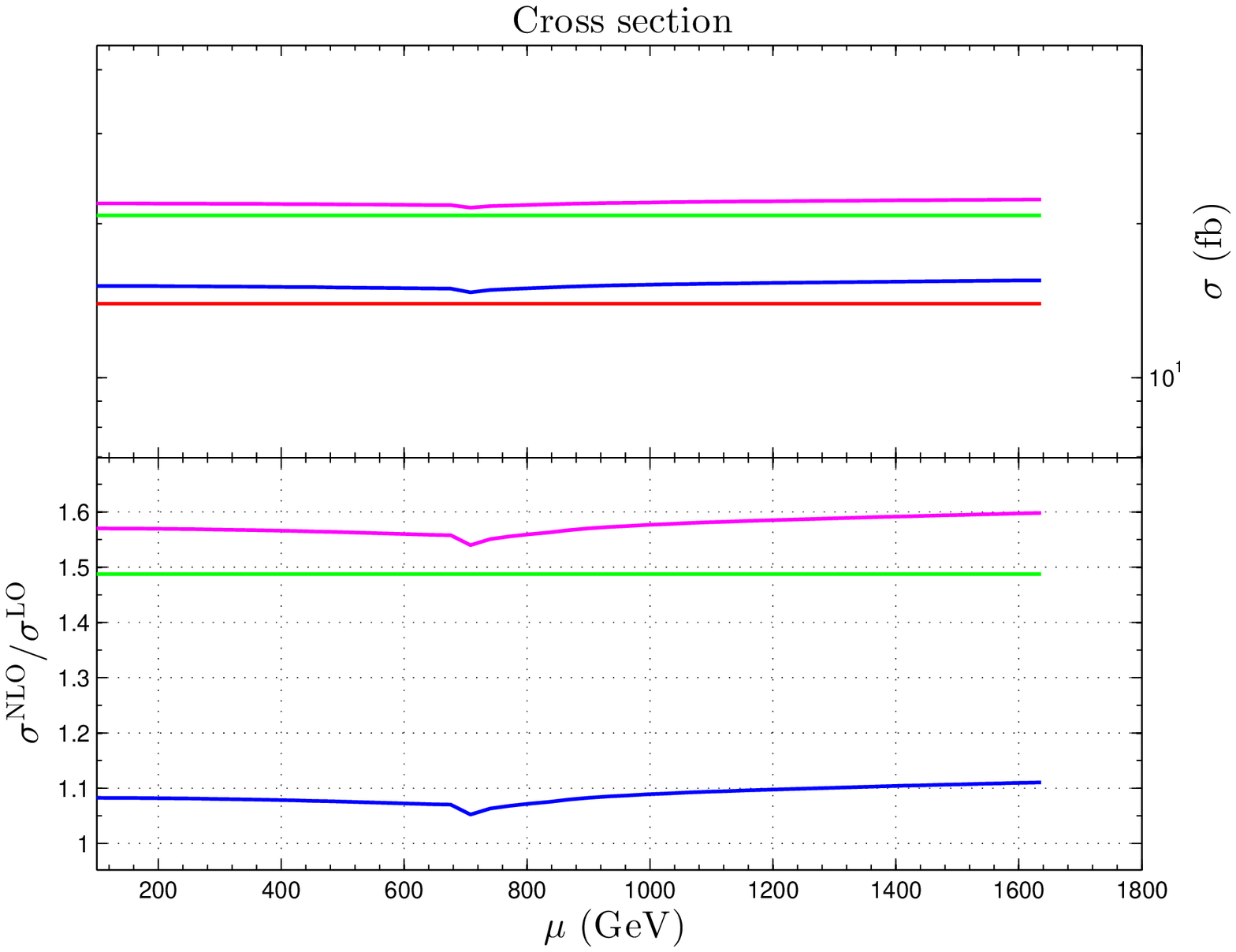}
\caption{}
\end{subfigure}
\phantom{pic}  \\
\begin{subfigure}[b]{0.5\textwidth}
\includegraphics[width=7.4cm,height=6.3cm]{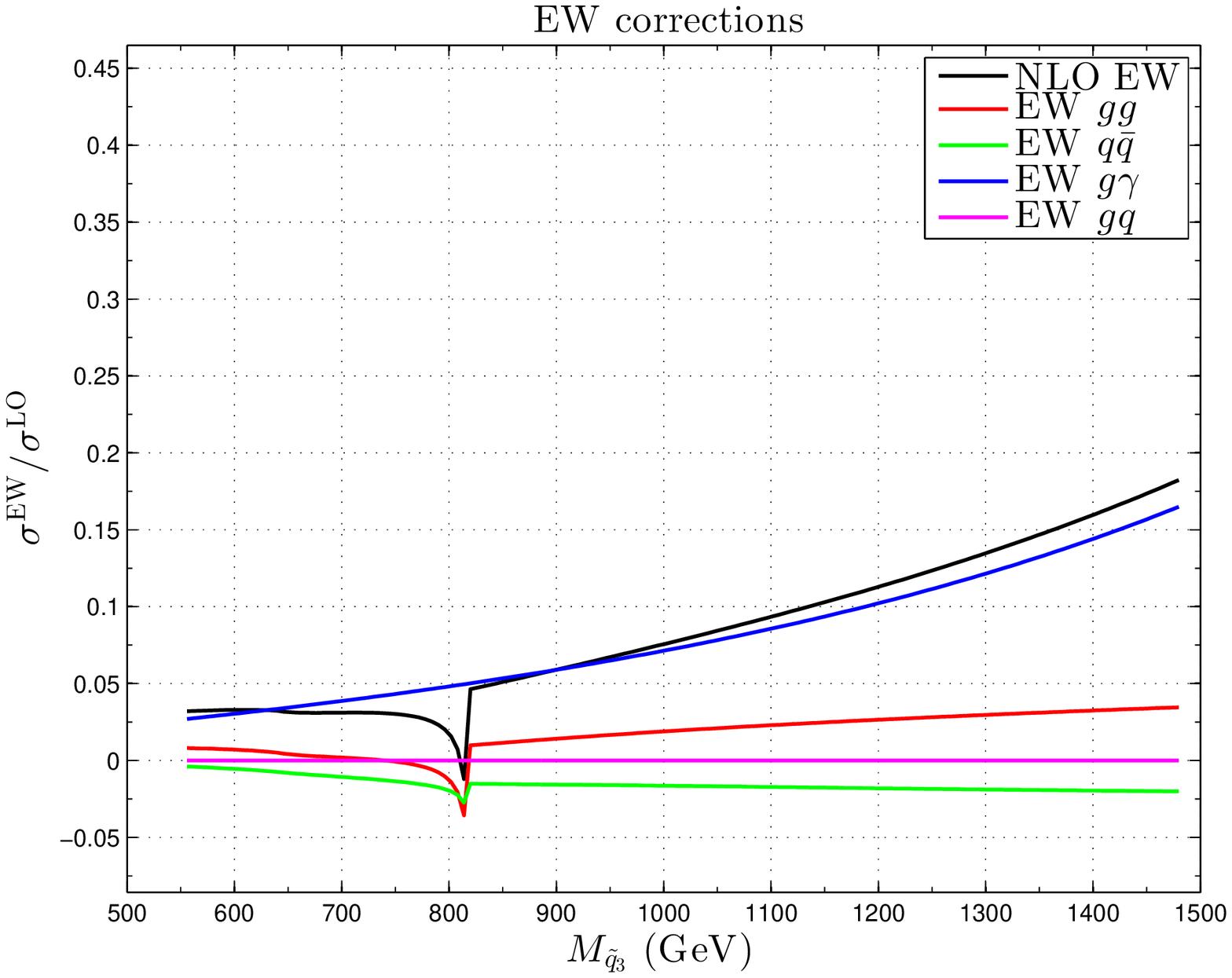}
\caption{}
\end{subfigure}
\begin{subfigure}[b]{0.5\textwidth}
\includegraphics[width=7.4cm,height=6.3cm]{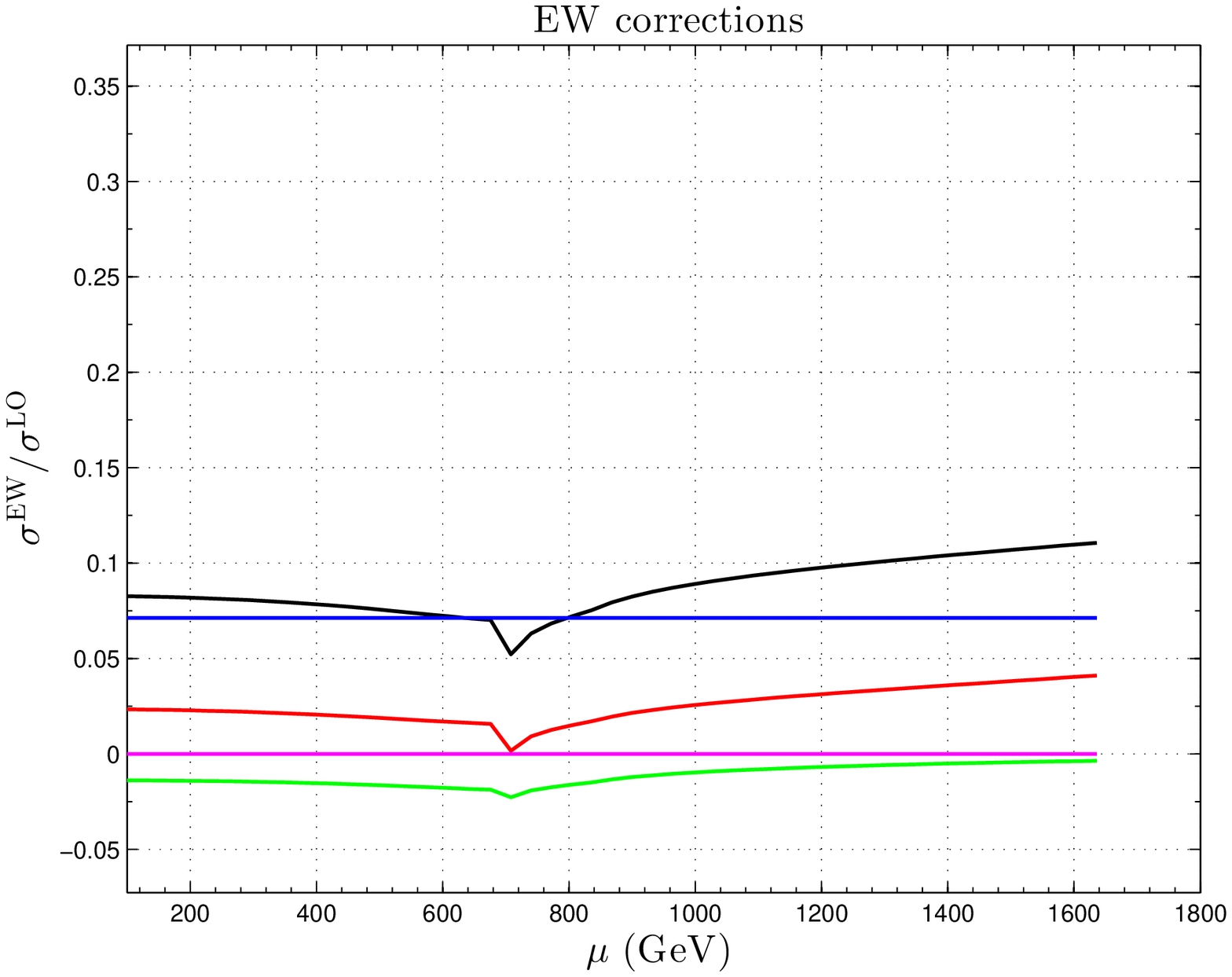}
\caption{}
\end{subfigure}
\phantom{pic}  \\
\begin{subfigure}[b]{0.5\textwidth}
\includegraphics[width=7.4cm,height=6.3cm]{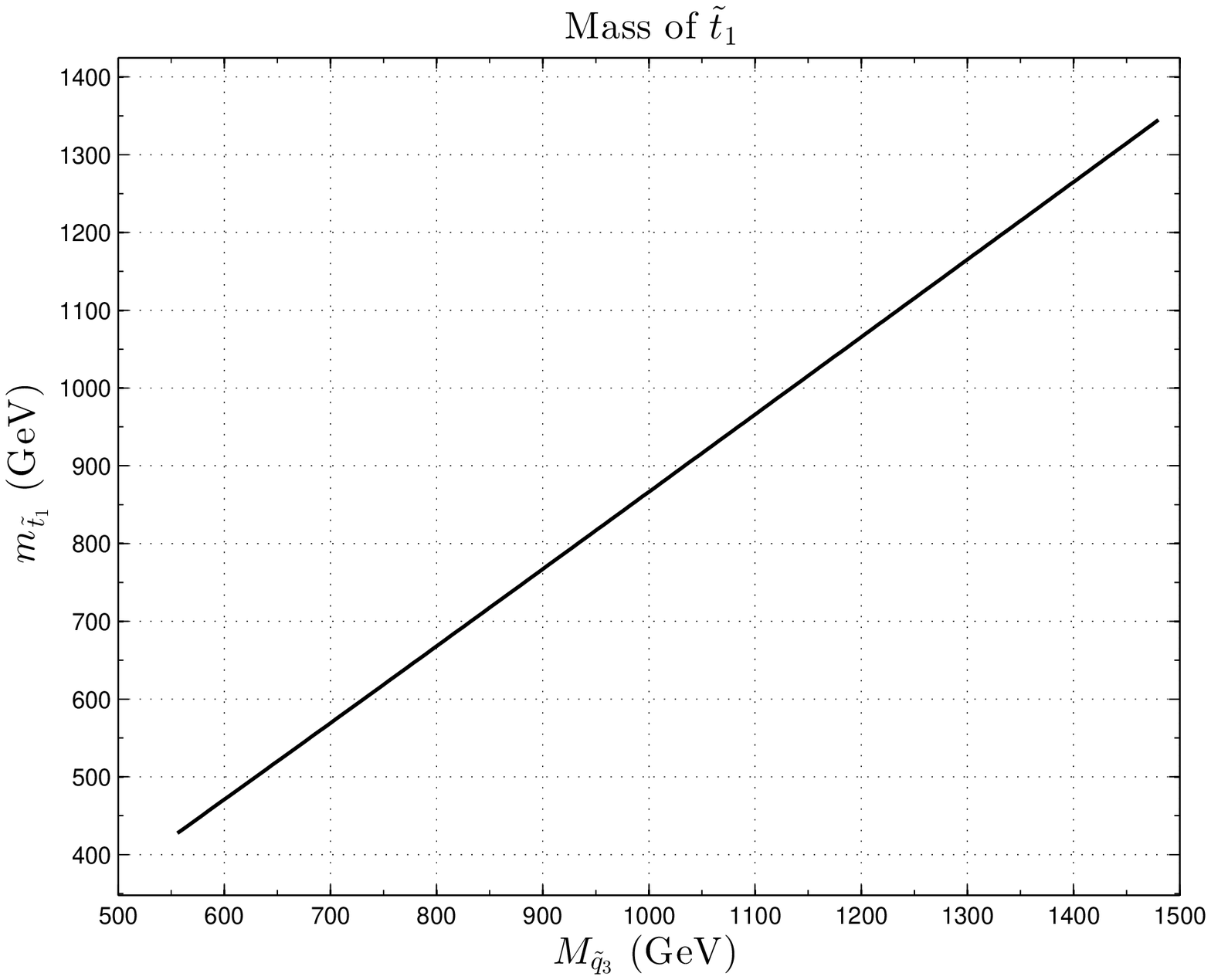}
\caption{}
\end{subfigure}
\begin{subfigure}[b]{0.5\textwidth}
\includegraphics[width=7.4cm,height=6.3cm]{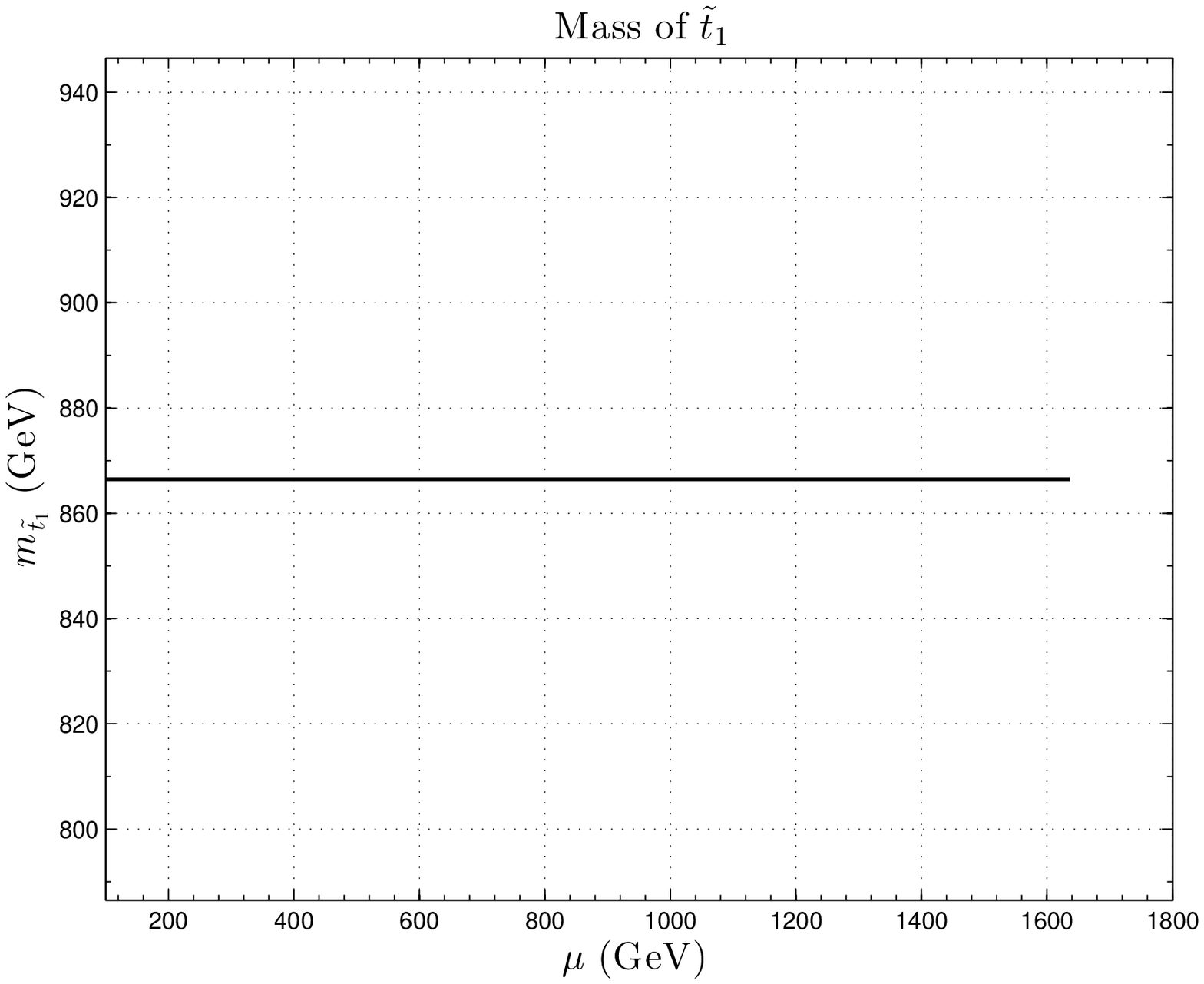}
\caption{}
\end{subfigure}
\caption[.]{Same as Fig.~\ref{fig:LightStop}, but for the light-stau scenario. The value of the  parameters
not involved in the scans are collected in Table~\ref{Tab:Bench}(c).}
\label{fig:LightStau}
 \end{figure}

\begin{figure}[t]
\begin{subfigure}[b]{0.5\textwidth}
\includegraphics[width=7.4cm,height=6.3cm]{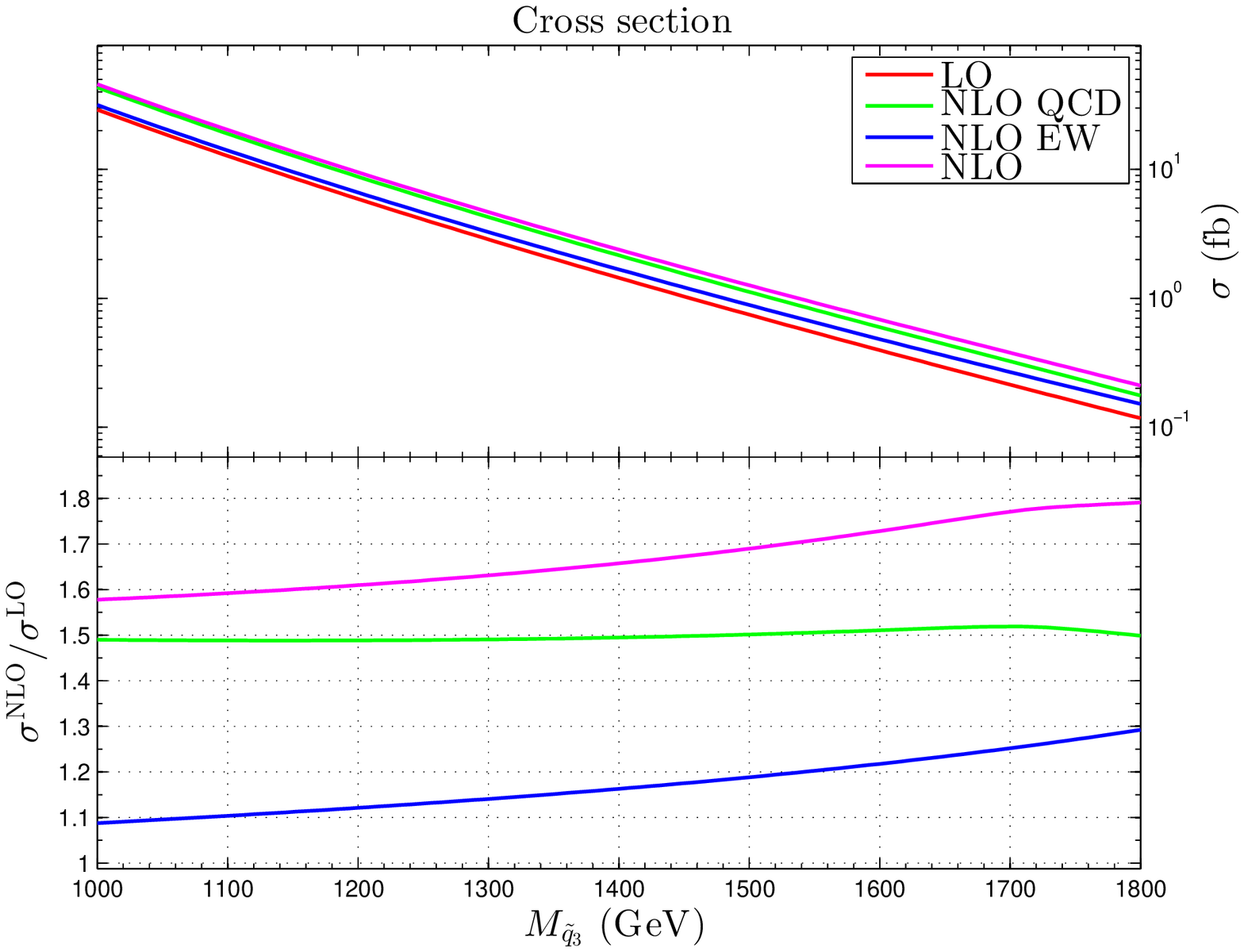}
\caption{}
\end{subfigure}
\begin{subfigure}[b]{0.5\textwidth}
\includegraphics[width=7.4cm,height=6.3cm]{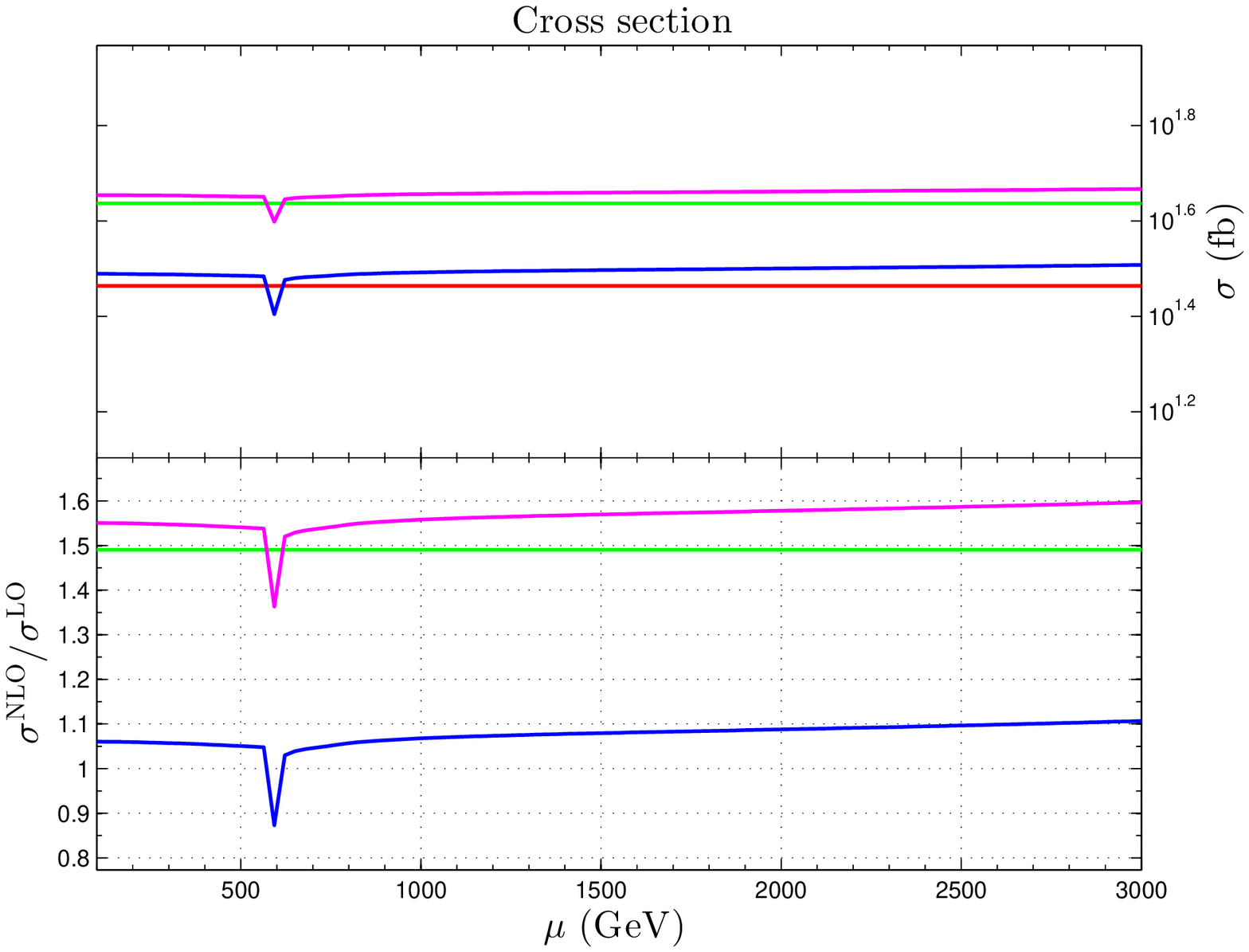}
\caption{}
\end{subfigure}
\phantom{pic}  \\
\begin{subfigure}[b]{0.5\textwidth}
\includegraphics[width=7.4cm,height=6.3cm]{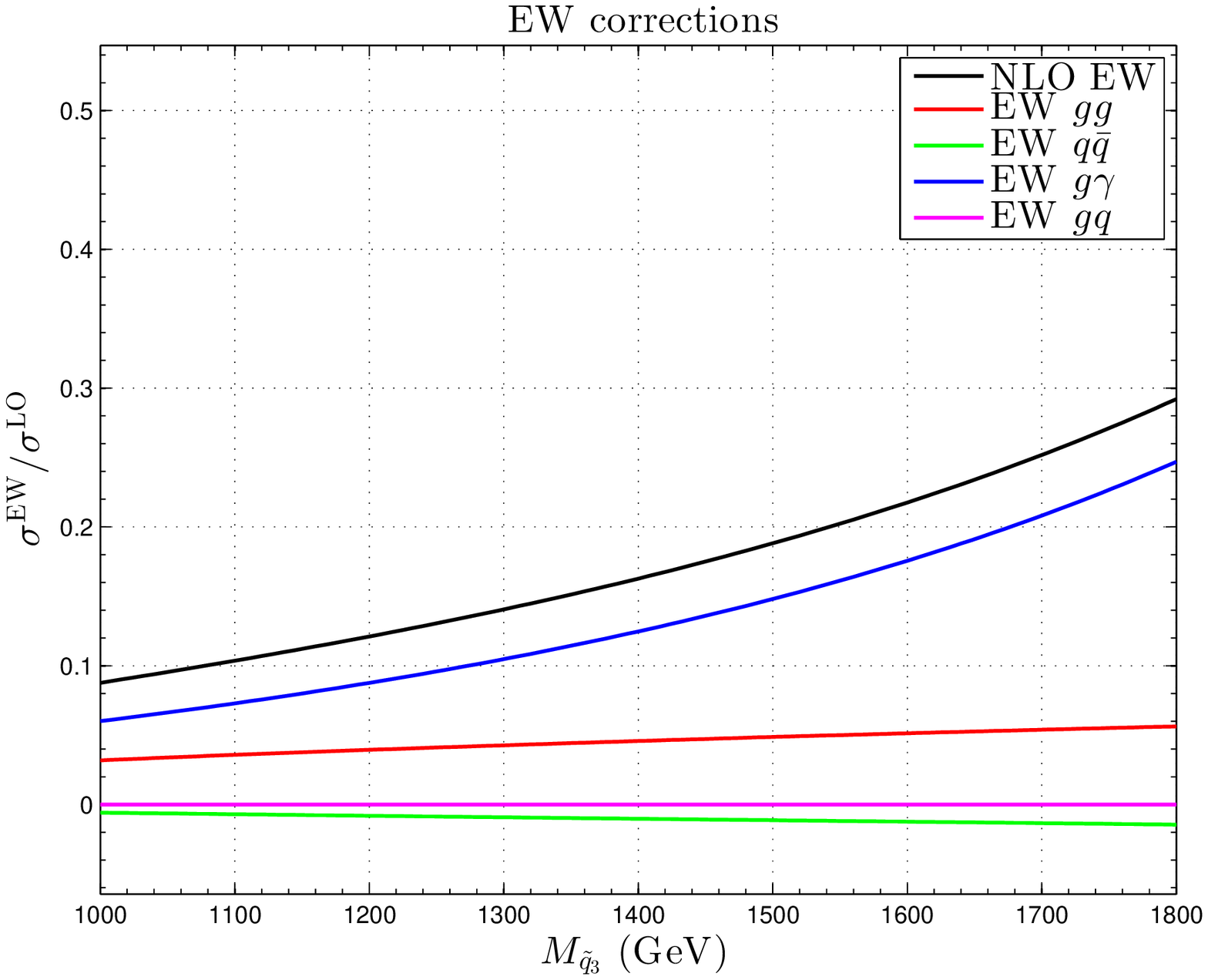}
\caption{}
\end{subfigure}
\begin{subfigure}[b]{0.5\textwidth}
\includegraphics[width=7.4cm,height=6.3cm]{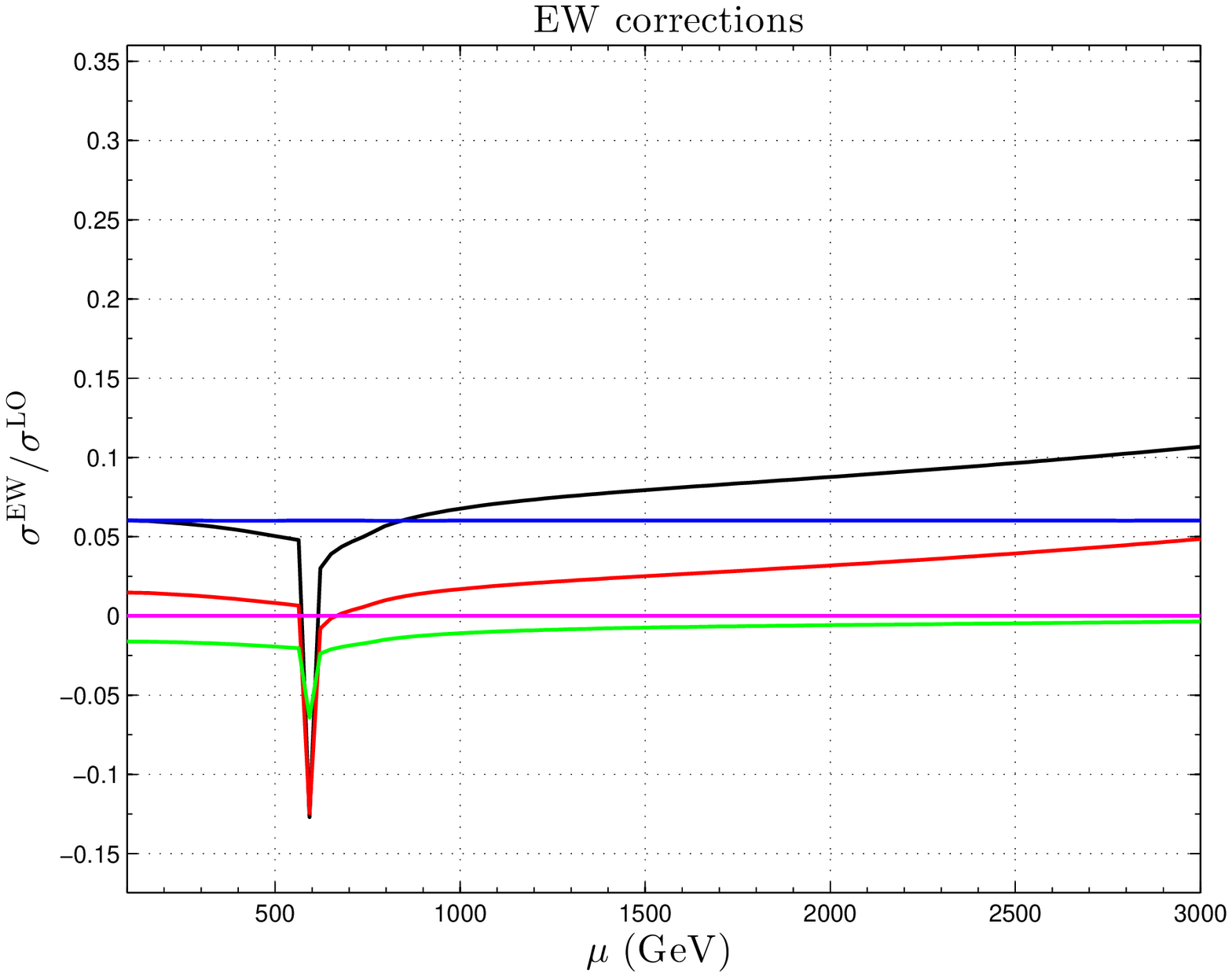}
\caption{}
\end{subfigure}
\phantom{pic}  \\
\begin{subfigure}[b]{0.5\textwidth}
\includegraphics[width=7.4cm,height=6.3cm]{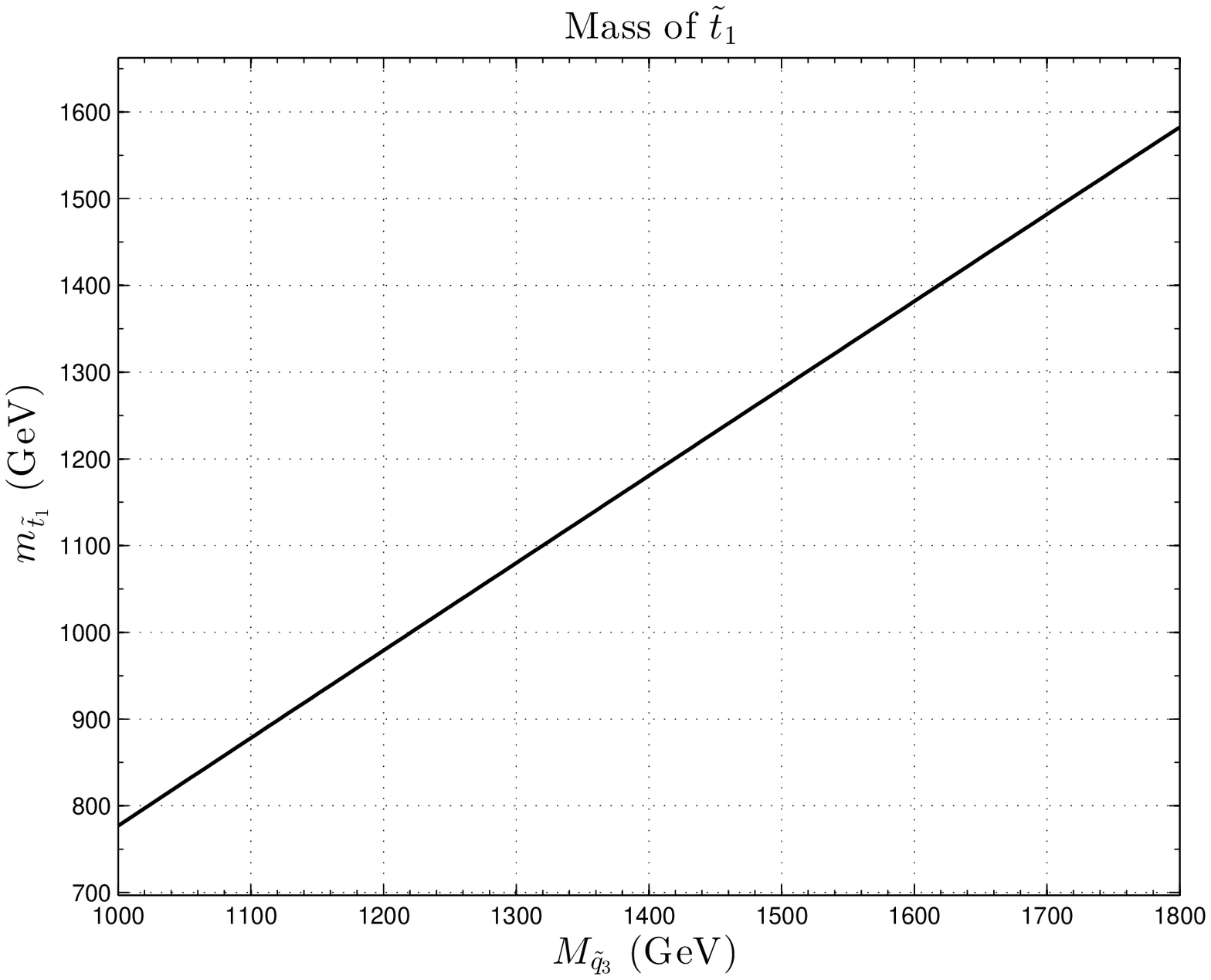}
\caption{}
\end{subfigure}
\begin{subfigure}[b]{0.5\textwidth}
\includegraphics[width=7.4cm,height=6.3cm]{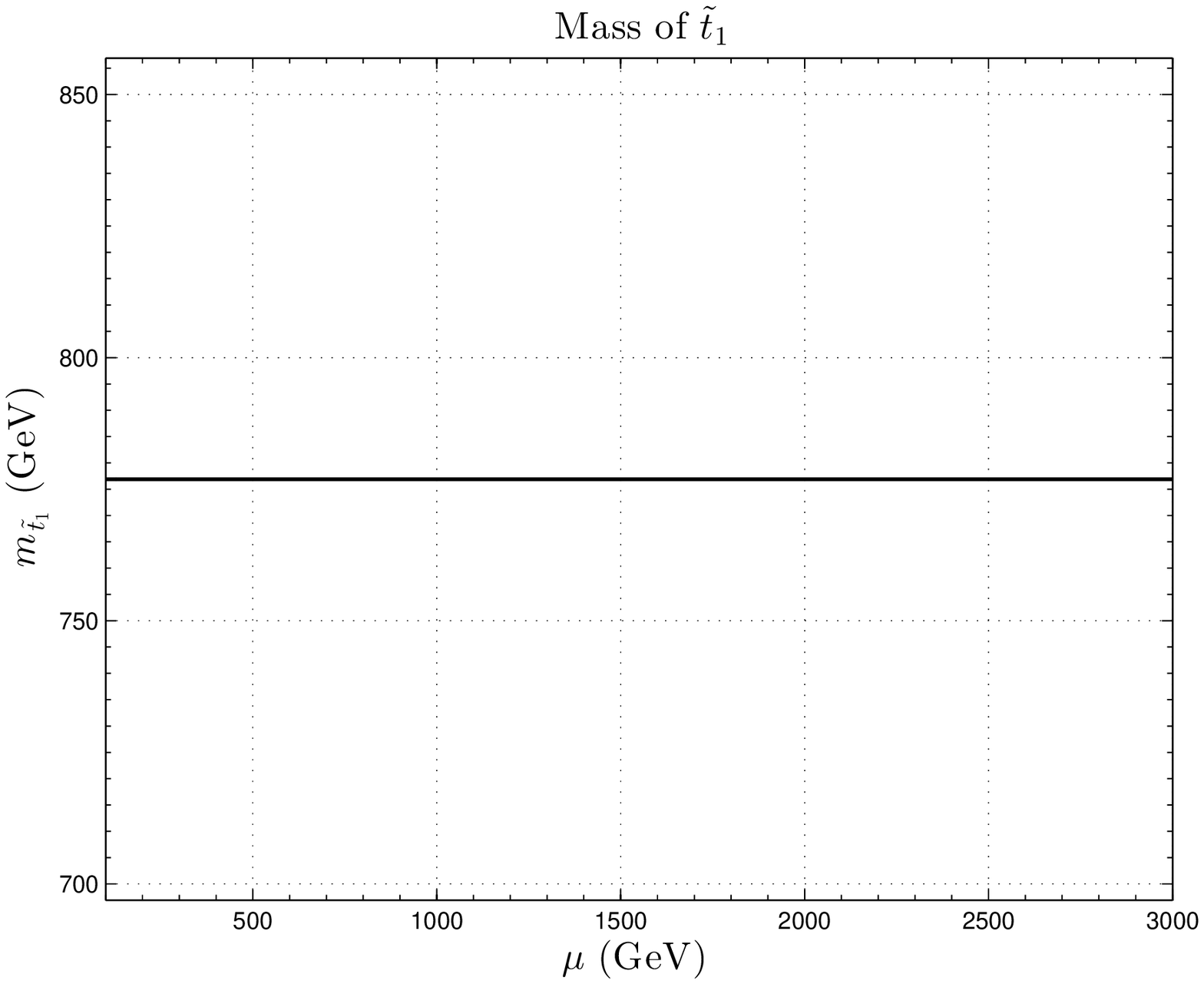}
\caption{}
\end{subfigure}
\caption[.]{Same as Fig.~\ref{fig:LightStop}, but for the tau-phobic scenario. The value of the  parameters
not involved in the scans are collected in Table~\ref{Tab:Bench}(d).}
\label{fig:Tauphobic}
 \end{figure}

\begin{figure}[t]
\begin{subfigure}[b]{0.5\textwidth}
\includegraphics[width=7.4cm,height=6.3cm]{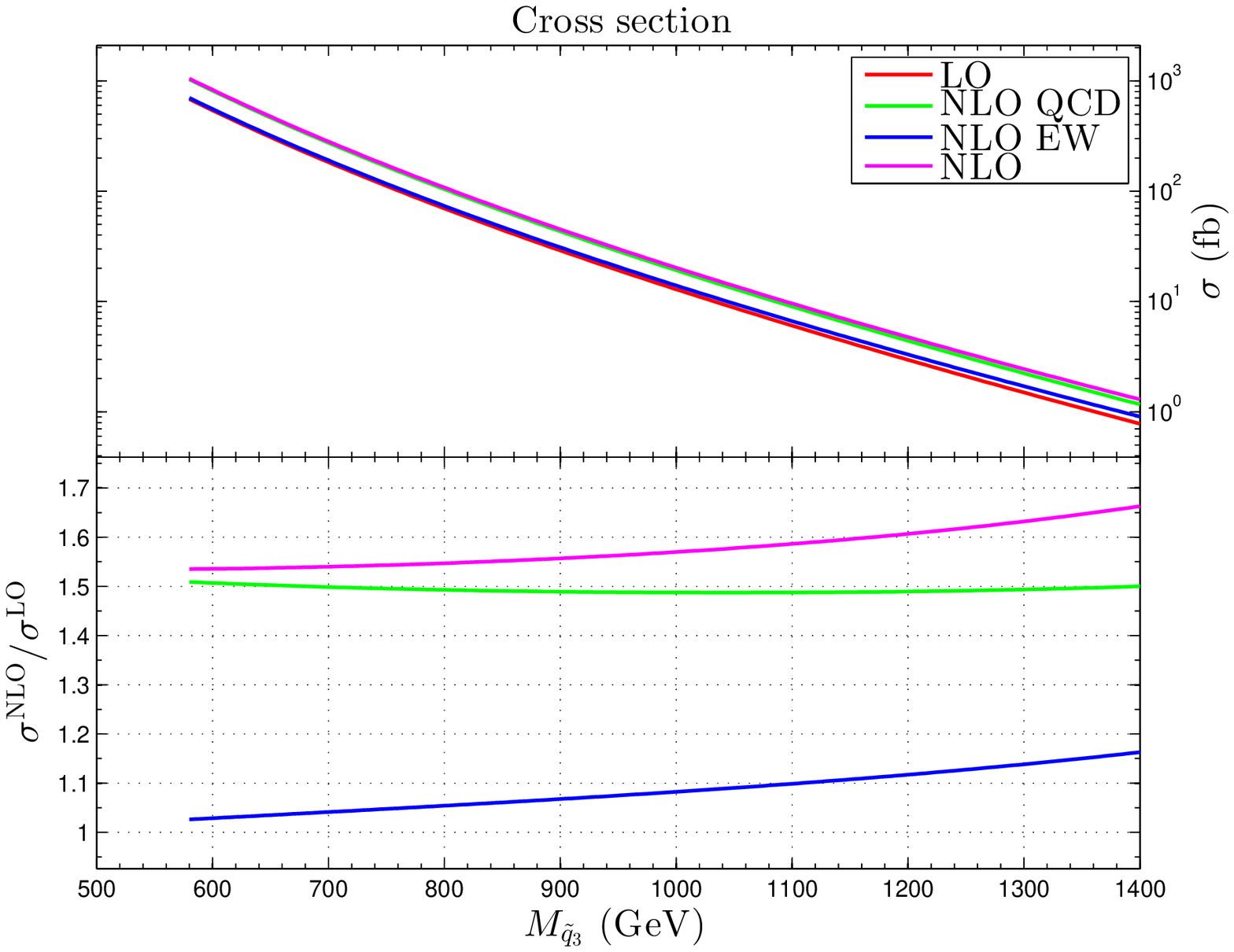}
\caption{}
\end{subfigure}
\begin{subfigure}[b]{0.5\textwidth}
\includegraphics[width=7.4cm,height=6.3cm]{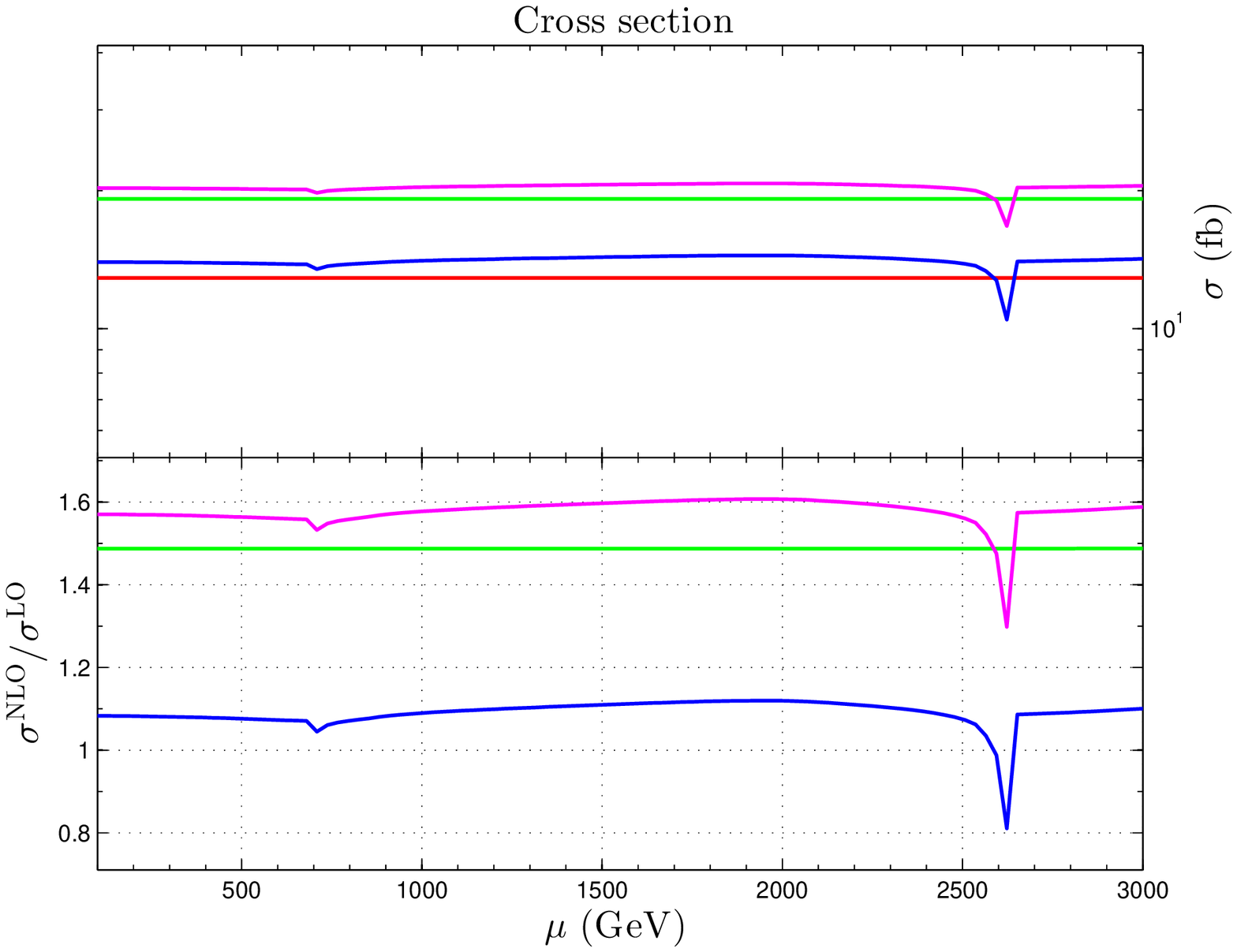}
\caption{}
\end{subfigure}
\phantom{pic}  \\
\begin{subfigure}[b]{0.5\textwidth}
\includegraphics[width=7.4cm,height=6.3cm]{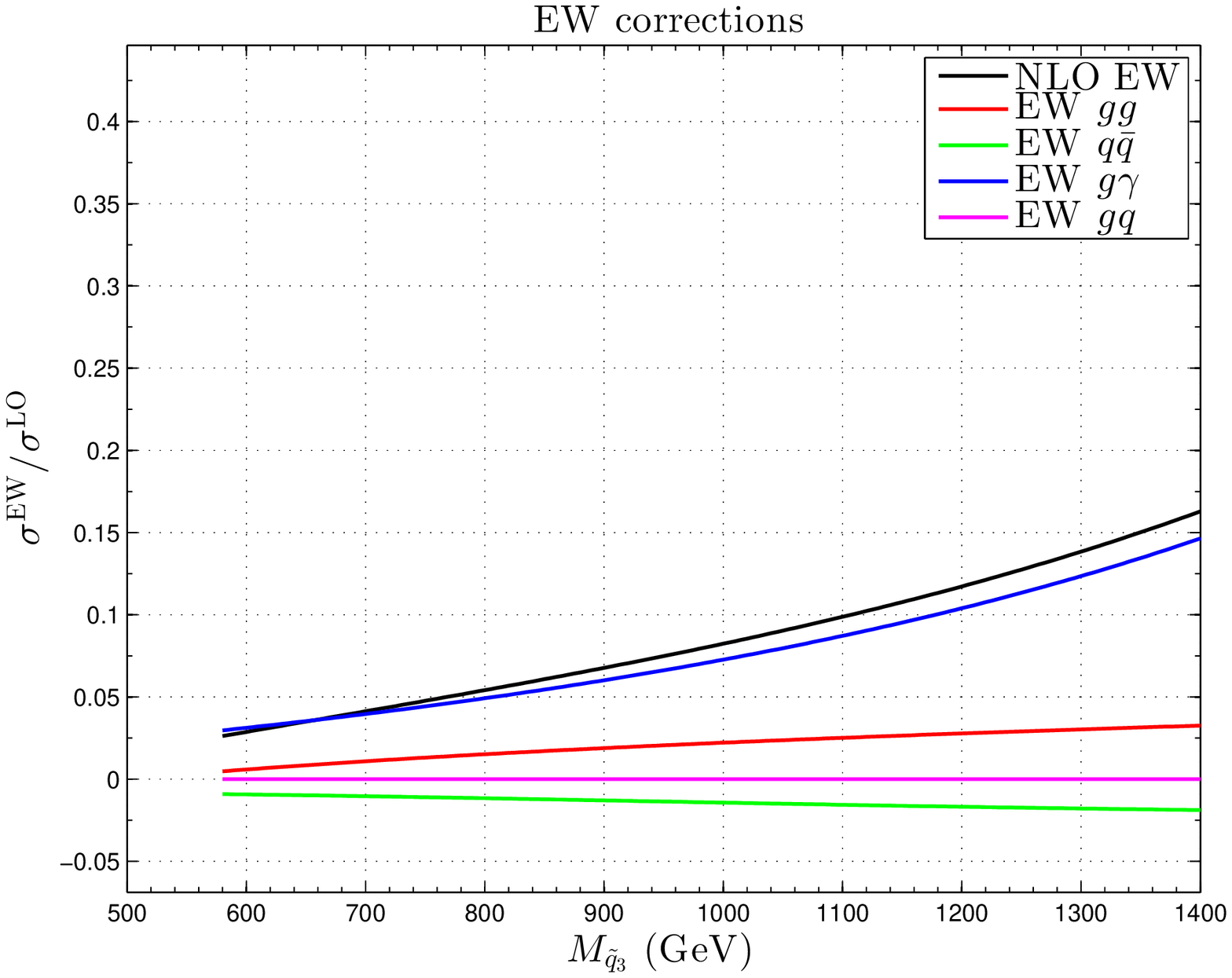}
\caption{}
\end{subfigure}
\begin{subfigure}[b]{0.5\textwidth}
\includegraphics[width=7.4cm,height=6.3cm]{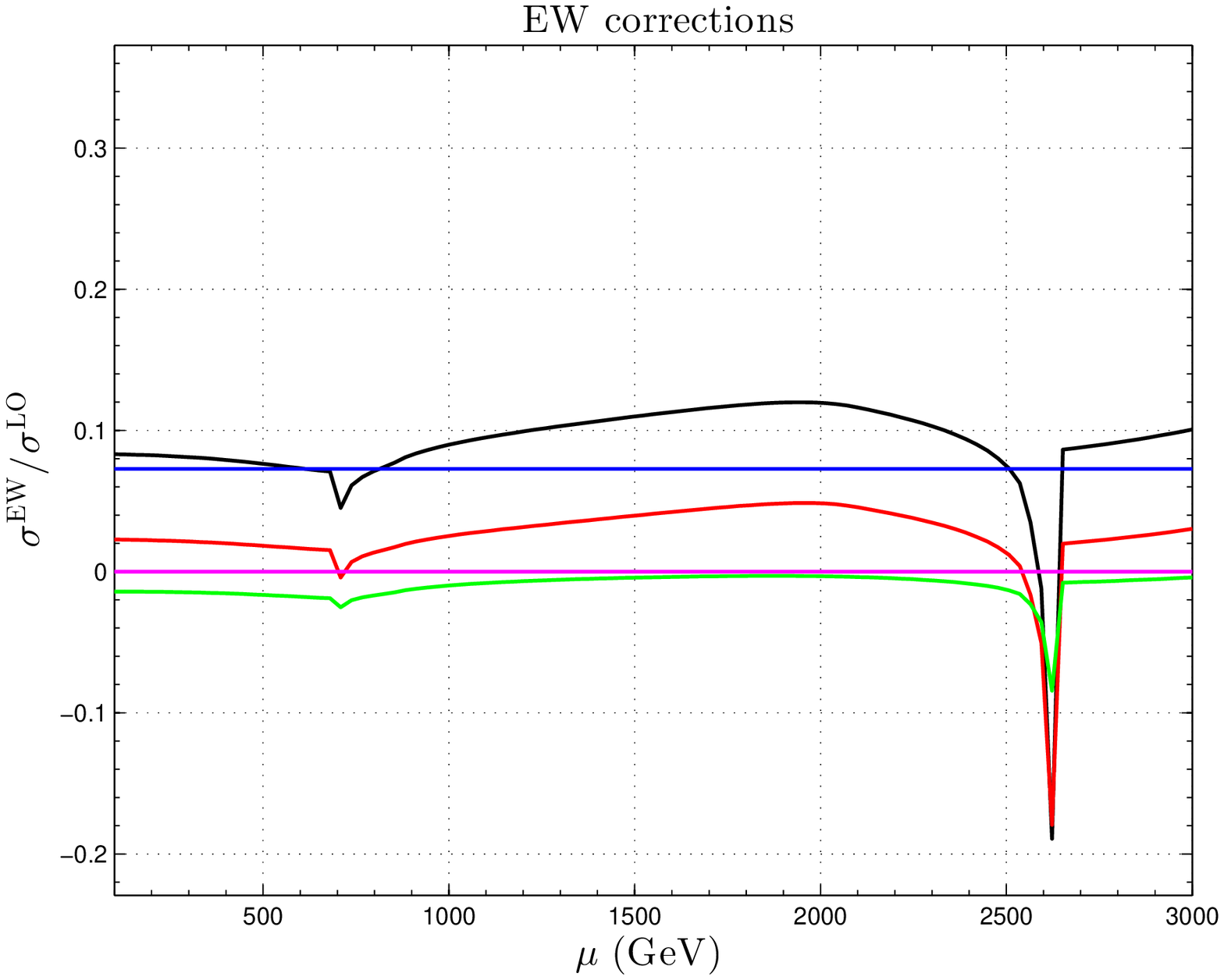}
\caption{}
\end{subfigure}
\phantom{pic}  \\
\begin{subfigure}[b]{0.5\textwidth}
\includegraphics[width=7.4cm,height=6.3cm]{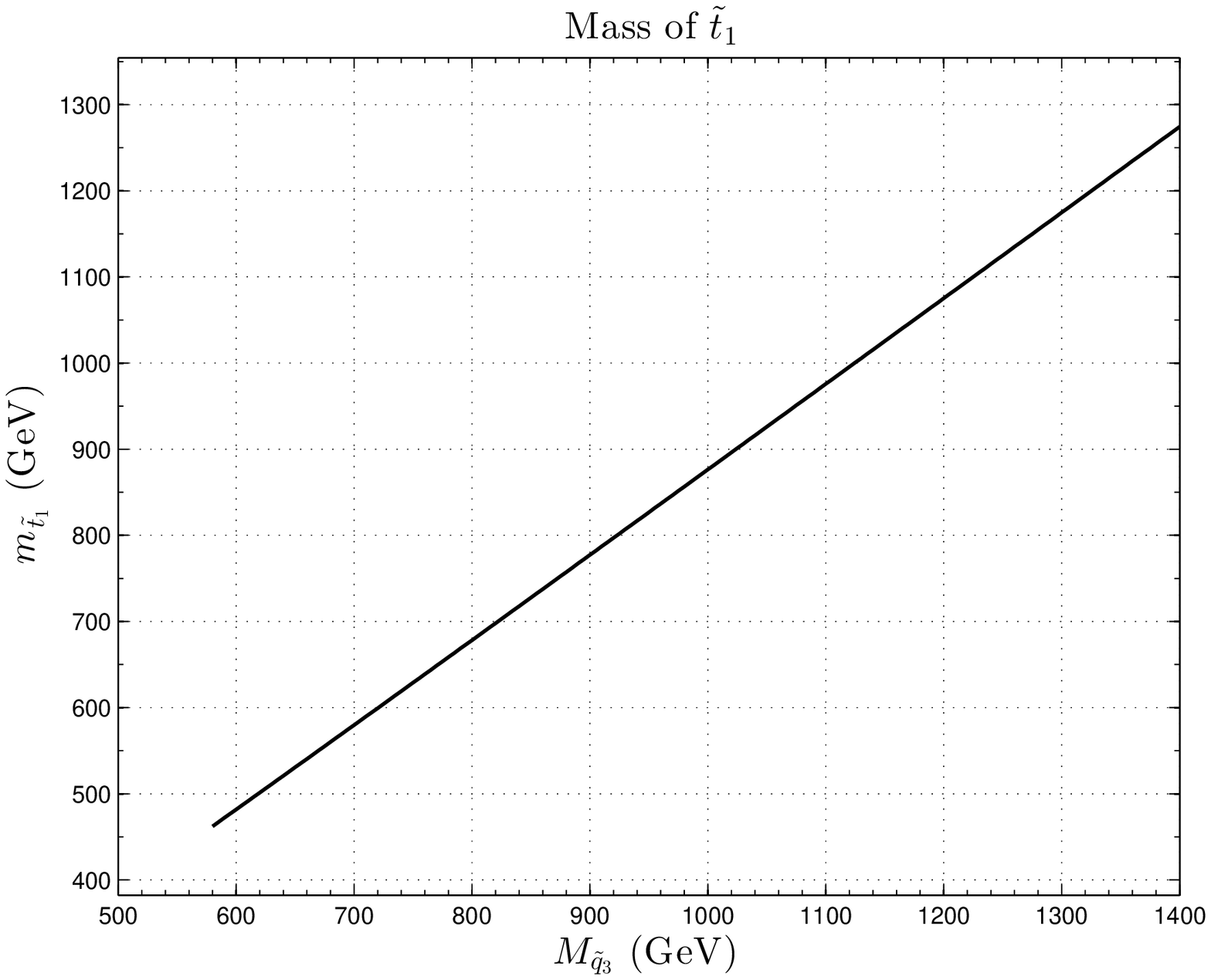}
\caption{}
\end{subfigure}
\begin{subfigure}[b]{0.5\textwidth}
\includegraphics[width=7.4cm,height=6.3cm]{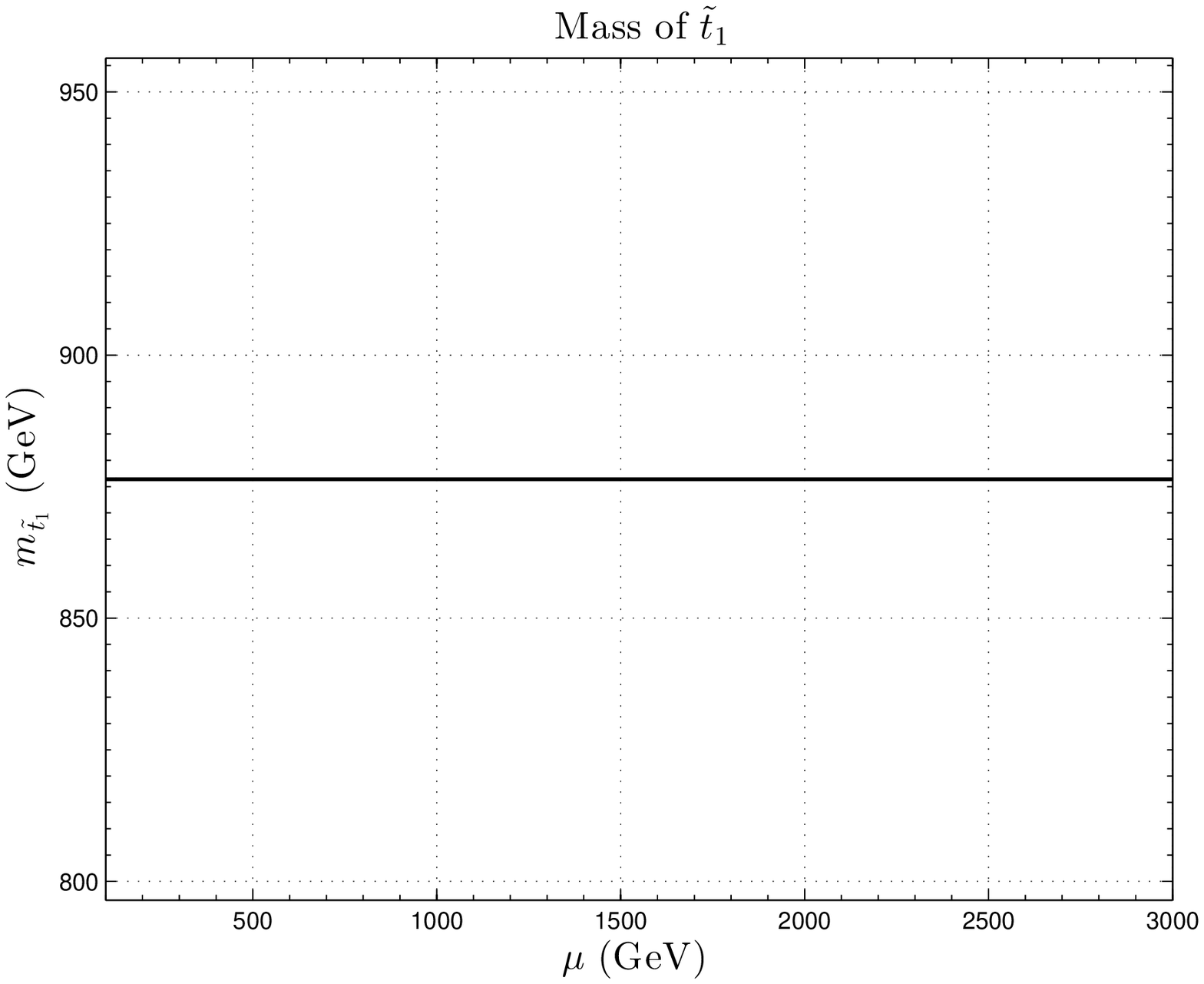}
\caption{}
\end{subfigure}
\caption[.]{Same as Fig.~\ref{fig:LightStop}, but for the $m^{\mbox{\tiny mod }+}_h$ scenario. The value of the  parameters
not involved in the scans are collected in Table~\ref{Tab:Bench}(e).}
\label{fig:Modp}
 \end{figure}

\begin{figure}[t]
\begin{subfigure}[b]{0.5\textwidth}
\includegraphics[width=7.4cm,height=6.3cm]{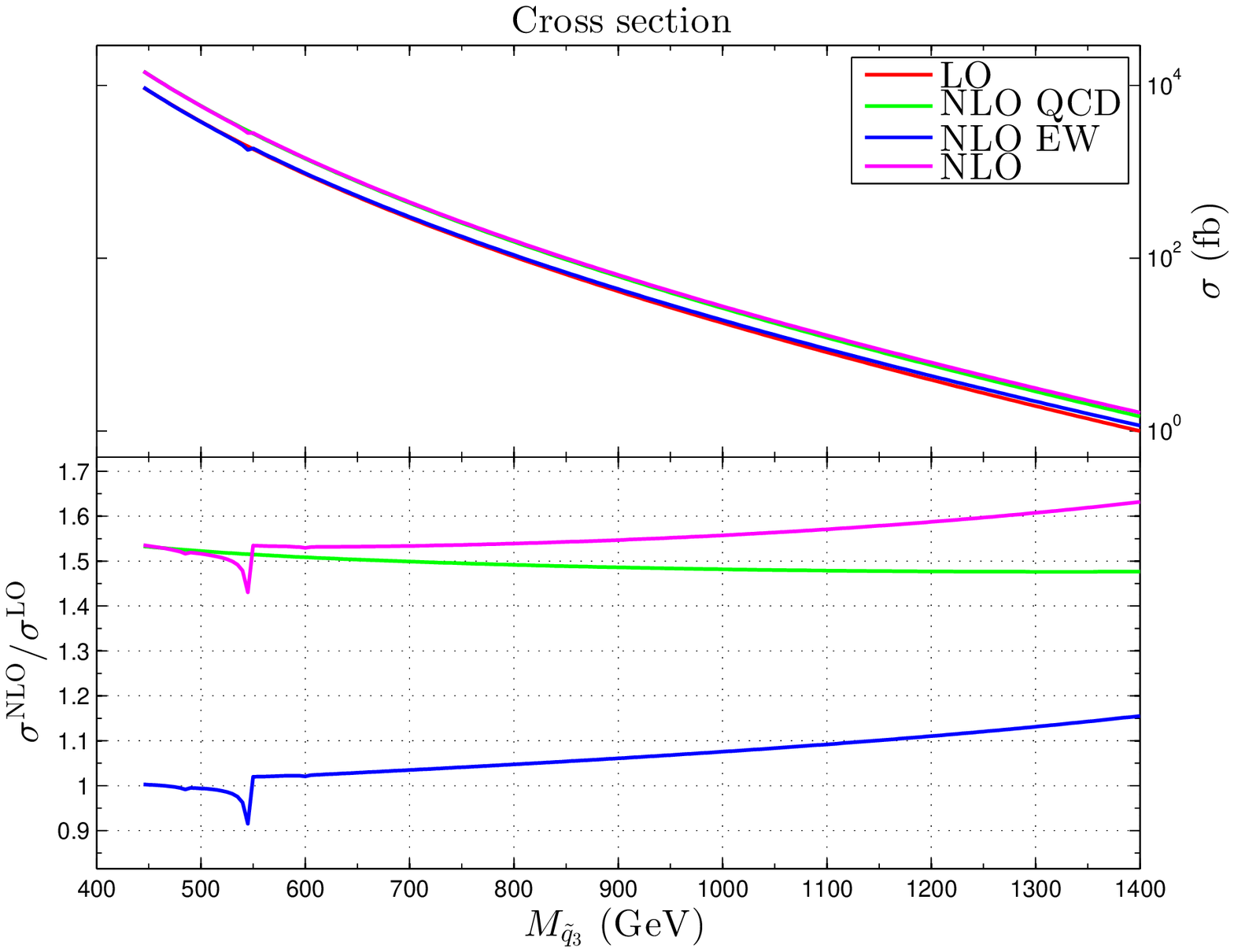}
\caption{}
\end{subfigure}
\begin{subfigure}[b]{0.5\textwidth}
\includegraphics[width=7.4cm,height=6.3cm]{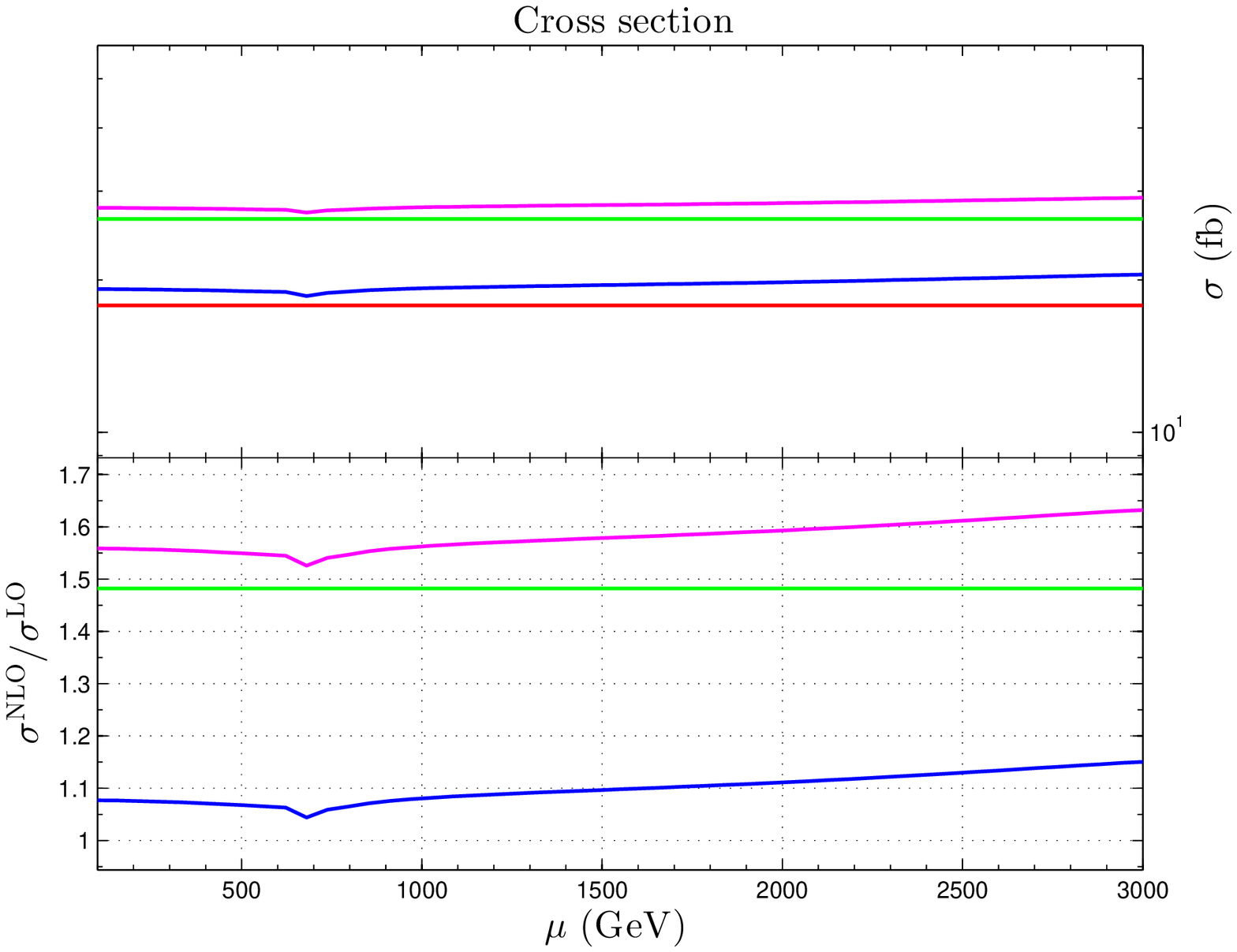}
\caption{}
\end{subfigure}
\phantom{pic}  \\
\begin{subfigure}[b]{0.5\textwidth}
\includegraphics[width=7.4cm,height=6.3cm]{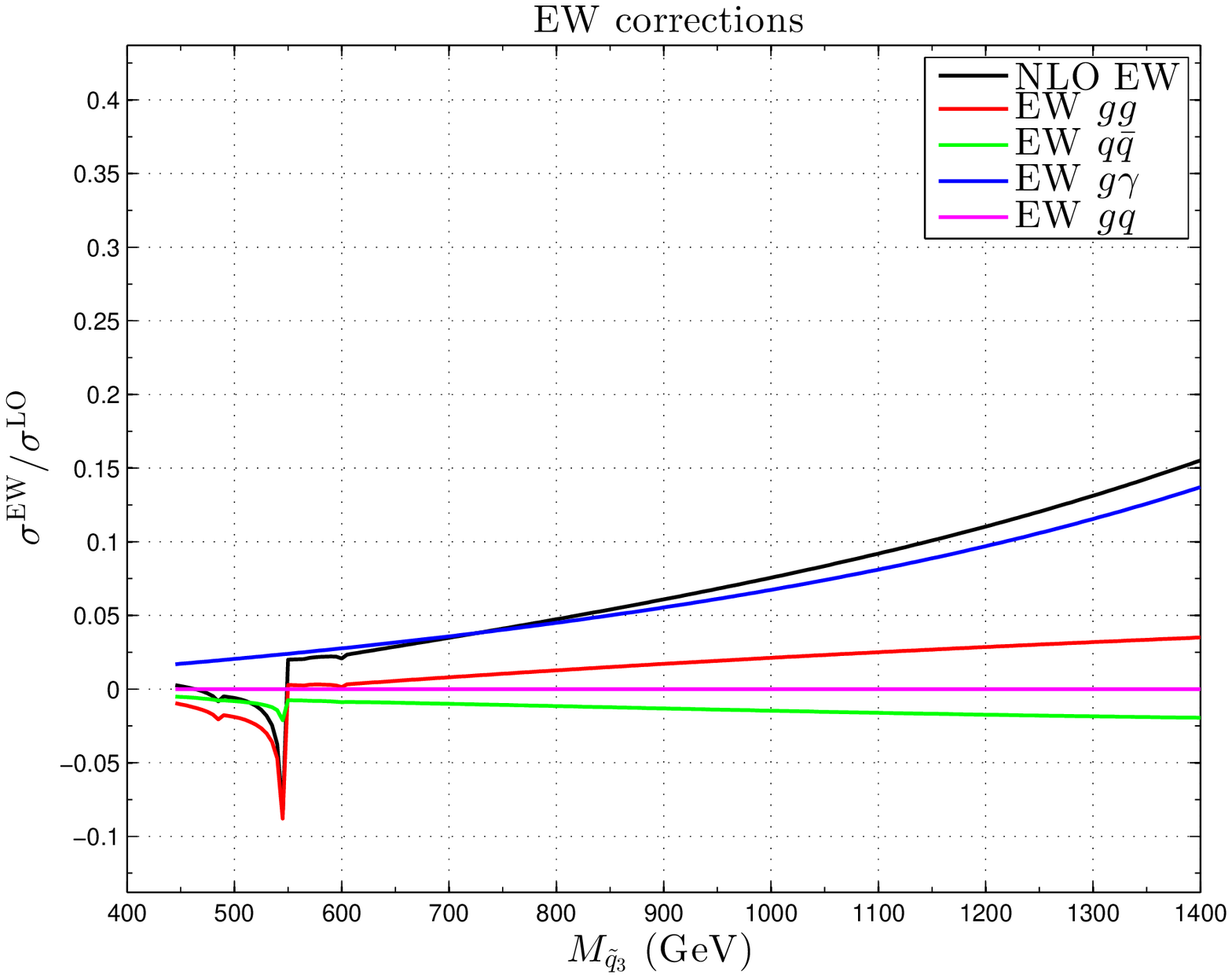}
\caption{}
\end{subfigure}
\begin{subfigure}[b]{0.5\textwidth}
\includegraphics[width=7.4cm,height=6.3cm]{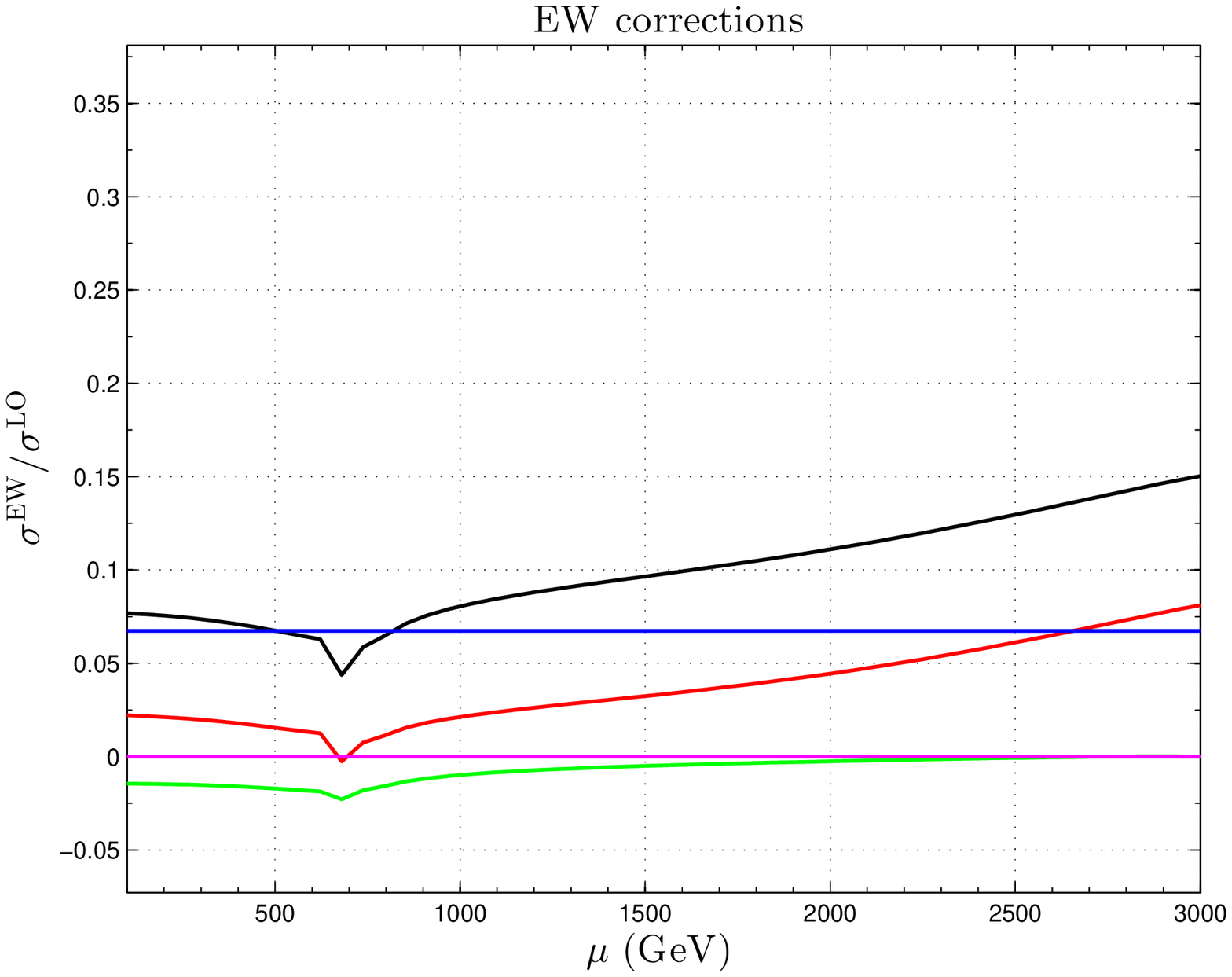}
\caption{}
\end{subfigure}
\phantom{pic}  \\
\begin{subfigure}[b]{0.5\textwidth}
\includegraphics[width=7.4cm,height=6.3cm]{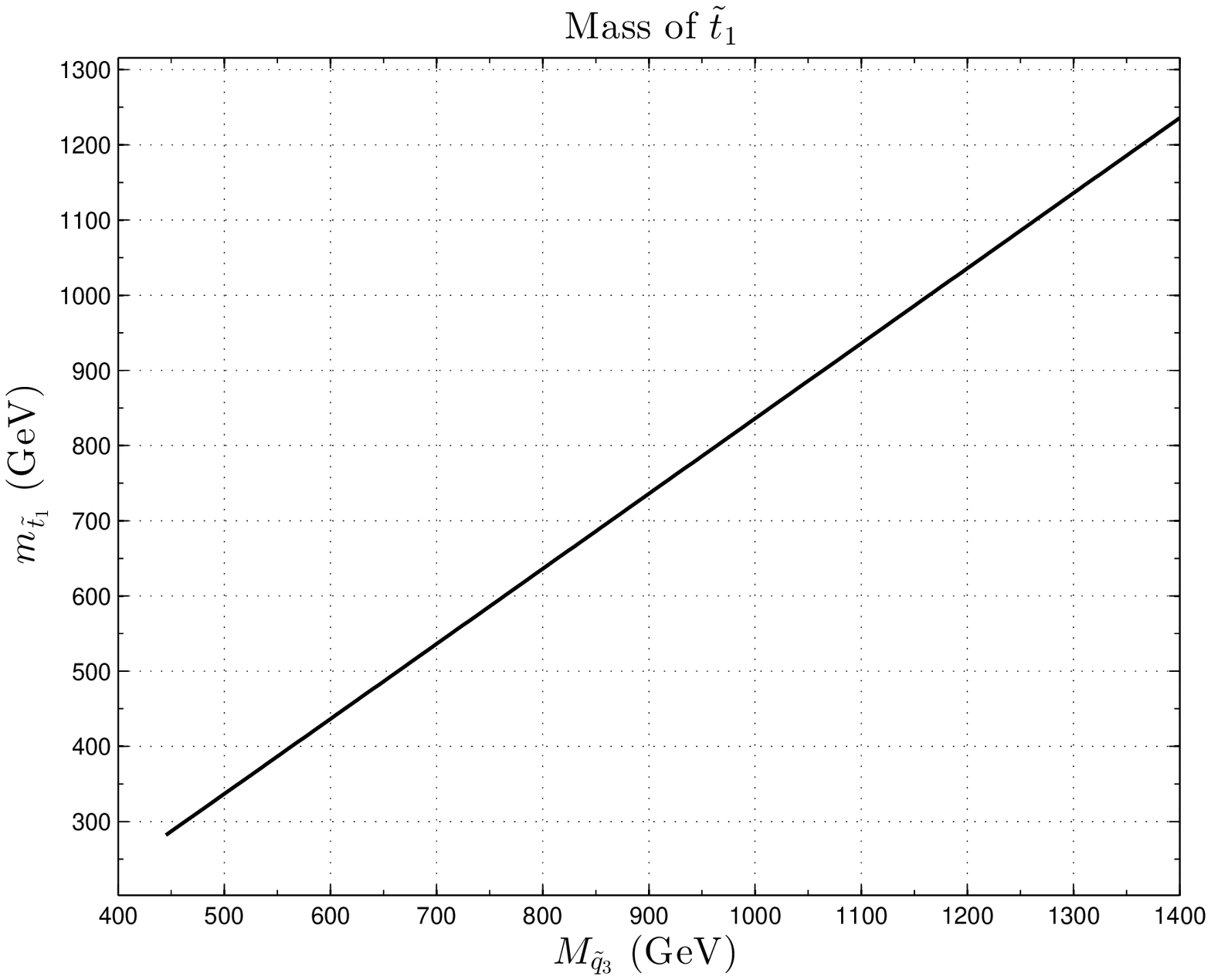}
\caption{}
\end{subfigure}
\begin{subfigure}[b]{0.5\textwidth}
\includegraphics[width=7.4cm,height=6.3cm]{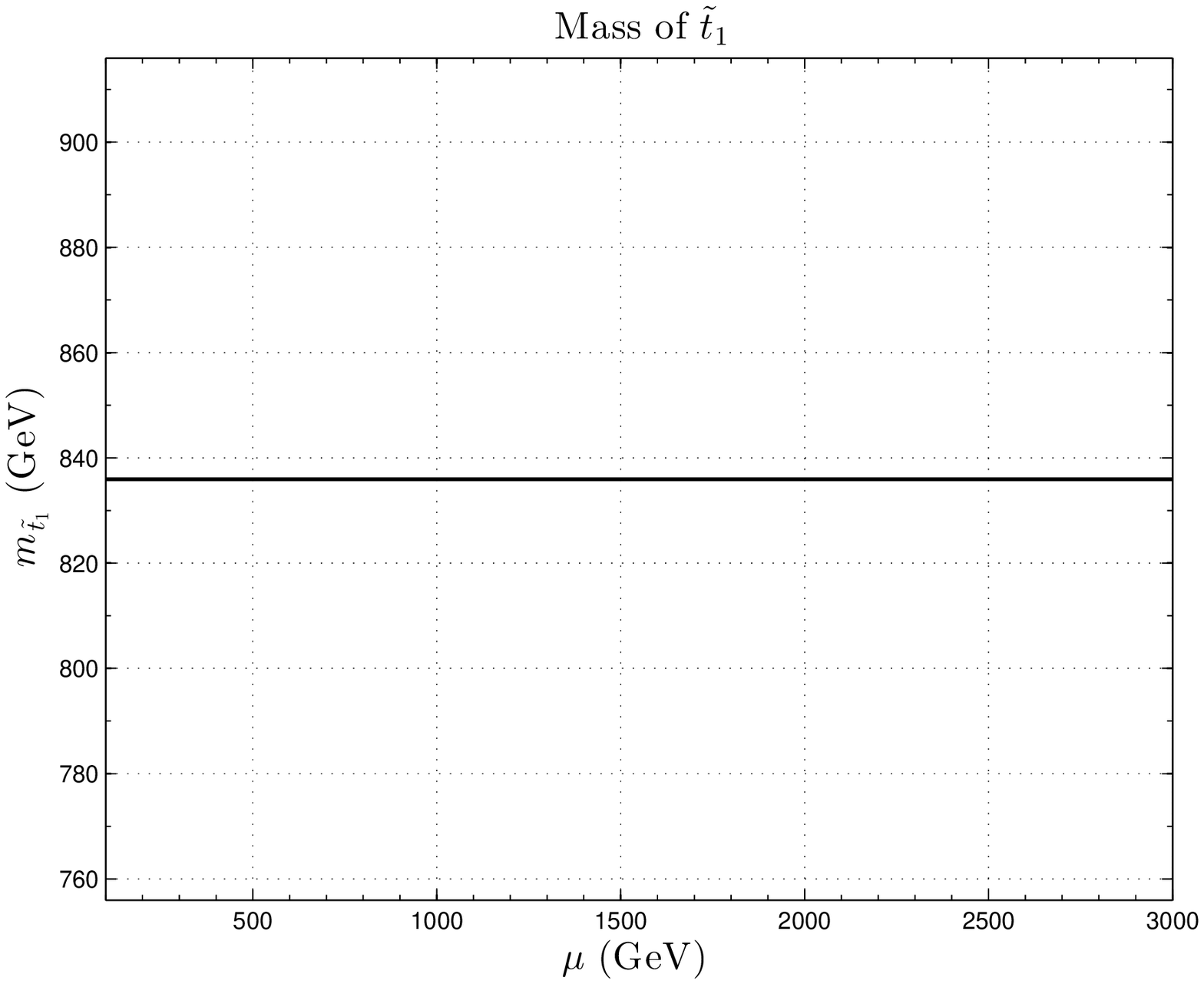}
\caption{}
\end{subfigure}
\caption[.]{Same as Fig.~\ref{fig:LightStop}, but for the $m^{\mbox{\tiny mod }-}_h$ scenario. The value of the  parameters
not involved in the scans are collected in Table~\ref{Tab:Bench}(f).}
\label{fig:Modm}
 \end{figure}

The results of the scans are shown in Figs.~\ref{fig:LightHiggs1}-\ref{fig:Modm}. 
Each Figure collects six plots related to  two different scans.  In 
the upper plots,  (a) and (b), we show the inclusive LO cross section defined in 
Eq.~(\ref{Eq:LO}) and the NLO cross section predictions 
defined in Eqs.~(\ref{Eq:NLO}). The yield of the NLO contribution relative to the LO cross section is displayed as well. 
The panels (c) and (d) show the  individual
EW contributions of the various channels defined in Eqs.~(\ref{Eq:NLOEWchannels}) as well
as the total NLO EW contribution, Eq.~(\ref{Eq:NLOEW}).  
They also show the impact of the various channels relative to the LO cross section.  
Panels (e) and (f) display the variation of  the mass of the produced $\stop$.

\begin{table}[t]
\begin{subtable}[l]{0.5\textwidth}
\centering
\begin{tabular}{ c | c   ||  c |  c }
\hline
\hline
 $m_{A_0}$  & $669$  GeV    & $\tan \beta$ & $16.5$ \\
 $X_t$ & $2000$ GeV  &  $\mu$ & $2640$ GeV \\
 $M_2$          & $201$ GeV     &  $m_{\tilde g}$ & $1000$ GeV \\
 $M_{\tilde q_{1,2}}$ & $1000$  GeV& $M_{\tilde \ell_{1,2}}$ & $300$ GeV\\
  $M_{\tilde q_{3}}$ & $1000$ GeV   & $M_{\tilde \ell_3}$ & $285$  GeV \\
 \hline
 \hline
\end{tabular}
\subcaption{Light-Higgs scenario}
\end{subtable}
\begin{subtable}[l]{0.5\textwidth}
\centering
\begin{tabular}{ c | c   ||  c |  c }
\hline
\hline
 $m_{A_0}$  & $700$    GeV & $\tan \beta$ & $20$ \\
  $X_t$ & $2 M_{\tilde q_3}$       &   $\mu$ & $350$ GeV  \\
 $M_2$          & $350$     GeV   &  $m_{\tilde g}$ & $1500$   GeV  \\
 $M_{\tilde q_{1,2}}$ & $1500$ GeV &     $M_{\tilde \ell_{1,2}}$ & $500$    GeV \\
  $M_{\tilde q_{3}}$ & $500$ GeV   &  $M_{\tilde \ell_3}$ & $1000$    GeV \\
\hline
\hline
\end{tabular}
\subcaption{Light-stop scenario}
\end{subtable} 
\phantom{plot} \\
\begin{subtable}[l]{0.5\textwidth}
\centering
\begin{tabular}{ c | c   ||  c |  c }
\hline
\hline
 $m_{A_0}$  & $700$ GeV    & $\tan \beta$ & $20$ \\
 $X_t$ & $1.6 \,M_{\tilde q_3}$     & $\mu$ & $500$ GeV   \\
 $M_2$          & $200$  GeV    &  $m_{\tilde g}$ & $1500$ GeV \\
 $M_{\tilde q_{1,2}}$ & $1500$  GeV  & $M_{\tilde \ell_{1,2}}$ & $500$ GeV \\
  $M_{\tilde q_{3}}$ & $1000$ GeV   &    $M_{\tilde \ell_3}$ & $245$ GeV \\
\hline
\hline
\end{tabular}
\subcaption{Light-stau scenario}
\end{subtable}
\begin{subtable}[l]{0.5\textwidth}
\centering
\begin{tabular}{ c | c   ||  c |  c }
\hline
\hline
 $m_{A_0}$  & $800$  GeV    & $\tan \beta$ & $15$ \\
$X_t$ & $2.45 \, M_{\tilde q_3}$        &       $\mu$ & $2000$ GeV  \\
 $M_2$          & $200$   GeV     &  $m_{\tilde g}$ & $1500$  GeV   \\
 $M_{\tilde q_{1,2}}$ & $1500$  GeV   & $M_{\tilde \ell_{1,2}}$ & $500$  GeV   \\
  $M_{\tilde q_{3}}$ & $1000$ GeV   &   $M_{\tilde \ell_3}$ & $500$  GeV   \\
 \hline
 \hline
\end{tabular}
\subcaption{Tau-phobic scenario}
\end{subtable}
\phantom{plot} \\
\begin{subtable}[l]{0.5\textwidth}
\centering
\begin{tabular}{ c | c   ||  c |  c }
\hline
\hline
 $m_{A_0}$  & $800$  GeV      & $\tan \beta$ & $30$ \\
 $X_t$ &  $1.5 \, M_{\tilde q_3}$   &     $\mu$ & $200$ GeV \\
 $M_2$          & $200$  GeV      &  $m_{\tilde g}$ & $1500$ GeV    \\
 $M_{\tilde q_{1,2}}$ & $1500$ GeV    & $M_{\tilde \ell_{1,2}}$ & $500$GeV    \\
  $M_{\tilde q_{3}}$ & $1000$ GeV   &    $M_{\tilde \ell_3}$ & $1000$ GeV   \\
 \hline
 \hline
\end{tabular}
\subcaption{$m^{\mbox{\tiny mod }+}_h$ scenario}
\end{subtable}
\begin{subtable}[l]{0.5\textwidth}
\centering
\begin{tabular}{ c | c   ||  c |  c }
\hline
\hline
 $m_{A_0}$  & $800$  GeV     & $\tan \beta$ & $30$   \\
  $X_t$ &  $-1.9 \, M_{\tilde q_3}$   &   $\mu$ & $200$ GeV    \\
 $M_2$          & $200$    GeV   &  $m_{\tilde g}$ & $1500$ GeV  \\
 $M_{\tilde q_{1,2}}$ & $1500$ GeV   & $M_{\tilde \ell_{1,2}}$ & $500$ GeV  \\
 $M_{\tilde q_{3}}$ & $1000$ GeV   &   $M_{\tilde \ell_3}$ & $1000$ GeV  \\
\hline
\hline
\end{tabular}
\subcaption{$m^{\mbox{\tiny mod }-}_h$ scenario}
\end{subtable}
\caption{Benchmark scenarios within our ten-parameter phenomenological MSSM.}
\label{Tab:Bench}
\end{table}

\paragraph{Light-Higgs scenario} This scenario is inspired by the best-fit point of  Ref.~\cite{Bechtle:2012jw}. 
In that study a fit of phenomenological MSSM scenarios compatible with electroweak precision observables and with 
experimental searches at the LHC and the Tevatron was performed.  
The values of the parameters characterizing this scenario are shown in Table~\ref{Tab:Bench}(a). 
The left (right) panels of Fig.~\ref{fig:LightHiggs1} show the results of the scan over $X_t$ ($M_{\tilde q_3}$),
and Fig.~\ref{fig:LightHiggs3}  displays the results of two scans over $\mu$.
Additional scans over $M_2$ and $m_{\tilde{g}}$ are very insensitive to the values of these parameters and we do not
include them here.
 
The dependence of the total cross section on $X_t$ is mild and basically related to the dependence of $m_{\tilde t_1}$ on $X_t$. On the other hand, the total cross section depends strongly on $M_{\tilde q_3}$ and varies over 
four   orders of magnitude in the considered parameter range, cfr. Fig.~\ref{fig:LightHiggs1}(b). For $M_{\tilde q_3} \ge 1400$ GeV ($m_{\stop} \ge 1300$ GeV) the NLO cross section drops below $1$~fb. Correspondingly 
high luminosities  are required  to exclude this  scenario. In the high $M_{\tilde q_3}$ region the relative impact of the EW corrections become more important, 
up to $20-30 \%$ of the LO cross section. 
The relative impact of the EW corrections is enhanced by the behavior of the QCD corrections whose relative yield  decreases from $50$ to $40 \%$ above the $\stop \to \tilde g t$ threshold, 
$M_{\tilde q_3} \simeq 1300$~GeV.  As  can be inferred from Fig.~\ref{fig:LightHiggs1}(d), the  EW corrections from the $q\bar q$ and $qg$ channels are  negligible, while the ones from the $gg$-channel are  of the order of 
$5 \%$ of the LO cross section, irrespective to the value of $M_{\tilde q_3}$.  In this scan the $g \gamma$-channel dominates the EW corrections for $M_{\tilde q_3} \ge 1000$ GeV. Its inclusion in experimental studies would 
significantly reduce the theoretical uncertainties, i.e. the systematic uncertainty originated from neglecting the EW corrections.

%
The LO and NLO QCD cross section is independent of $\mu$, see Fig.~\ref{fig:LightHiggs3}(a) and  Fig.~\ref{fig:LightHiggs3}(b). 
However, the NLO EW and thus also the full NLO cross section 
do depend on $\mu$. The NLO EW corrections double their impact on the LO cross section when $\mu$ varies from $500$ to $2500$~GeV. 
In the left part of Fig.~\ref{fig:LightHiggs3}, where we take $M_{\tilde q_3} = 1000$ GeV, 
their contributions are below $15 \%$, i.e. four times smaller than those of the QCD corrections. 
The right scan, where we assume $M_{\tilde q_3} = 1250$~GeV, 
exhibits larger EW corrections: they amount to more than $15 \%$ of 
the LO cross section for $\mu \ge 1500$~GeV. 
The relative impact of the various channels of the EW corrections for
$M_{\tilde q_3} = 1000$  GeV and $M_{\tilde q_3} = 1260$~GeV  is shown 
in Fig.~\ref{fig:LightHiggs3}(c) and Fig.~\ref{fig:LightHiggs3}(d)
respectively.  Again, the $q\bar q$ and $g q$-channel are negligible, while the contribution of the $g \gamma$-channel is constant and larger for $M_{\tilde q_3} = 1260$~GeV.
In the $M_{\tilde q_3} = 1000$  ($1250$)~GeV scan this channel contributes 
about $7 \%$ ($11 \%$) of the  LO cross section. Instead, the contributions from the $gg$-channel strongly depend on $\mu$ and increase as this parameter increases. For $M_{\tilde q_3} = 1000$~GeV it dominates the EW 
corrections for $\mu \ge 2700$~GeV. It is worth to note that   in both cases the inclusion of the tree-level, model-independent $g\gamma$-channel
would significantly reduce the EW theoretical uncertainty. In these scans we observe a $\stop \to \tilde \chi^0_3 t$ threshold at $\mu\simeq  650~(950)$ GeV for $M_{\tilde q_3} = 1000$  ($1250$)~GeV.

\paragraph{Light-stop scenario}  This scenario has been defined in Ref.~\cite{Carena:2013qia}  
in correspondence to the values of the parameters summarized in Table~\ref{Tab:Bench}(b).
The value of  $M_{\tilde q_3}$ is significantly below the TeV-scale, while $X_t$ is chosen to 
maximize the mass of the lightest CP-even Higgs boson $h$,  leading to a phenomenologically-viable scenario with a light stop.

The dependence of the total cross section on $M_{\tilde q_3}$ is shown in Fig.~\ref{fig:LightStop}(a). Both the LO and the NLO
cross section depend strongly on this parameter: their values span three orders of magnitude when
$M_{\tilde q_3}$ varies from $400$ to $1000$~GeV.  The relative impact of the QCD corrections is large, while its dependence on $M_{\tilde q_3}$ is mild: over the entire 
$M_{\tilde q_3}$ interval they are  $50-55 \%$  of the LO cross section.  Outside  the $\stop \to \tilde\chi_4^0 t$ threshold region, the EW corrections are positive and 
small, i.e. they are $8 \%$ of the LO cross section at most. The relative impact of the EW corrections of the different channels as a function of $M_{\tilde q_3} $ is shown 
in Fig.~\ref{fig:LightStop}(c). Outside of the threshold region the 
bulk of the EW contributions originate from the $g\gamma$-channel, owing to
mutual cancellations between the $gg$- and the $q \bar q$-channels.

Fig.~\ref{fig:LightStop}(b) shows the dependence of the total cross section on $\mu$. The LO and NLO total cross sections are independent of the value of $\mu$ in a vast portion of the scanned region. The total NLO cross 
section decreases in the high $\mu$ region, i.e. $\mu \ge 1200$~GeV, as a consequence of the decrease of the  EW corrections.  The variation 
is below $10\%$ and it takes place when approaching the $\stop \to \tilde b_1 W$  threshold  located in the phenomenologically excluded region, 
$M_{\tilde q_3} \ge 1780$~GeV.

As can be inferred from Fig.~\ref{fig:LightStop}(d), the $g\gamma$-channel dominates in the $\mu=300-600$~GeV range. For higher values of this parameter, $\mu = 700-1200$~GeV, the  $gg$- and the $g \gamma$-channel 
dominate and result in an overall negative NLO EW correction.  The inclusion of the $g\gamma$-channel only  would thus overestimate the impact of the  EW corrections.

\paragraph{Light-stau scenario} This scenario exhibits a sizable
mixing in the $\tilde{\tau}$ sector and a rather low value of $M_{\tilde  \ell_3}$, 
allowing for a light $\tilde{\tau}$ state and a possible enhancement of the  $h \to \gamma\gamma$ 
decay rate compared to its SM prediction~\cite{Carena:2013qia}.  The value of the parameters characterizing this scenario are listed in Table~\ref{Tab:Bench}(c).

In the left (right) plots of Fig.~\ref{fig:LightStau} we investigate the dependence on $M_{\tilde q_3}$ ($\mu$) within this  scenario.
The NLO QCD corrections to the total cross section do only marginally depend on $M_{\tilde q_3}$ and negligibly on $\mu$. They amount to $50 \%$ of the LO cross section, cfr. Fig.~\ref{fig:LightStau}(a) 
and~\ref{fig:LightStau}(b).

The dependence of the EW contributions on $M_{\tilde q_3}$ is less trivial. Below the $\stop \to  \tilde\chi^0_3 t$  threshold 
at $M_{\tilde q_3} \simeq 800$~GeV  the EW corrections are flat and small, below $5\%$ of 
the LO cross section. Above the threshold the corrections grow as $M_{\tilde q_3}$ grows reaching $10\%$ ($15 \%$) of the LO cross section for $M_{\tilde q_3} \simeq 1150$  ($1200$)~GeV, i.e. for 
$m_{\tilde t_1} \simeq 1050 $ ($1200$)~GeV. In the high $M_{\tilde q_3}$ region the cross section prediction is thus substantially modified by the inclusion of the EW corrections.  As shown in 
Fig.~\ref{fig:LightStau}(c), outside the threshold region the EW contributions are decently approximated by the $g \gamma$-channel only; 
indeed, the other channels contribute only up to $2\%$ of the LO cross section.

Here, the EW corrections are only mildly affected by the variation of $\mu$, cfr. Fig.~\ref{fig:LightStau}(b). Far from the threshold region, $\mu \simeq 700$~GeV, 
they are flat and of the order of $7-10 \%$ of the LO cross section.  The leading contribution  to the EW corrections is the $\mu-$independent contribution of the $g \gamma$-channel, which  below (above) the threshold is slightly 
(considerably) enhanced by the other channels.

\paragraph{Tau-phobic scenario}  This scenario belongs to the two-parameter family of scenarios also introduced in~\cite{Carena:2013qia};
here we define it by the parameters listed in  Table~\ref{Tab:Bench}(d). An intermediate value of $M_{\tilde \ell_3}$ together with a large value of $\mu$ lead 
to a somewhat suppressed $h \to \tau^+\tau^-$ decay rate, while all other Higgs observables remain SM-like.

The relative size of the QCD corrections, about $50 \%$ of the LO the cross section,
is not affected by the variation of either $M_{\tilde q_3}$ or $\mu$,
see Figs.~\ref{fig:Tauphobic}(a)  and~\ref{fig:Tauphobic}(b). 
The EW contributions, on the other hand, do depend on $M_{\tilde q_3}$; their relative size 
is tripled when $M_{\tilde q_3}$ varies from $1000$ to $1800$~GeV. 
Hence, the EW corrections are important and not negligible;
they are more than $10 \%$ of the LO cross section over almost the entire  region of $M_{\tilde q_3}$ considered, i.e.  $M_{\tilde q_3} \ge 1100$~GeV or, equivalently, 
$m_{\tilde t_1} \ge 900$~GeV. Moreover they can be as large as $25 \%$  of the LO
cross section in the large $M_{\tilde q_3}$ region, $M_{\tilde q_3} \ge 1700$~GeV.

The dependence of the EW corrections on $\mu$ is milder but nevertheless significant. The 
EW contributions relative to the LO cross section increase for increasing $\mu$ 
and double their value as $\mu$ varies from $100$ to $3000$~GeV reaching $10\%$ at $\mu \simeq 2500$~GeV, 
cfr. Fig.~\ref{fig:Tauphobic}(b).

In the considered parameter region mutual cancellations between the $gg$ and $q \bar q$ channels render the $g \gamma$-channel the dominant one.
Its inclusion would considerably lower the remaining EW theoretical uncertainty down to $5\%$ of the LO cross section.
The detailed  behavior, however, strongly depends on $\mu$,  as can be inferred from Fig.~\ref{fig:Tauphobic}(d).
In particular approximating the EW contributions by just the $g\gamma$-channel  becomes less and less accurate 
for increasing $\mu$: the  $g \gamma$-channel yields less than half of the EW 
corrections for $\mu \ge 2000$~GeV.

\paragraph{${\mathbf m^{\mbox{\tiny mod}\pm}_h}$ scenarios}  These scenarios are  defined in Ref.~\cite{Carena:2013qia} 
as  modifications of the $m^{\mbox{\tiny max }}_h$ scenario~\cite{Heinemeyer:1999zf}. They lead to 
a smaller mass of the lightest CP-even Higgs boson by reducing the ratio
\begin{displaymath}
r_t \equiv  \left | \frac{X_t }{M_{\tilde q_3} } \right | \, .
\end{displaymath}
The input parameters of the $ m^{\mbox{\tiny mod }+}_h$ and of the  $m^{\mbox{\tiny mod }-}_h$ scenario are collected in 
Table~\ref{Tab:Bench}(e) and Table~\ref{Tab:Bench}(f), respectively. 
They  differ in  the value of $r_t$ and in the sign of $X_t$.

The dependence of the NLO corrections on  $M_{\tilde q_3}$ in both scenarios is similar   
to the scenarios investigated before, see Fig.~\ref{fig:Modp}(a) and Fig.~\ref{fig:Modm}(a).   
The QCD corrections stay at  $50 \%$ of the LO cross section over the entire $M_{\tilde q_3}$ interval. 
The relative impact of the EW corrections increases with $M_{\tilde q_3}$. They
are larger than $10\%$ of the LO cross section for $M_{\tilde q_3} \ge 1100$~GeV,
that is for $m_{\tilde t_1}$ larger than $950$~GeV.  As shown in 
Fig.~\ref{fig:Modp}(c) and Fig.~\ref{fig:Modm}(c) mutual cancellations between the 
$q\bar q$ and $gg$-channel effects make the EW corrections effectively
equal to the contribution of the $g \gamma$-channel.

Fig.~\ref{fig:Modp}(b) shows the $\mu$-dependence of the NLO contributions  
in the $m^{\mbox{\tiny mod }+}_h$ scenario.  
Outside the $\stop \to \tilde\chi_4^0 t$ and the $\stop \to \tilde b_1 W$   threshold region, 
located at $\mu\simeq  710$ GeV and $\mu\simeq 2620$ GeV respectively,  the EW corrections relative 
to the LO cross section are almost flat and vary from $8\%$ to $12\%$ as $\mu$ varies in the considered interval. 
As can be inferred from Fig.~\ref{fig:Modp}(d), the largest EW contribution originates from the $g\gamma$-channel.
Outside the threshold regions the combined effect of the other channels are $2 \%$ of the LO cross section at 
most.

In the $m^{\mbox{\tiny mod }-}_h$ scenario the relative yield of the  
EW corrections doubles its value from $7\%$ to $15\%$ as $\mu$ varies from $100$ to $3000$ GeV.  
In the low $\mu$ region, i.e. below the  $\stop \to \tilde\chi_4^0 t$
threshold at  $\mu \simeq 700$~GeV,  the  $g\gamma$-channel 
is a good approximation of the EW  corrections. 
Above the threshold  its  importance  is reduced by 
the increase of the corrections of the $gg$-channel, 
which become dominant in the region $\mu \ge 2500$~GeV.

\subsection{Differential distributions}
\label{ssec:distributions}
In the previous subsection we studied the numerical impact of the next-to-leading order contributions 
to the total cross section for stop--anti-stop  production at the LHC combining 
contributions of EW and QCD origin.  Although  the current experimental studies account for  the 
higher order corrections by re-weighting the events generated at LO with the NLO(+NLL) predictions for the total rate, 
it is well known that higher-order contributions can significantly alter the shape of kinematic distributions.

\begin{figure}[t]
\begin{subfigure}[b]{0.5\textwidth}
\includegraphics[width=7.4cm,height=6.3cm]{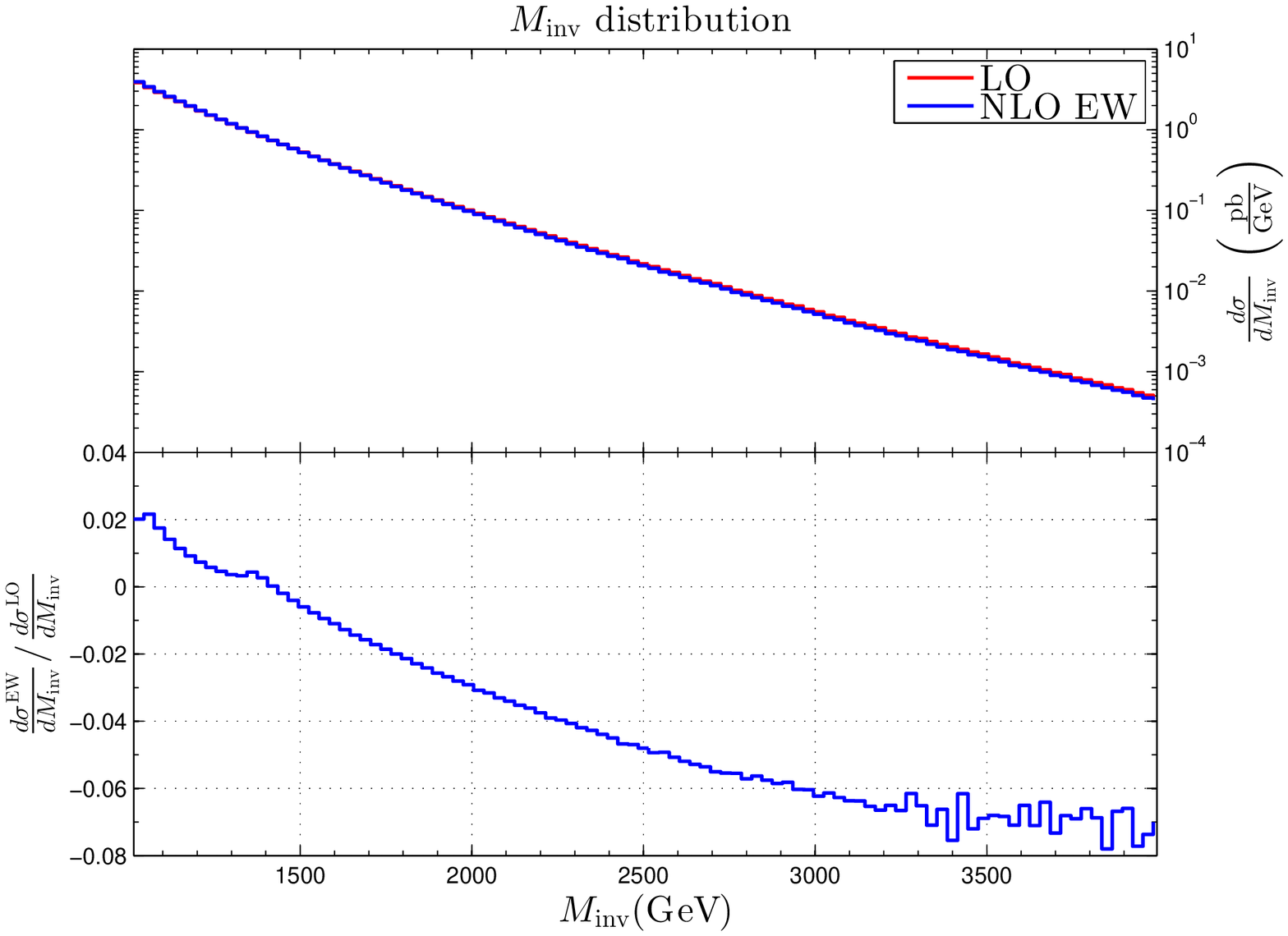}
\caption{}
\end{subfigure}
\begin{subfigure}[b]{0.5\textwidth}
\includegraphics[width=7.4cm,height=6.3cm]{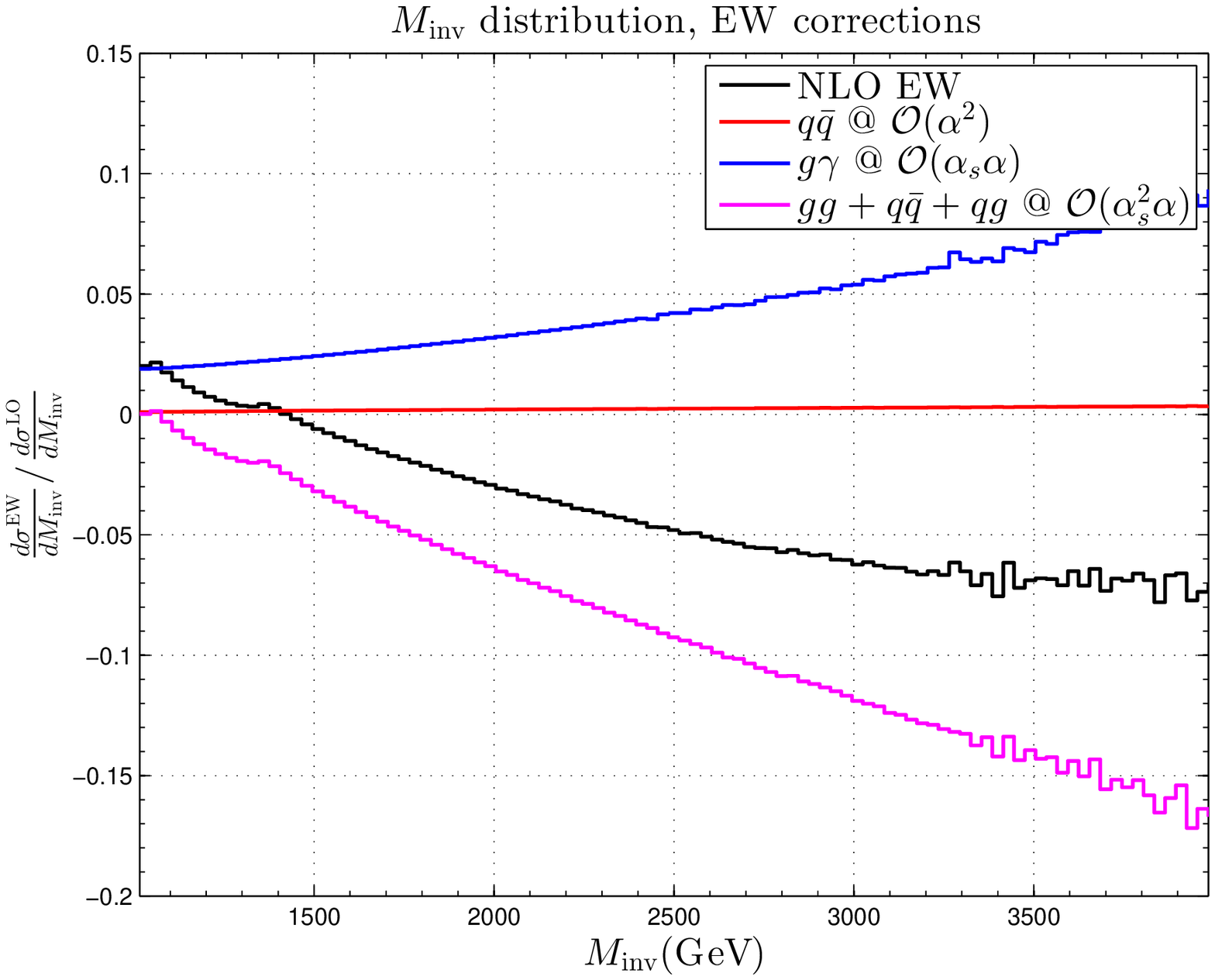}
\caption{}
\end{subfigure}
\phantom{pic}  \\
\begin{subfigure}[b]{0.5\textwidth}
\includegraphics[width=7.4cm,height=6.3cm]{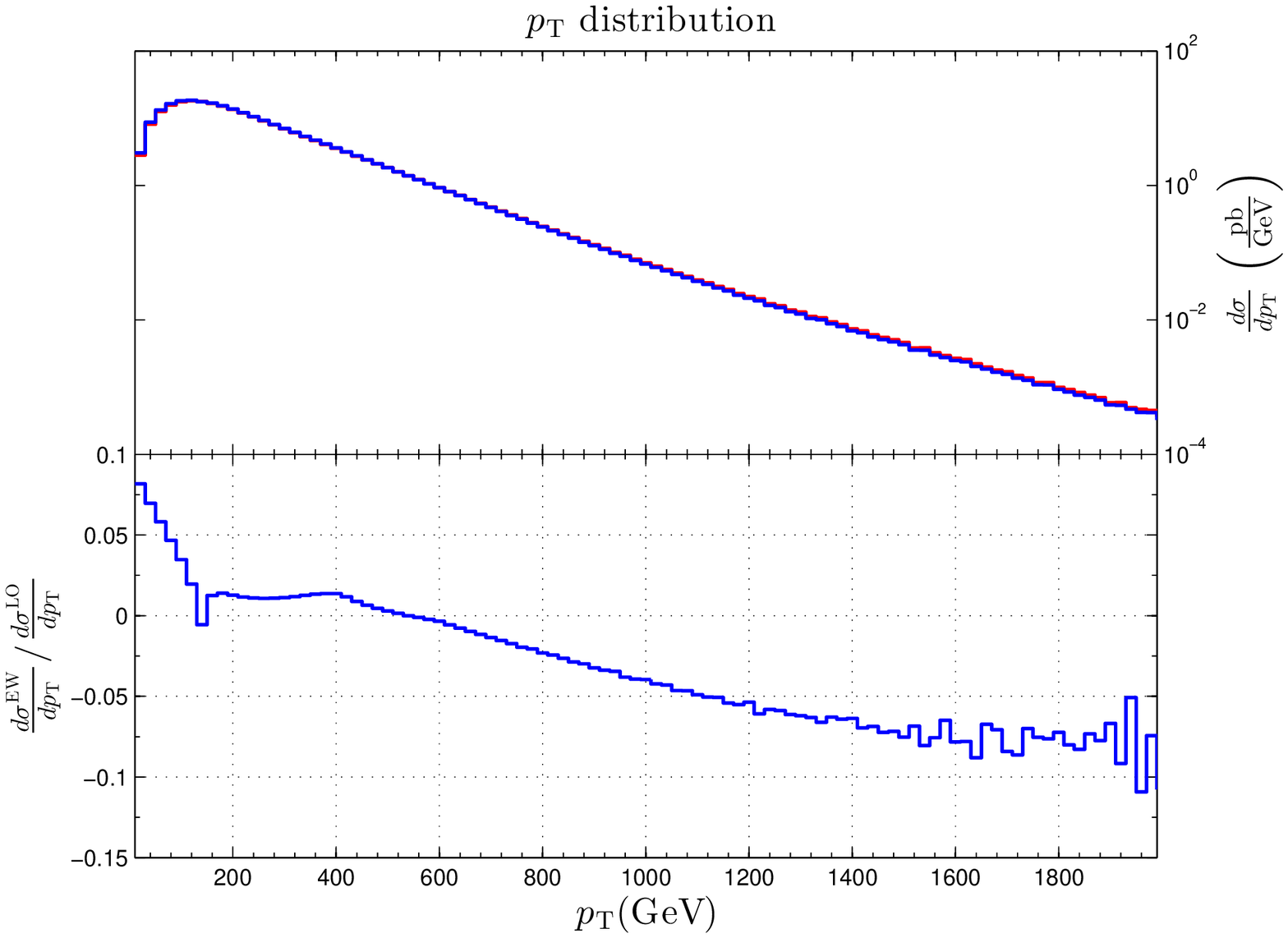}
\caption{}
\end{subfigure}
\begin{subfigure}[b]{0.5\textwidth}
\includegraphics[width=7.4cm,height=6.3cm]{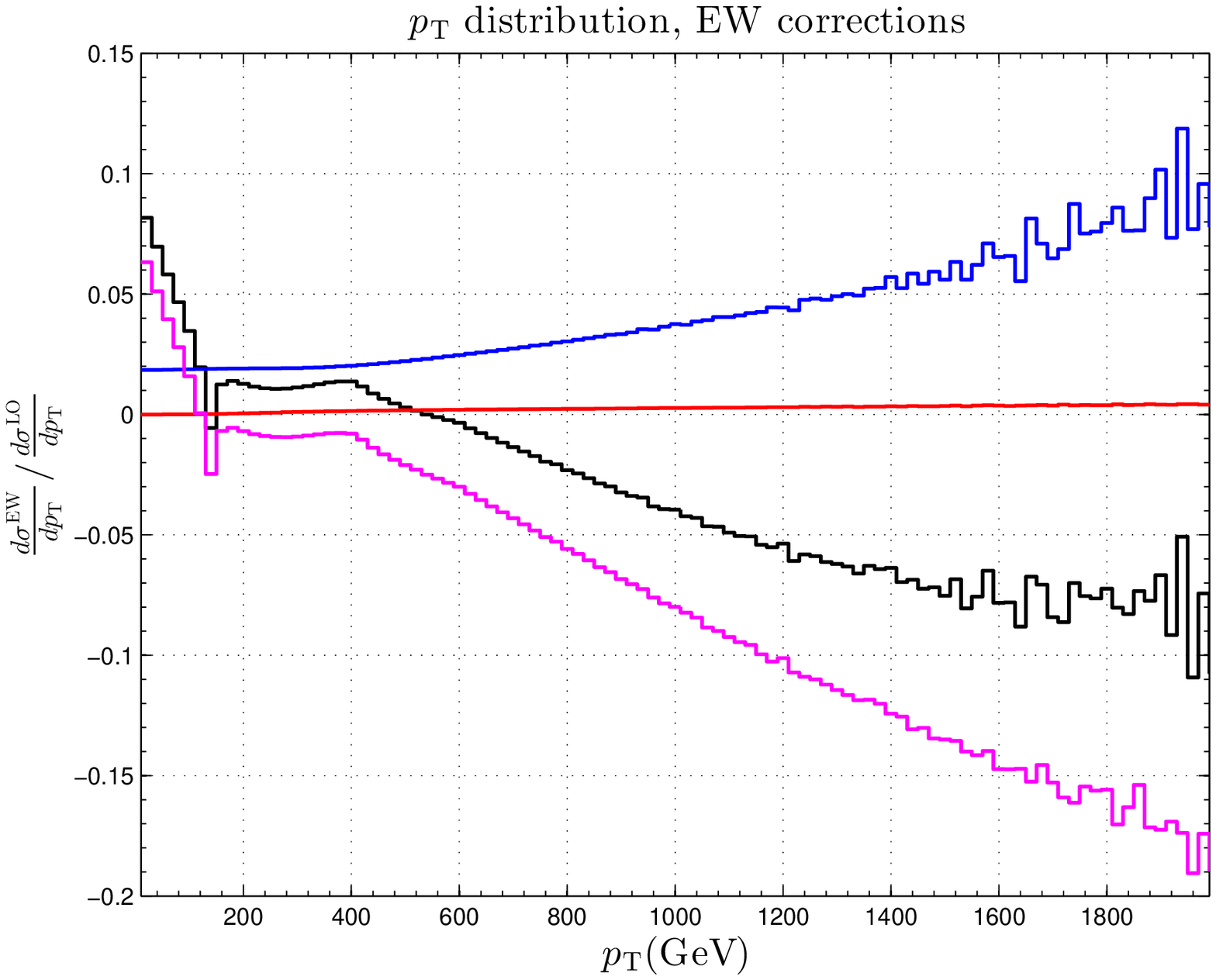}
\caption{}
\end{subfigure}
\phantom{pic}  \\
\begin{subfigure}[b]{0.5\textwidth}
\includegraphics[width=7.4cm,height=6.3cm]{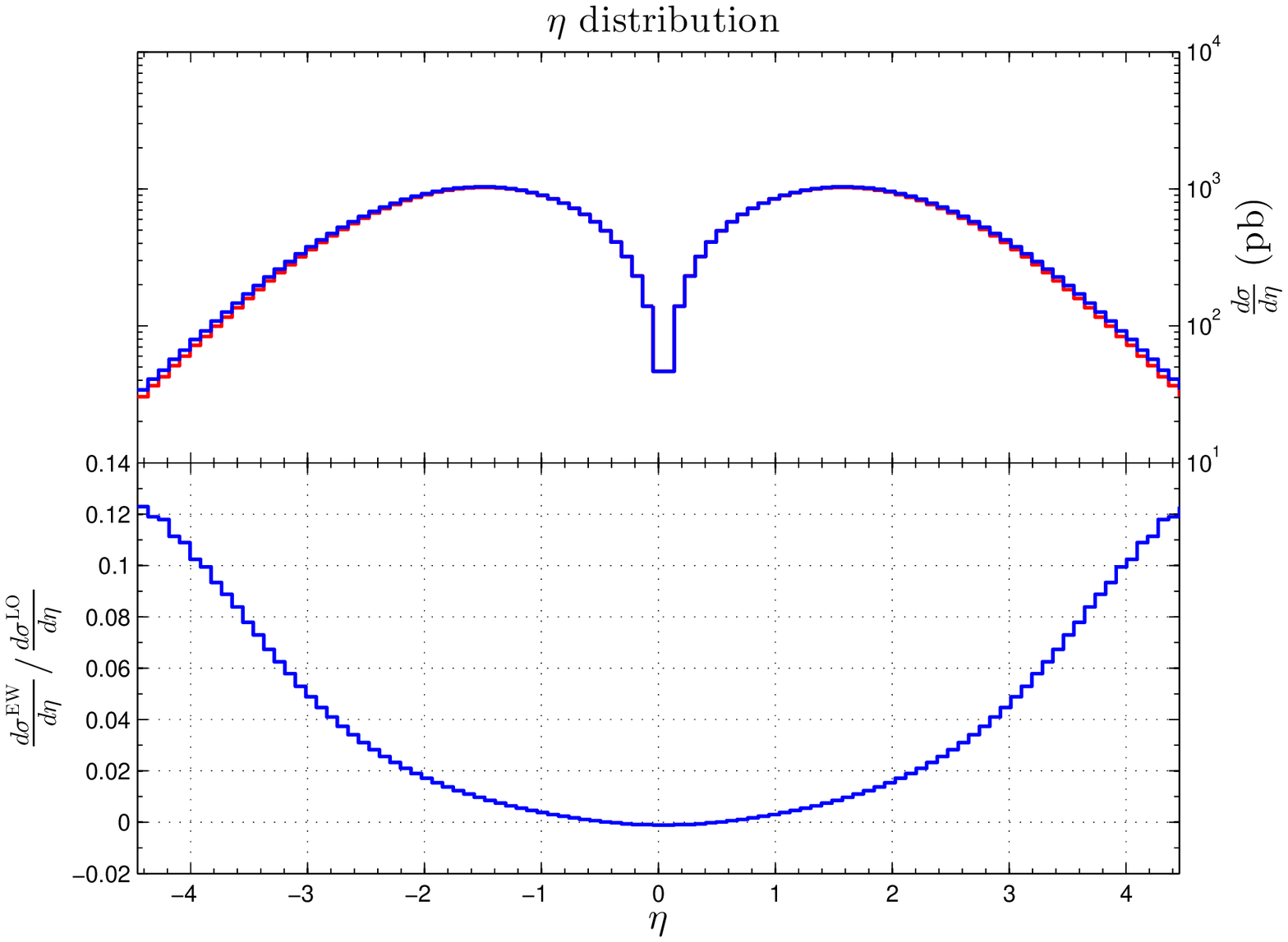}
\caption{}
\end{subfigure}
\begin{subfigure}[b]{0.5\textwidth}
\includegraphics[width=7.4cm,height=6.3cm]{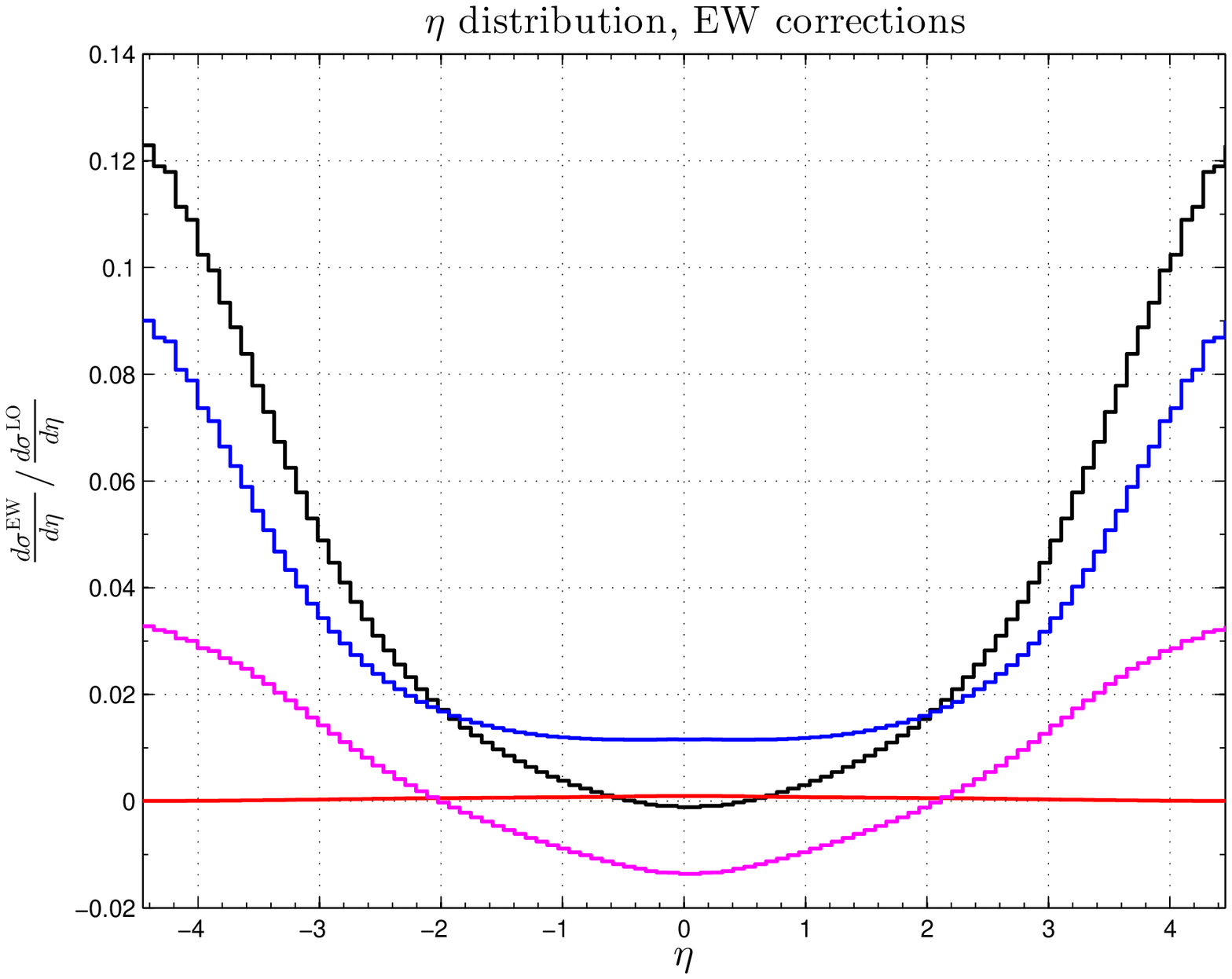}
\caption{}
\end{subfigure}
\caption[.]{Invariant mass, transverse momentum and pseudo-rapidity distribution in the light-stop scenario defined in Table~\ref{Tab:Bench}(b).}
\label{fig:Dlstop}
 \end{figure}

\begin{figure}[t]
\begin{subfigure}[b]{0.5\textwidth}
\includegraphics[width=7.4cm,height=6.3cm]{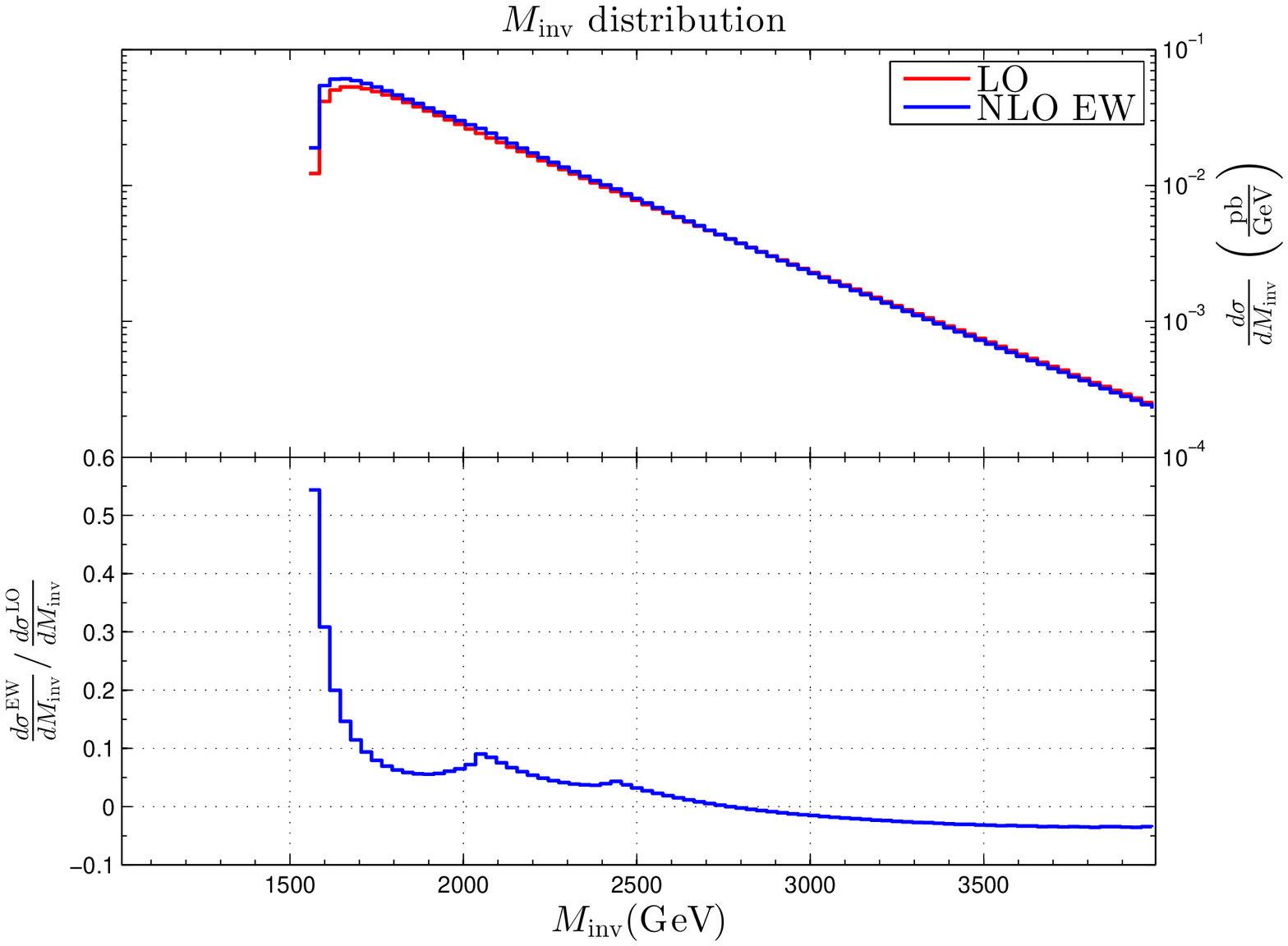}
\caption{}
\end{subfigure}
\begin{subfigure}[b]{0.5\textwidth}
\includegraphics[width=7.4cm,height=6.3cm]{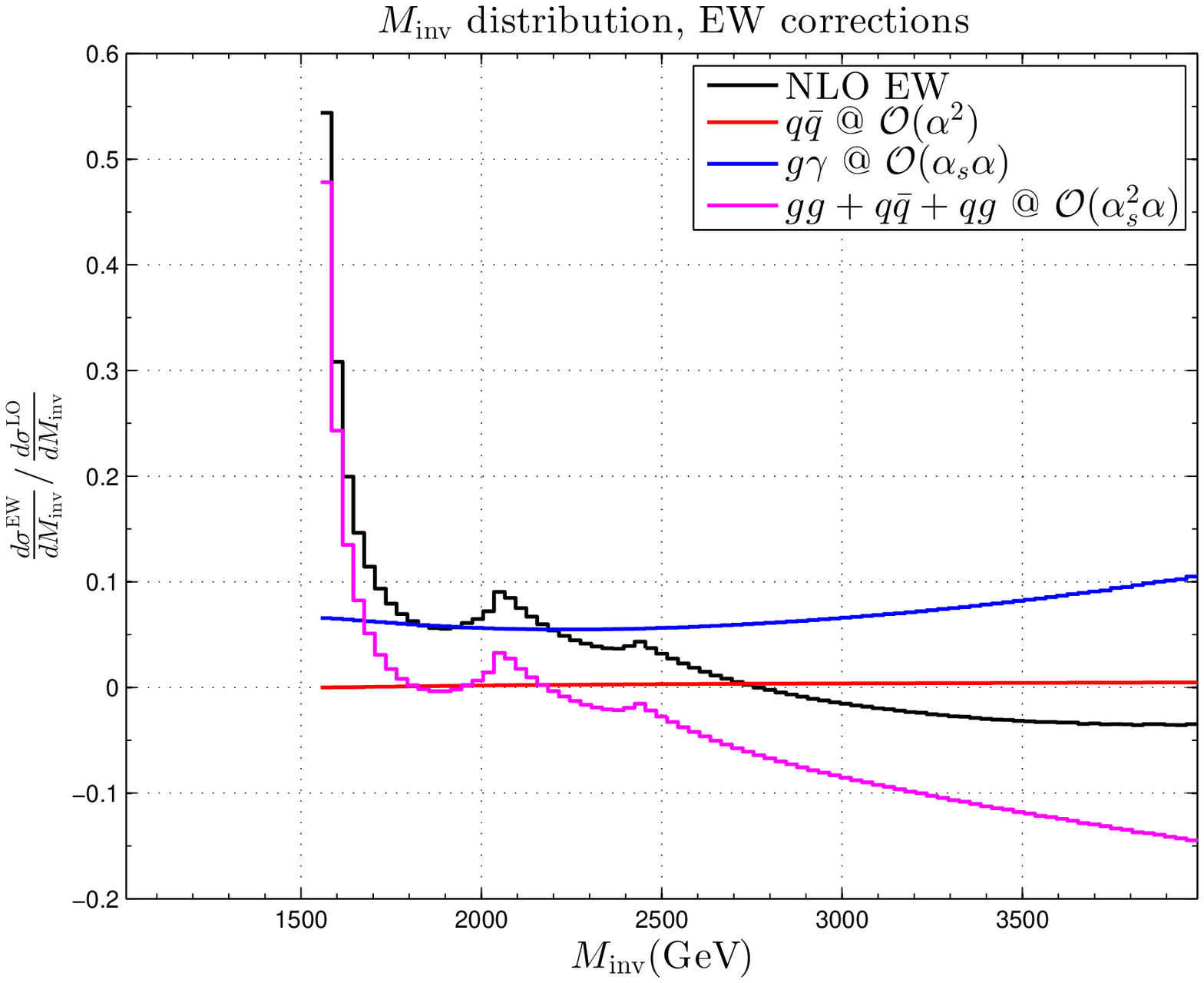}
\caption{}
\end{subfigure}
\phantom{pic}  \\
\begin{subfigure}[b]{0.5\textwidth}
\includegraphics[width=7.4cm,height=6.3cm]{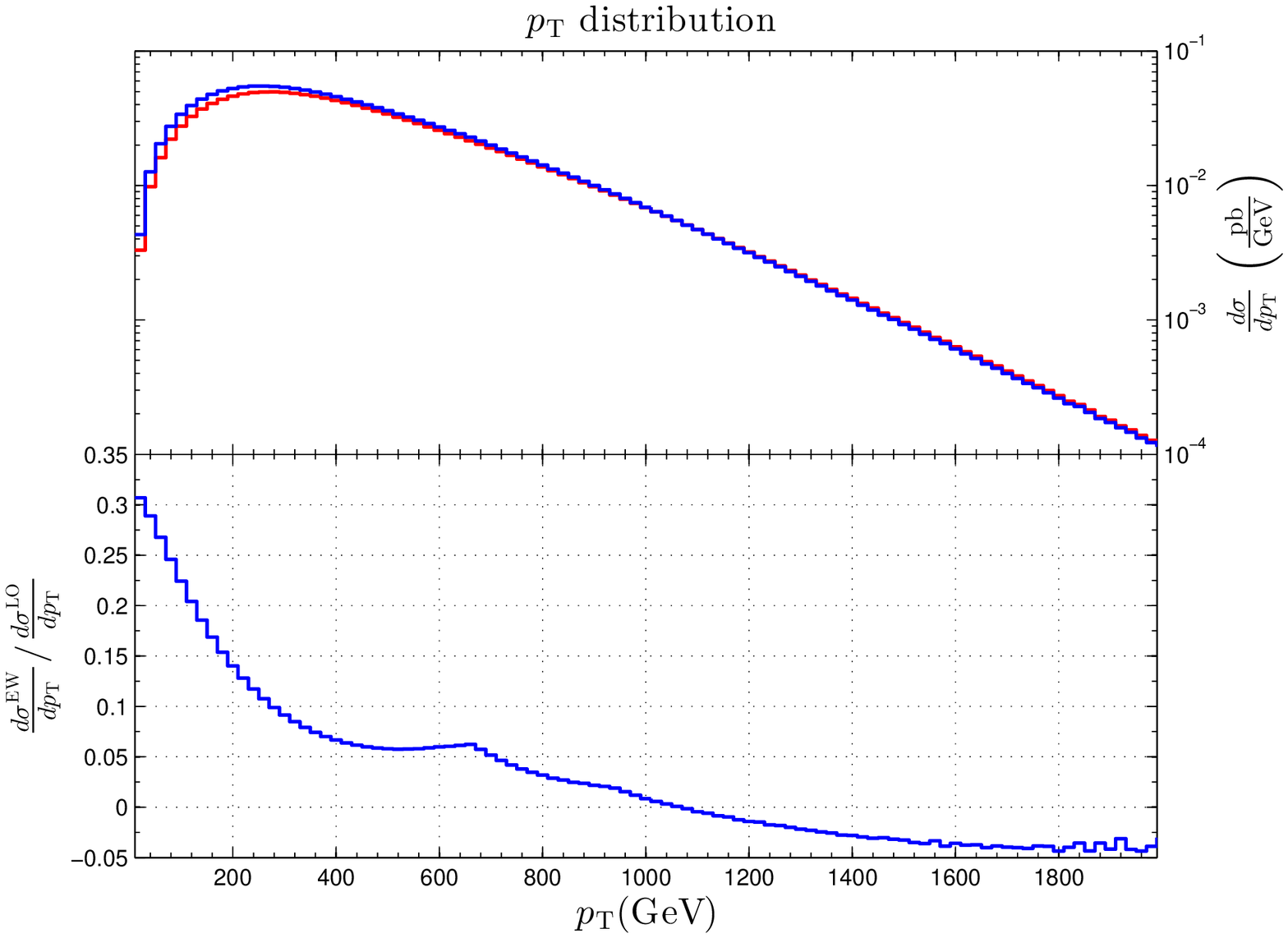}
\caption{}
\end{subfigure}
\begin{subfigure}[b]{0.5\textwidth}
\includegraphics[width=7.4cm,height=6.3cm]{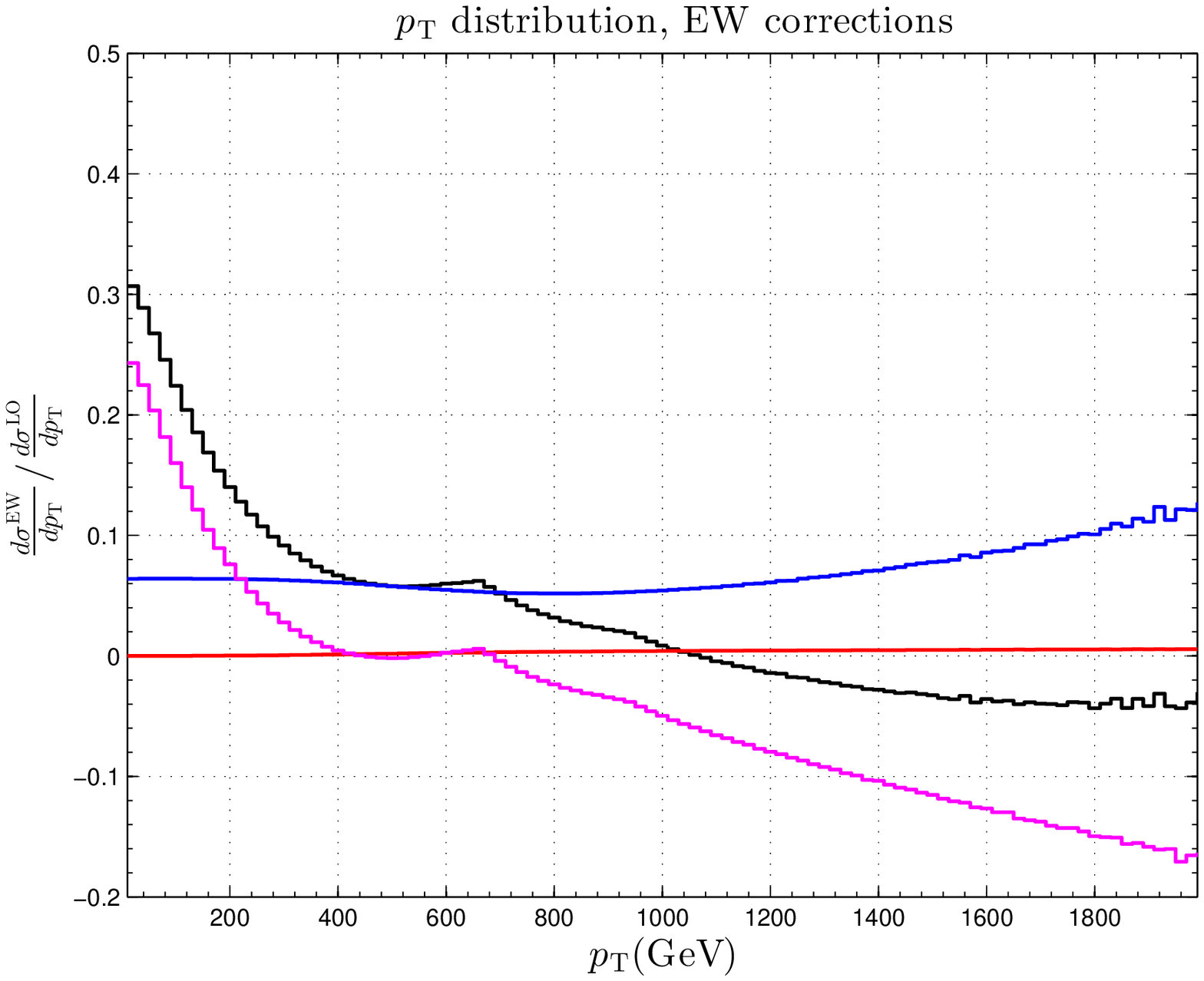}
\caption{}
\end{subfigure}
\phantom{pic}  \\
\begin{subfigure}[b]{0.5\textwidth}
\includegraphics[width=7.4cm,height=6.3cm]{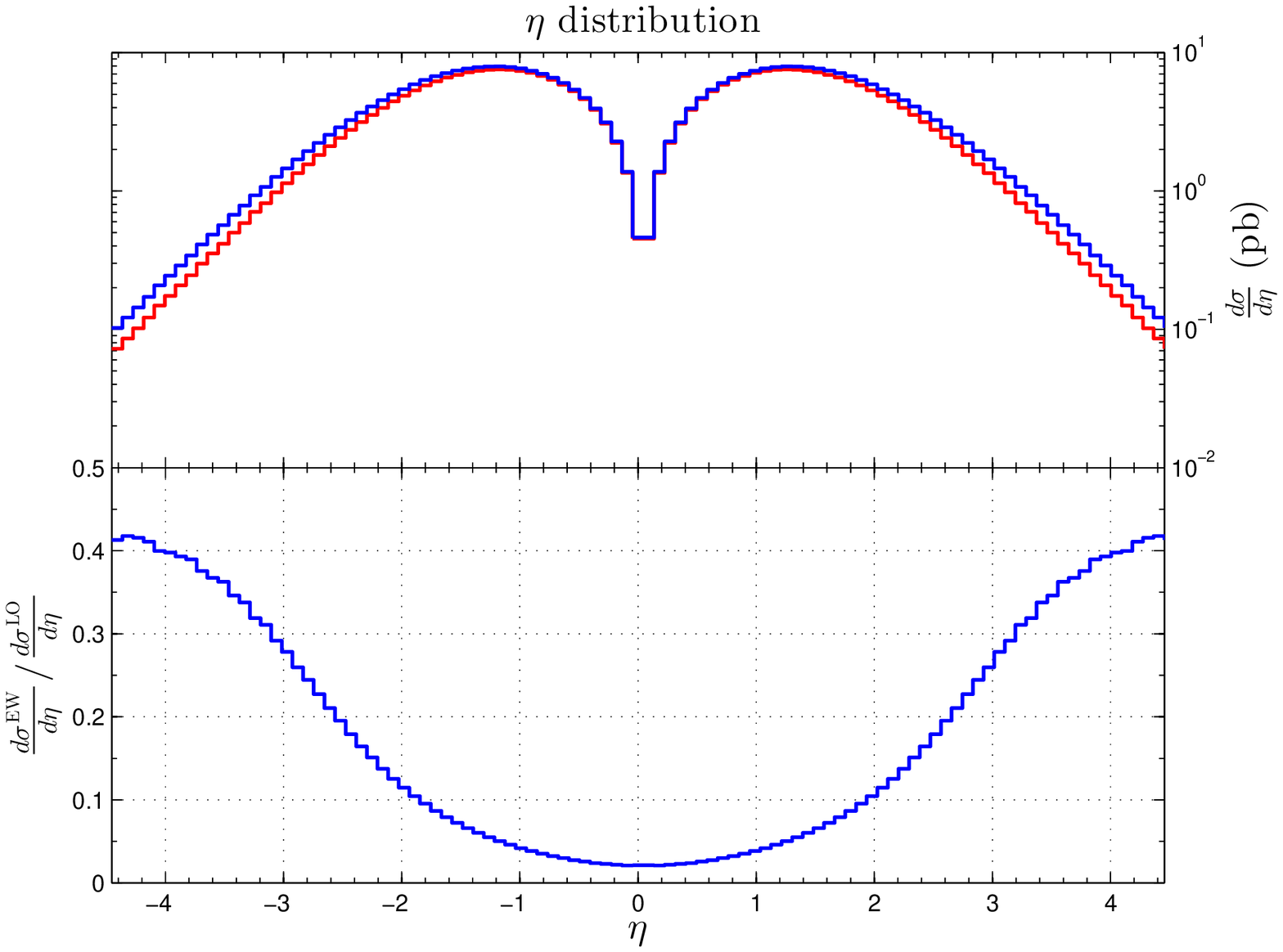}
\caption{}
\end{subfigure}
\begin{subfigure}[b]{0.5\textwidth}
\includegraphics[width=7.4cm,height=6.3cm]{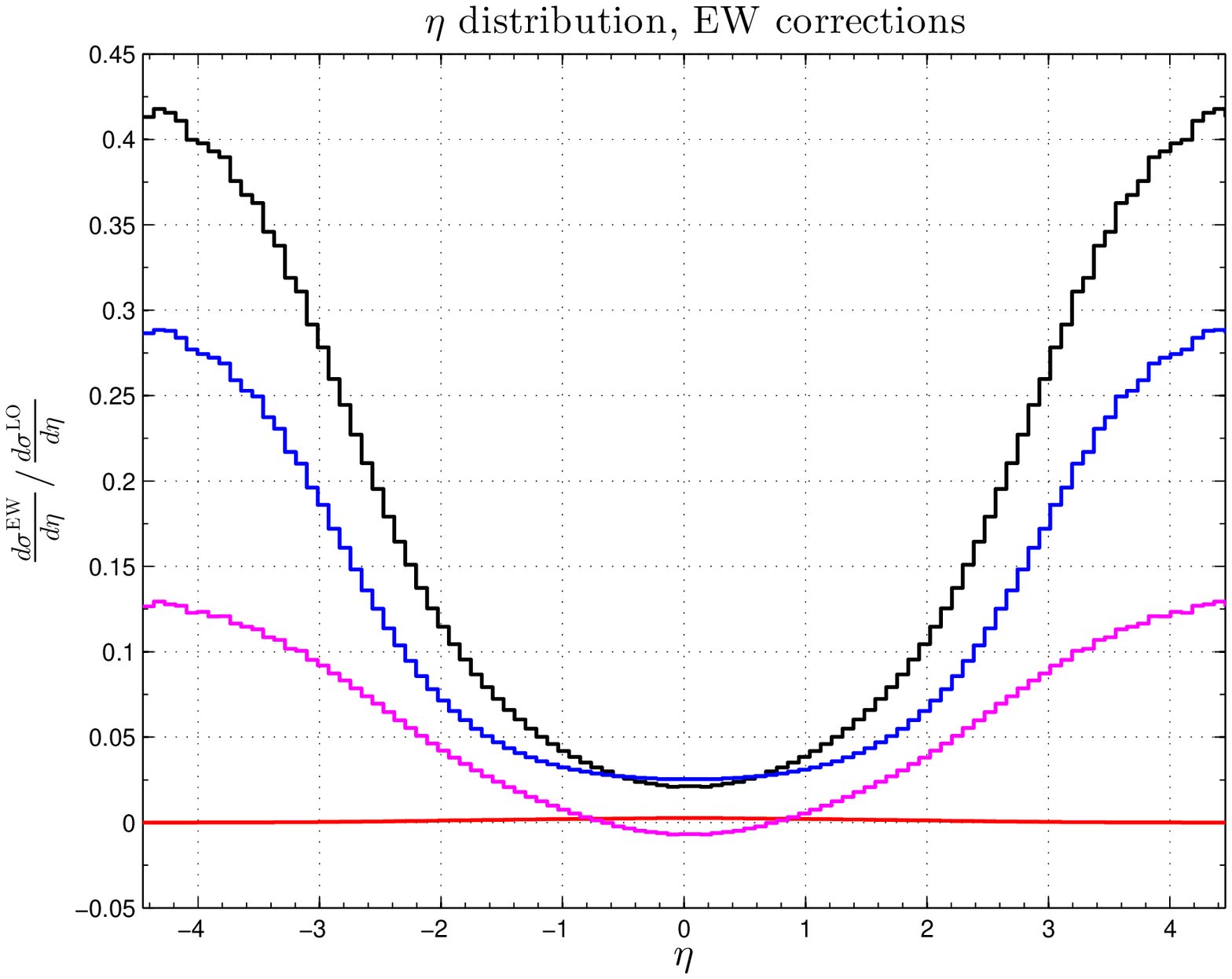}
\caption{}
\end{subfigure}
\caption[.]{Same as Fig.~\ref{fig:Dlstop} but for the tau-phobic scenario defined in Table~\ref{Tab:Bench}(d).}
\label{fig:Dtauph}
 \end{figure}

In this subsection we study the impact of the EW corrections on differential distributions with respect
to the  invariant mass, the transverse momentum and the pseudo-rapidity,  defined as
 \begin{align}
 M_{\rm inv} &\equiv \sqrt{\left ( p_{\tilde t_1^{\phantom{\ast}}} + p_{\tilde t_1^\ast}\right)^2 }\, , \hspace{0.4cm}  & 
 p_T  & \equiv \max\left ( p_{T \, \tilde t_1^{\phantom{\ast}}} \, , \; p_{T\, \tilde t_1^\ast} \right)\, ,  \nonumber \hspace{0.4cm} 
 \eta &\equiv  \left \{ \begin{array}{ll}
 \eta_{\tilde t_1^{\phantom{\ast}}}  \quad \mbox{if} \quad    |\eta_{\tilde t_1^{\phantom{\ast}}} | \ge |\eta_{ \tilde t_1^\ast}|   \\ [1.0ex]
\eta_{ \tilde t_1^\ast}  \quad \mbox{if} \quad    |\eta_{\tilde t_1^{\phantom{\ast}}} | < |\eta_{ \tilde t_1^\ast}|   \end{array}  \right. \, , & & 
%
 \end{align}
respectively. The quantities $p_j$, $p_{T\, j}$ and $\eta_j$ are the four-momentum, the transverse momentum and 
the pseudo-rapidity  of the particle $j$, respectively.  We have considered the six SUSY scenarios 
described in Table~\ref{Tab:Bench}, but here we present results for the light-stop and for the tau-phobic scenario only, 
since the other scenarios exhibit similar features of the latter one. The results  
are collected in Fig.~\ref{fig:Dlstop}  and in Fig.~\ref{fig:Dtauph}.  
In the upper plots,  (a) and (b), we show the  LO  and NLO invariant mass 
distribution [panel (a)],  and the  yield of the 
$\mathcal{O}(\alpha^2)$, $\mathcal{O}(\alpha_s\alpha)$ and
$\mathcal{O}(\alpha_s^2\alpha)$ corrections [panel (b)].  
Panels (c), (d) and panels (e), (f) show the same information for the transverse momentum 
and the pseudo-rapidity distributions, respectively.

\paragraph{Invariant mass distribution}  In scenarios with a heavy stop 
exemplified by the tau-phobic scenario in Fig.~\ref{fig:Dtauph},
the relative impact of the  EW corrections to the invariant mass distributions decreases as  $M_{\rm inv}$ increases. 
As can be inferred  from Fig.~\ref{fig:Dtauph}(a) the EW  corrections are positive and sizable 
in the small $M_{\rm inv}$ region while in the large $M_{\rm inv}$ region the EW corrections become negative. 
In this regime the relative impact of the  corrections is almost flat and their size is small. 
%
The behavior of the EW corrections can be explained looking at the contributions of the different EW corrections collected in Fig.~\ref{fig:Dtauph}(b).  
The $\mathcal{O}(\alpha_s\alpha)$  and the $\mathcal{O}(\alpha^2_s\alpha)$ corrections  are the numerically
dominant ones. In the small $M_{\rm inv}$ region they are both positive and sum up enhancing the EW corrections. In the large $M_{\rm  inv}$ region, instead,  
the  positive $\mathcal{O}(\alpha_s\alpha)$ corrections partially compensate the large negative contributions of  the $\mathcal{O}(\alpha^2_s\alpha)$ corrections.

Also in the light stop-scenario the relative yield  of the EW corrections  decreases as  $M_{\rm inv}$ increases but, immediately above the production threshold, they become negative, 
see Fig.~\ref{fig:Dlstop}(a). As one can see in Fig.~\ref{fig:Dlstop}(b), 
this  behavior is again related to mutual cancellation between the $\mathcal{O}(\alpha_s\alpha)$ and the 
$\mathcal{O}(\alpha^2_s\alpha)$ corrections. 
In the entire $M_{\rm inv}$ region the $\mathcal{O}(\alpha^2)$ corrections are irrelevant for all considered scenarios.

\paragraph{Transverse momentum distribution}  The EW corrections to the transverse momentum distribution 
exhibit features similar to those for the invariant mass distribution. 

In the low $p_{\rm T}$ region the EW corrections are positive and sizable in all scenarios but the  light-stop one, i.e. their relative yield is above $10\%$. For high values of  $p_{\rm T}$, instead,  the absolute value of the EW corrections
decreases and  they become negative, see Fig.~\ref{fig:Dtauph}(c).  
As shown in Fig.~\ref{fig:Dtauph}(d), this behavior  is originated by the same interplay between 
the  $\mathcal{O}(\alpha_s\alpha)$ and $\mathcal{O}(\alpha^2_s\alpha)$ corrections 
present in the  $M_{\rm inv}$ distribution. 

As shown in Fig.~\ref{fig:Dlstop}(c),  in the light-stop scenario the relative impact of the EW corrections is below $10\%$. 
This feature can be understood by looking at the behavior of the various EW contributions, Fig.~\ref{fig:Dlstop}(d). In the low $p_{\rm T}$ region the $\mathcal{O}(\alpha_s\alpha)$ and $\mathcal{O}(\alpha^2_s\alpha)$ contributions are both positive but small, 
i.e. below $5 \%$, while in the high $p_{\rm T}$ region  the two contributions cancel against each other.

\paragraph{ Pseudo-rapidity distribution}  As illustrated in Fig.~\ref{fig:Dtauph}(e), in all heavy-stop scenarios the EW corrections are important for large pseudo-rapidities, 
i.e. they are above $10 \%$ for $|\eta| \ge 2$.
In the  low pseudo-rapidity region the EW corrections are positive but small, and almost negligible in the central region. 
The  behavior of the EW corrections can be understood by looking at   Fig.~\ref{fig:Dtauph}(f). The  large and positive $\mathcal{O}(\alpha_s\alpha)$ corrections originate from the $g \gamma$ partonic process. This process proceeds via a $t$- and $u$-channel tree-level diagram, thus it produces preferably final particles in the forward region. In this region its large contribution is further enhanced by the positive $\mathcal{O}(\alpha^2_s\alpha)$ corrections, cfr. Fig.~\ref{fig:Dtauph}(f). In the low pseudo-rapidity region, instead, the positive contributions of   $\mathcal{O}(\alpha_s\alpha)$  are canceled by negative corrections of $\mathcal{O}(\alpha^2_s\alpha)$.

As shown in  Fig.~\ref{fig:Dlstop}(e),  in the light-stop scenario the EW corrections are generally small, i.e. below $10\%$ for $|\eta| \le 4$.   
  
\bigskip
\noindent
The importance of the EW  corrections considerably depends on the kinematics of the produced $\stop\stop^\ast$. Thus, the yield on experimental event rates can be substantially affected by the application of kinematical 
cuts. Furthermore, NLO QCD corrections to the decay of the produced top-squarks may further alter significantly the shape of the kinematical 
distributions~\cite{Boughezal:2012zb,Boughezal:2013pja}. Therefore, such NLO corrections to the decay should also be investigated systematically for the EW corrections, including EW corrections also in the decay.



%
%
%
%
%
%
%
%
%
%
%


\section{Conclusions}
\label{sec:conclusions}
In this paper we have presented the first  phenomenological study for $\stop\stop^*$ production at the LHC  
including both the complete NLO QCD and NLO EW contributions. 
We have used the most recent PDF sets including QED effects, 
presenting a thorough  study
of parameter regions compatible with a SM-like Higgs boson  observed at the LHC. 
The allowed MSSM  regions 
are characterized by a rather heavy  $\stop$   and/or by a large mixing in the stop sector, i.e. large $X_t$.

Our analysis has shown 
that NLO EW contributions to $\stop\stop^*$ production 
are not always negligible even on an inclusive level;
they can be sizable, as large as $15-20\%$ of the LO cross section, 
particularly in the parameter regions with heavy stops. This 
is mainly due to the  contribution of the $g \gamma$-channel which increases with the mass of the produced top-squarks.
This contribution can be easily included in the MSSM predictions for experimental investigations.  
Moreover,  it  could reduce the residual  theoretical uncertainty related to the EW corrections below
$5\%$  of the LO cross section, at least in the parameter configurations where $\mu$ is not too large.  
The dependence on the remaining parameters of the model is found to be rather weak.

The presented study was performed using the Monte Carlo integrator  {\tt SusyHell}. 
This code includes the NLO EW corrections to all squark and gluino production channels and it will be made public in the future. It will allow for detailed phenomenological studies of the EW corrections to colored SUSY 
particle production, on an inclusive and also a fully differential level. Using this tool also the electroweak contributions should eventually be combined with higher order  corrections to the decay processes.

\begin{acknowledgments}
We thank Oscar St\r{a}l for useful discussions on the code  {\tt HiggsSignals}. 
JML was supported by the European Commission through the ``LHCPhenoNet'' Initial Training Network PITN-GA-2010-264564.
\end{acknowledgments}

\bibliographystyle{JHEP}

\bibliography{references}

\end{document}